\documentclass[review,sort&compress,10pt]{elsarticle}
\usepackage{amsmath}
\usepackage{rotating}
\usepackage{graphicx}
\usepackage{pstricks}
\usepackage{color}

\usepackage{lineno,hyperref}

\newcommand{\ba}{\begin{eqnarray}}
\newcommand{\ea}{\end{eqnarray}}

\journal{Progress in Particle and Nuclear Physics}

\begin{document}

\begin{frontmatter}

\title{{Cluster structure of light nuclei}}

\author[address1]{R. Bijker\corref{mycorrespondingauthor}}
\cortext[mycorrespondingauthor]{Corresponding author}
\ead{bijker@nucleares.unam.mx}
\author[address2]{F. Iachello} 
\ead{francesco.iachello@yale.edu}

\address[address1]{Instituto de Ciencias Nucleares, Universidad Nacional Aut\'onoma de M\'exico, 04510 Ciudad de M\'exico, M\'exico}
\address[address2]{Center for Theoretical Physics, Sloane Laboratory, Yale University, New Haven, Connecticut 06520-8120, U.S.A.}

\begin{abstract}
We review recent studies of the cluster structure of light nuclei within the framework of the algebraic cluster model (ACM) for nuclei composed of $k$ $\alpha$-particles and within the framework of the cluster shell model (CSM) for nuclei composed of $k$ $\alpha$-particles plus $x$ additional nucleons. The calculations, based on symmetry considerations and thus for the most part given in analytic form, are compared with experiments in light cluster nuclei. The comparison shows evidence for $Z_2$, $D_{3h}$ and $T_d$ symmetry in the even-even nuclei $^{8}$Be ($k=2$), $^{12}$C ($k=3$) and $^{16}$O ($k=4$), respectively, and for the associated double groups $Z'_2$ and $D'_{3h}$ in the odd nuclei $^{9}$Be, $^{9}$B  ($k=2$, $x=1$) and $^{13}$C ($k=3$, $x=1$), respectively.
\end{abstract}

\begin{keyword}
Cluster models \sep Alpha-cluster nuclei \sep Symmetries \sep Algebraic models 
\end{keyword}

\end{frontmatter}

\section{Introduction}

The cluster structure of light nuclei has a long history dating back to the 1930's with early studies of $\alpha$-cluster models by Wheeler \cite{Wheeler:1937zza} and Hafstad and Teller \cite{PhysRev.54.681}, followed by Dennison \cite{Dennison:1954zz} and Kameny \cite{Kameny:1956zz}. Soon afterwards, the connection between the cluster model and the shell-model was investigated by Wildermuth and Kallenopoulos \cite{WILDERMUTH1958150}. In 1965, Brink \cite{Brink,Brink:1970ufk} suggested specific geometric configurations for nuclei composed of $k$ $\alpha$-particles, here referred as $k\alpha$ nuclei. In particular, the suggested configurations of the ground state were, for $k=2$ a dumbbell configuration with $Z_2$ symmetry ($^{8}$Be), for $k=3$ an equilateral triangle with $D_{3h}$ symmetry ($^{12}$C) and for $k=4$ a tetrahedron with $T_d$ symmetry ($^{16}$O), as shown in Fig.~\ref{Fig1}. Brink's suggestion stimulated a considerable amount of work in an attempt to derive cluster properties from the shell model, especially by the Japanese school \cite{10.1143/PTP.51.1266,10.1143/PTP.53.447,10.1143/PTPS.68.1,10.1143/PTPS.68.29} and from mean field theories \cite{Eichler:1970nqw}. Also, the cluster structure of specific nuclei was extensively investigated, as for example in $^{16}$O \cite{Robson:1978vh,ROBSON1982257}, and Brink's model was applied to a wide range of cluster nuclei from $^{12}$C to $^{44}$Ti in \cite{Rae_1993,Rae_1994}. A review of cluster models up to 2006 can be found in \cite{VONOERTZEN200643}, and more recent ones in \cite{Schuck_2016} and \cite{RevModPhys.90.035004}.

In recent years, there has been considerable renewed interest in the structure of $\alpha$-cluster nuclei, especially for the nucleus $^{12}$C \cite{Freer:2014qoa}. The observation of new rotational states built on the ground state \cite{Freer:2007zz,Kirsebom:2010zz,Freer:2011zza,Marin-Lambarri:2014zxa} and the Hoyle state  \cite{Itoh:2011zz,Freer:2012se,Zimmerman:2013cxa} has stimulated a large effort to understand the structure of $^{12}$C ranging from studies based on Antisymmetric Molecular Dynamics (AMD) \cite{10.1143/PTP.117.655}, Fermion Molecular Dynamics (FMD) \cite{Chernykh:2007zz}, BEC-like cluster model \cite{Funaki:2009fc}, ab initio no-core shell model \cite{PhysRevLett.84.5728,Roth:2011ar,Maris_2012}, lattice EFT \cite{Epelbaum:2011md,Epelbaum:2012qn,PhysRevLett.112.102501}, no-core symplectic model \cite{Dreyfuss:2012us} and the Algebraic Cluster Model (ACM) \cite{Bijker:2000fw,Bijker:2002ac,Bijker:2014tka,Bijker:2016bpb}. In the first part of this paper, we review the ACM as applied to $k\alpha$ nuclei with $k=2$, $3$, $4$. 

An important question is the extent to which cluster structures survive the addition of nucleons (protons and neutrons). We refer to nuclei composed of $k$ $\alpha$-particles plus $x$ nucleons as $k\alpha+x$  nuclei. This question has also been addressed in the past, especially in the case of the Be isotopes seen as $^{8}\mbox{Be}+x$ nucleons, with a variety of methods \cite{KUNZ1960275,PhysRev.128.1343,10.1143/PTP.30.585,Neudatchin1969,Golovanova1971,10.1143/PTP.49.800} culminating, in the 1970's, in the extensive work of Okabe, Abe and Tanaka \cite{10.1143/PTP.57.866,10.1143/PTP.61.1049} using the Linear Combination of Atomic Orbitals (LCAO) method and its generalizations. 
In recent years, FMD \cite{Feldmeier:2000cn,Roth:2004ua,Neff:2003ib,Neff:2005pvm} and AMD  \cite{10.1143/PTPS.142.205,KANADAENYO2003497,Kanada-Enyo:2003fhn} calculations have provided very detailed and accurate microscopic descriptions of the Be isotopes with large overlap with the Brink model \cite{Brink:1970ufk}. In another seminal development, Von Oertzen \cite{vonOertzen:1970ecu,VONOERTZEN19751,IMANISHI198729,Oertzen} has discussed the structure of $^9$B, $^9$Be, $^{10}$B and $^{10}$Be in a two-center shell model, in which these isotopes are seen as $^{8}$Be plus neutrons and protons. In very recent years, a description of $k\alpha+x$ nuclei has been suggested in terms of the Cluster Shell Model (CSM) \cite{DellaRocca:2017qkx,DellaRocca:2018mrt} which builds on the algebraic description of $k\alpha$ nuclei in terms of the ACM \cite{Bijker_2016}. In the second part of this paper, we review the CSM as applied to $k\alpha+x$ nuclei with $k=2$, $3$, $4$ and $x=1$.

We note in this connection that the Cluster Shell Model (CSM) takes fully into account the Pauli principle, as discussed in Section~\ref{sec8}. Individual nucleons are placed in the single-particle orbitals described in Section~\ref{CSM} according to the Pauli principle. The treatment of the Pauli principle in CSM is thus identical to that in the Nilsson model \cite{Nilsson:1955fn}, in the Brink model \cite{Brink} and the LCAO method \cite{Golovanova1971,10.1143/PTP.49.800}. The question of the Pauli principle in the Brink model is also discussed in \cite{Rae_1993,Rae_1994} and in the molecular model with $Z_2$ symmetry dumbbell in \cite{Golovanova1971,10.1143/PTP.49.800}. We emphasize here the point that in the CSM one is able to take into account the Pauli principle not only for the dumbbell configuration (two-center shell model) but also for the triangular configuration (three-center shell model) and for the tetrahedral configuraton (four-center shell model). The latter two have not been discussed before within the context of nuclear physics. 

The figures of Section~\ref{CSM} also show the occurrence of ``magic'' numbers for protons and neutrons at 4 for the dumbbell configuration, at 6 for the triangular configuration and at 8 for the tetrahedral configuration. The stability of the $k\alpha$ nuclei is thus inherent in the approach described here and justifies the ACM. A remaining question, however, is to what extent the reduction from the spherical shell model to the cluster model induces two- or higher order terms in the effective $\alpha$-$\alpha$ interaction. The algebraic approach when written in terms of coordinates and momenta corresponds to an effective $\alpha$-$\alpha$ interaction of the Morse type \cite{Morse_1929}, as discussed in \cite{IachelloLevine}. The Morse type interaction has a ``hard core'' which effectively mimics the Pauli principle, since it does not allow for two $\alpha$-particles to get close and overlap strongly. Also, as shown in the figures in Section~\ref{ACM} where the matter density is plotted, the situation encountered in light nuclei is that in which the $\alpha$-particles are in a close-packing situation, that is they just touch but do not overlap. Taking into account effectively the Pauli principle in the $\alpha$-$\alpha$ interaction within the framework of the algebraic method ACM is therefore a good approximation to the full microscopic approach in terms of nucleons.

The aim of this paper is to present all formulas and calculations to compare with experimental data and, as a result, to show evidence for the occurrence of the geometric symmetries $Z_2$, $D_{3h}$, $T_d$ and $Z'_2$, $D'_{3h}$ in the structure of $k\alpha$ and $k\alpha+1$ nuclei. 

The ACM model reviewed in the first part of this article is purely phenomenological, as the collective model of Bohr and Mottelson and does not attempt a microscopic description in terms of nucleon-nucleon interactions, but rather exploits symmetry considerations to derive most of the observables in explicit analytic form that can be compared with experiment. Conversely, the   CSM reviewed in the second part is a microscopic model that makes use of a symmetry-adapted basis, the cluster basis with $Z_2$, $D_{3h}$ and $T_d$ symmetry instead of the spherical basis. 

One important question is the extent to which these symmetries emerge from microscopic calculations. Extensive calculations have been done for $^{8}$Be and $^{9}$Be ($Z_2$ and $Z'_2$ symmetry) within the framework of microscopic approaches mentioned in the paragraphs above. For these nuclei, microscopic approaches appear to produce cluster features correctly, although effective charges still need to be introduced in the analysis of electromagnetic transition rates in shell model based calculations 
\cite{PhysRevLett.84.5728}. A detailed comparison between the algebraic approach and microscopic approaches for $^{8}$Be and $^{9}$Be is given in \cite{DellaRocca:2018mrt}. Also for $^{12}$C and $^{13}$C ($D_{3h}$ and $D'_{3h}$ symmetry) extensive calculations exist, especially for $^{12}$C. The AMD and FDM microscopic calculations produce results in quantitative agreement with the symmetry. Also, lattice EFT produces results in $^{12}$C and $^{16}$O which support $D_{3h}$ and $T_d$ symmetry. It would be of great interest to understand whether the cluster structure of $^{12}$C and $^{13}$C emerges from {\it ab initio} calculations, such as the no-core shell model (NCCI) \cite{PhysRevLett.84.5728,Roth:2011ar,Maris_2012} for which calculations are planned. The results presented here, based on purely symmetry concepts, provide benchmarks for microscopic studies of cluster structure of light nuclei.

\section{The algebraic cluster model}
\label{ACM}

The algebraic cluster model is based on the algebraic theory of molecules introduced in 1981 \cite{IACHELLO1981581} and reviewed in \cite{IachelloLevine}. It amounts to a bosonic quantization of the Jacobi variables according to the general quantization scheme for problems with $\nu$ degrees of freedom in terms of the Lie algebra $U(\nu+1)$ \cite{kamran1994lie}. For $k\alpha$ structures, the number of degrees of freedom, after removing the center-of-mass motion, is $\nu=3(k-1)$, leading to the Lie algebra of $U(3k-2)$.

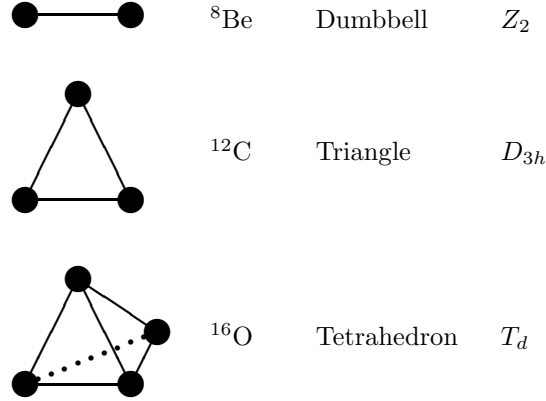
\begin{figure}
\centering
\setlength{\unitlength}{1pt} 
\begin{picture}(220,180)(0,0)
\thicklines
\put(20, 20) {\circle*{10}} 
\put(60, 20) {\circle*{10}}
\put(70, 40) {\circle*{10}}
\put(40, 60) {\circle*{10}}
\put(20, 20) {\line( 1,0){40}}
\put(20, 20) {\line( 1,2){20}}
\put(60, 20) {\line(-1,2){20}}
\put(60, 20) {\line( 1,2){10}}
\put(70, 40) {\line(-3,2){30}}
\multiput(20, 20)(5,2){11}{\circle*{2}}
\put( 90, 35) {$^{16}$O}
\put(130, 35) {Tetrahedron}
\put(200, 35) {$T_d$}

\put(20, 90) {\circle*{10}} 
\put(60, 90) {\circle*{10}}
\put(40,130) {\circle*{10}}
\put(20, 90) {\line( 1,0){40}}
\put(20, 90) {\line( 1,2){20}}
\put(60, 90) {\line(-1,2){20}}
\put( 90,105) {$^{12}$C}
\put(130,105) {Triangle}
\put(200,105) {$D_{3h}$}

\put(20,160) {\circle*{10}} 
\put(60,160) {\circle*{10}}
\put(20,160) {\line( 1,0){40}}
\put( 90,155) {$^{8}$Be}
\put(130,155) {Dumbbell}
\put(200,155) {$Z_2$}
\end{picture}
\caption[Cluster configurations]
{Cluster configurations for $k=2$, $3$ and $4$ $\alpha$-particles.}
\label{Fig1}
\end{figure}

The ACM is a model which describes the relative motion of a cluster system. We start by 
introducing the relative Jacobi coordinates for a $k$-body system (see Fig.~\ref{Fig1}) 
\ba
\vec{\rho} &=& \left( \vec{r}_{1} - \vec{r}_{2} \right) /\sqrt{2} ~,  
\nonumber\\
\vec{\lambda} &=& \left( \vec{r}_{1} + \vec{r}_{2} - 2\vec{r}_{3} \right)
/\sqrt{6} ~, 
\nonumber\\
\vec{\eta} &=& \left( \vec{r}_{1} + \vec{r}_{2} + \vec{r}_{3} - 3\vec{r}_{4} \right)
/\sqrt{12} ~,  
\nonumber\\
\vdots && 
\label{Jacobi}
\ea
together with their conjugate momenta. Here $\vec{r}_i$ represent the coordinates of the 
constituent particles. The relevant Jacobi coordinates are $\vec{\rho}$ for $k=2$,  
$\vec{\rho}$ and $\vec{\lambda}$ for $k=3$, and $\vec{\rho}$, $\vec{\lambda}$ and $\vec{\eta}$ for $k=4$ 
(see Table~\ref{kacm}). The ACM uses the method of bosonic quantization which consists in quantizing 
the Jacobi coordinates and momenta with vector boson operators and adding an additional scalar boson 
\ba 
b^{\dagger}_{\rho m} ~, \; b^{\dagger}_{\lambda m} ~, \; b^{\dagger}_{\eta m} ~, \; 
s^{\dagger} ~, \hspace{1cm} (m=-1,0,1) ~,
\label{bb}
\ea
under the constraint that the total number of bosons $N$ is conserved. 

Cluster states are described in the ACM in terms of a system of $N$ interacting 
bosons with angular momentum and parity $L^P=1^-$ (dipole or vector bosons) 
and $L^P=0^+$ (monopole or scalar bosons). The $3(k-1)$ components of the vector 
bosons together with the scalar boson span a $(3k-2)$-dimensional space with group 
structure $U(3k-2)$. The many-body states are classified according to the totally 
symmetric irreducible representation $[N]$ of $U(3k-2)$, where $N$ represents the 
total number of bosons. 

An explicit construction of the algebra and derivation of analytic formulas for energy levels, 
electromagnetic transition rates, matter and charge densities and associated form factors in 
electron scattering has been completed for cases $k=2$ \cite{IACHELLO1981581}, 
$k=3$ \cite{Bijker:2000fw,Bijker:2002ac} and $k=4$ \cite{Bijker:2014tka,Bijker:2016bpb}. 
It is summarized in Table~\ref{kacm} and results will be reviewed in the following subsections.

\begin{table}
\centering
\caption[Algebraic cluster model]
{Algebraic cluster model.}
\vspace{10pt}
\label{kacm}
\begin{tabular}{ccccc}
\hline
\noalign{\smallskip}
$k$ & Nucleus & $U(3k-2)$ & Discrete symmetry & Jacobi variables \\
\noalign{\smallskip}
\hline
\noalign{\smallskip}
2 & $^{8}$Be & $U(4)$ & $Z_2$ & $\vec{\rho}$ \\
3 & $^{12}$C & $U(7)$ & $D_{3h} \supset D_3$ & $\vec{\rho}$, $\vec{\lambda}$ \\
4 & $^{16}$O & $U(10)$ & $T_{d}$ & $\vec{\rho}$, $\vec{\lambda}$, $\vec{\eta}$ \\
\noalign{\smallskip}
\hline
\end{tabular}
\end{table}

\subsection{Classification of states}

The discrete symmetry of clusters imposes conditions on the allowed quantum states. 
The mathematical method for determining the allowed states ({\it i.e.} constructing 
representations of the discrete group $G$) is by means of the use of so-called symmetry adapter 
operators. For cases $k=2$, $3$, $4$ and identical constituents, one can exploit the 
isomorphism of the discrete point group with the permutation group $S_k$. The associated 
symmetry adapter operators are the transposition $P(12)$ and the cyclic permutation 
$P(12 \cdots k)$, see Table~\ref{sk}. All other permutations can be expressed in 
terms of these elementary ones \cite{KRAMER1966241}. 

\begin{table}
\centering
\caption[Symmetry adapter operators]{Symmetry adapter operators of the permutation group.}
\vspace{10pt}
\label{sk}
\begin{tabular}{lcc}
\hline
\noalign{\smallskip}
& \multicolumn{2}{c}{Symmetry adapter} \\
Group $G$ & Transposition & Cyclic permuation \\
\noalign{\smallskip}
\hline
\noalign{\smallskip}
$S_2 \sim Z_2 \sim P$ & $P(12)$ & $P(12)$ \\
$S_3 \sim D_{3}$ & $P(12)$ & $P(123)$ \\
$S_4 \sim T_{d}$ & $P(12)$ & $P(1234)$ \\
\noalign{\smallskip}
\hline
\end{tabular}
\end{table}

For the harmonic oscillator there exists a procedure for the explicit construction of states 
with good permutation symmetry \cite{KRAMER1966241}. However, in the application to the ACM the 
number of oscillator quanta may be large (up to 10) and moreover in general the oscillator shells 
are mixed. Therefore, a general procedure was developed in which the wave functions with good 
permutation symmetry $|\psi_t\rangle$ are generated numerically by diagonalizing $S_k$ invariant 
interactions. Subsequently, the permutation symmetry $t$ of a given wave function is determined 
by examining its transformation properties under the transposition $P(12)$ and the cyclic 
permutation $P(12 \cdots k)$. This procedure is explained in more detail in~\ref{app}. 

Representations can be labeled either by $S_k$ or by the isomorphic discrete group $G$, as shown 
in Table~\ref{irreps}. Here the representations of $S_k$ are labelled by the Young tableaux, while 
those of $G$ are labelled by the standard notation used in molecular physics 
\cite{herzberg1989molecular,wilson1955molecular}. 

\begin{table}
\centering
\caption[Representations]{Labelling of representations.}
\vspace{10pt}
\label{irreps}
\begin{tabular}{lccc}
\hline
\noalign{\smallskip}
Group $G$ & $S_k$ Label & $G$ Label & Degeneracy \\
\noalign{\smallskip}
\hline
\noalign{\smallskip}
$S_2 \sim Z_2 \sim P$ & $[2]$   & $A$ & Singly \\
                      & $[11]$  & $B$ & Singly \\
$S_3 \sim D_{3}$ & $[3]$   & $A_1$ & Singly \\
                 & $[21]$  & $E$   & Doubly \\
                 & $[111]$ & $A_2$ & Singly \\
$S_4 \sim T_{d}$ & $[4]$    & $A_1$ & Singly \\
                 & $[31]$   & $F_2$ & Triply \\
                 & $[22]$   & $E$   & Doubly \\
                 & $[211]$  & $F_1$ & Triply \\
                 & $[1111]$ & $A_2$ & Singly \\
\noalign{\smallskip}
\hline
\end{tabular}
\end{table}

In application to $\alpha$-cluster nuclei, like $^{8}$Be, $^{12}$C and $^{16}$O, in which 
the constuent parts are identical, the eigenstates of the Hamiltonian should transform 
according to the symmetric representations of the corresponding permutation group. 

\subsubsection{Dumbbell configuration}

\begin{figure}
\centering
\setlength{\unitlength}{1pt}
\begin{picture}(135,70)(0,30)
\thicklines
\put ( 25, 50) {\circle*{8}}
\put ( 75, 50) {\circle*{8}}
\put ( 25, 50) {\line (1,0){50}}
\put ( 25, 50) {\vector(-1,0){20}}
\put ( 75, 50) {\vector( 1,0){20}}
\put ( 40, 30) {$v(A)$}
\put ( 25, 80) {\circle*{8}}
\put ( 75, 80) {\circle*{8}}
\put ( 25, 80) {\line (1,0){50}}
\put ( 50, 80) {\vector(1,0){20}}
\put ( 45, 90) {$\vec{\rho}$}
\put ( 10, 77) {$1$}
\put ( 85, 77) {$2$}
\end{picture}
\caption[Jacobi vector for a dumbbell configuration]
{Jacobi vector $\vec{\rho}$ for a dumbbell configuration and its vibrations.}
\label{fundvib2}
\end{figure}
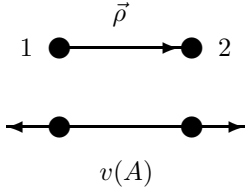

An algebraic description of this configuration is given by the algebra of $U(4)$ \cite{IACHELLO1981581}. 
This algebra is constructed with boson creation operators $b^{\dagger}_{\rho,m}$ with $m=0,\pm 1$ and  
$s^{\dagger}$, altogether denoted by $c^{\dagger}_{\alpha}$ with $\alpha=1,\ldots,4$, and annihilation 
operators $b_{\rho,m}$, $s$. Here $b^{\dagger}_{\rho,m}$ and $b_{\rho,m}$ are the 
quantization of the Jacobi variable $\vec{\rho}$ 
(see Fig.~\ref{fundvib2}) and its conjugate momentum, and $s^{\dagger}$, $s$ is an auxiliary scalar boson.
The bilinear products $G_{\alpha\beta}=c^{\dagger}_{\alpha}c_{\beta}$ with $\alpha,\beta=1,\ldots,4$ 
of creation and annihilation operators generate the Lie algebra of $U(4)$. Specifically these are
\ba
&& (s^{\dagger} \times \tilde{s})^{(0)} ~, \qquad (b^{\dagger}_{\rho} \times \tilde{s})^{(1)} ~, 
\qquad (s^{\dagger} \times \tilde{b}_{\rho})^{(1)} ~,
\nonumber\\
&& (b^{\dagger}_{\rho} \times \tilde{b}_{\rho})^{(L)} ~, \hspace{1cm} (L=0,1,2) ~,
\ea
where $\tilde{b}_{\rho,m}=(-1)^{1-m} b_{\rho,-m}$ and $\tilde{s}=s$. We consider here rotations 
and vibrations of the dumbbell configuration. States can be classified by a vibrational quantum 
number $v=0,1,2,\ldots,$ and a rotational quantum number $L$ and its projection $M$ as $| v,L,M \rangle$. 
In the case in which the two constituents are identical (two $\alpha$-particles) the dumbbell has 
$Z_2 \sim S_2 \sim P$ symmetry. All vibrational states $v$ have symmetry $A$ under $Z_2$ since 
the two particles are identical (Fig.~\ref{fundvib2}). The angular momentum content of each
vibrational band is $L^P=0^+$, $2^+$, $4^+$, $\ldots$, where the parity $P$ has been added, 
although here it is not an independent quantum number, $P=(-)^L$. 

\subsubsection{Equilateral triangle configuration}

An algebraic description of this configuration is given by the algebra of $U(7)$ 
\cite{Bijker:2000fw,Bijker:2002ac}. This algebra is constructed with boson creation operators  
$b^{\dagger}_{\rho,m}$, $b^{\dagger}_{\lambda,m}$ with $m=0,\pm 1$ and $s^{\dagger}$, altogether 
denoted by $c^{\dagger}_{\alpha}$ with $\alpha=1,\ldots,7$, and annihilation operators 
$b_{\rho,m}$, $b_{\lambda,m}$, $s$. The two vector boson operators $b^{\dagger}_{\rho,m}$, 
$b^{\dagger}_{\lambda,m}$ and $b_{\rho,m}$, $b_{\lambda,m}$ are the quantization of the two 
Jacobi variables
\ba
\vec{\rho} &=& \left( \vec{r}_{1} - \vec{r}_{2} \right) /\sqrt{2} ~,  
\nonumber\\
\vec{\lambda} &=& \left( \vec{r}_{1} + \vec{r}_{2} - 2\vec{r}_{3} \right)/\sqrt{6} ~. 
\ea
The bilinear products $G_{\alpha\beta}=c^{\dagger}_{\alpha}c_{\beta}$ with $\alpha,\beta=1,\ldots,7$ 
generate the Lie algebra of $U(7)$. A specific form is given in \cite{Bijker:2002ac}. 

\begin{figure}
\centering
\setlength{\unitlength}{1.5pt}
\begin{picture}(100,120)(0,20)
\thicklines
\put ( 25, 50) {\circle*{8}}
\put ( 75, 50) {\circle*{8}}
\put ( 50,100) {\circle*{8}}
\put ( 25, 50) {\line ( 1,0){50}}
\put ( 25, 50) {\line ( 1,2){25}}
\put ( 75, 50) {\line (-1,2){25}}
\put ( 50, 50) {\vector(1,0){20}}
\put ( 50, 50) {\vector(0,1){45}}
\put ( 47,110) {$3$}
\put ( 15, 35) {$1$}
\put ( 80, 35) {$2$}
\put( 45, 35) {$\vec{\rho}$}
\put( 55, 65) {$\vec{\lambda}$}
\end{picture}
\caption[Jacobi variables for an equilateral triangle configuration]
{Jacobi variables $\vec{\rho}$, $\vec{\lambda}$ for an equilateral triangle configuration.}
\label{Jacobi3}
\end{figure}
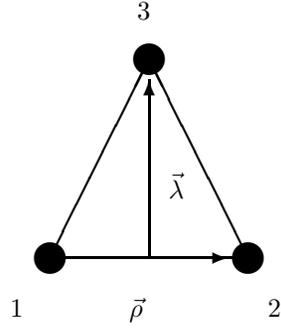

\begin{figure}
\centering
\setlength{\unitlength}{1.0pt}
\begin{picture}(300,140)(0,0)
\thicklines
\put ( 25, 50) {\circle*{10}}
\put ( 75, 50) {\circle*{10}}
\put ( 50,100) {\circle*{10}}
\put ( 25, 50) {\line ( 1,0){50}}
\put ( 25, 50) {\line ( 1,2){25}}
\put ( 75, 50) {\line (-1,2){25}}
\put ( 50,100) {\vector( 0, 1){20}}
\put ( 25, 50) {\vector(-1,-1){15}}
\put ( 75, 50) {\vector( 1,-1){15}}
\put (125, 50) {\circle*{10}}
\put (175, 50) {\circle*{10}}
\put (150,100) {\circle*{10}}
\put (125, 50) {\line ( 1,0){50}}
\put (125, 50) {\line ( 1,2){25}}
\put (175, 50) {\line (-1,2){25}}
\put (150,100) {\vector( 0, 1){20}}
\put (125, 50) {\vector( 1,-1){15}}
\put (175, 50) {\vector(-1,-1){15}}
\put (225, 50) {\circle*{10}}
\put (275, 50) {\circle*{10}}
\put (250,100) {\circle*{10}}
\put (225, 50) {\line ( 1,0){50}}
\put (225, 50) {\line ( 1,2){25}}
\put (275, 50) {\line (-1,2){25}}
\put (250,100) {\vector( 1, 0){20}}
\put (225, 50) {\vector(-1,-2){10}}
\put (275, 50) {\vector(-1, 2){10}}
\put ( 40, 15) {$v_1(A)$}
\put (137, 15) {$v_{2a}(E)$}
\put (237, 15) {$v_{2b}(E)$}
\end{picture}
\caption[Fundamental vibrations of a triangular configuration]
{Fundamental vibrations of a triangular configuration (point group $D_{3h}$). 
The $A$ vibration is singly degenerate, 
while $E$ is doubly degenerate with components $v_{2a}$, $v_{2b}$.}
\label{fundvib3}
\end{figure}
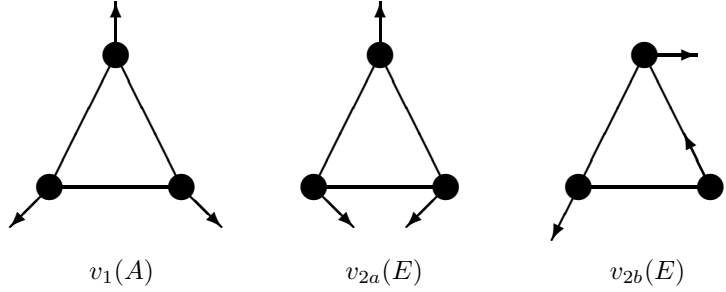

We consider here rotations and vibrations of an equilateral triangular configuration. States can 
be classified as 
\ba
\left| (v_1,v_2^{\ell_2});t,K,L^P,M \right> ~, 
\label{basis3}
\ea
where $t$ denotes the representations of $D_{3h}$, and $K$ the projection of the angular momentum $L$ on 
the symmetry axis. In the case in which the three constituents are identical (three $\alpha$-particles), 
the triangular configuration of Fig.~\ref{Jacobi3} has $D_{3h}$ symmetry. This imposes some conditions 
on the allowed values of $K$ and $L$. In Eq.~(\ref{basis3}), $(v_1,v_2^{\ell_2})$ label the vibrational 
states with $v_1=0,1,\ldots$, $v_2=0,1,\ldots$, and $\ell_2=v_2,v_2-2,\ldots,1$ or $0$ for $v_2$ odd or even. 
The fundamental vibrations of a triangular configuration are shown in Fig.~\ref{fundvib3}. 

For vibrational bands with $\ell_2=0$ and $1$, the allowed values of the angular momentum are  
\ba
\begin{array}{lll}
K=3n &\hspace{1cm}& n=0,1,2,\ldots, \\ 
L=0,2,4,\ldots, && \mbox{for } K=0 \\ 
L=K,K+1,K+2,\ldots && \mbox{for } K\neq 0 
\end{array}
\label{states3a}
\ea
for $(v_{1},v_{2}^{\ell_{2}=0})$, and 
\ba
\begin{array}{lll}
K=3n+1,3n+2 &\hspace{1cm}& n=0,1,2,\ldots, \\ 
L=K,K+1,K+2,\ldots, &  
\end{array}
\label{states3b}
\ea
for $(v_{1},v_{2}^{\ell_{2}=1})$. The parity is $P=(-)^{K}$. The vibrational band $(1,0^{0})$ 
has the same angular momenta $L^{P}=0^{+},2^{+},3^{-},4^{\pm },\ldots$~, as the ground
state band $(0,0^{0})$, while the angular momentum content of the doubly degenerate vibration 
$(0,1^{1})$ is given by $L^{P}=1^{-},2^{\mp },3^{\mp},\ldots$~. 

\subsubsection{Tetrahedral configuration}

\begin{figure}
\centering
\setlength{\unitlength}{2pt}
\begin{picture}(120,100)(0,20)
\thicklines
\put( 50, 30) {\circle*{5}} 
\put( 30, 50) {\circle*{5}}
\put( 80, 50) {\circle*{5}}
\put( 60, 90) {\circle*{5}}
\put( 50, 30) {\vector(-1, 1){18}} 
\thinlines
\multiput( 30, 50)( 3, 0){18}{\circle*{1}}
\put( 80, 50) {\vector(-4,-1){40}}
\put( 60, 90) {\vector( 0,-1){45}}
\thicklines
\put( 50, 30) {\line(-1, 1){20}}
\put( 50, 30) {\line( 3, 2){30}}
\put( 60, 90) {\line(-3,-4){30}}
\put( 60, 90) {\line( 1,-2){20}}
\put( 60, 90) {\line(-1,-6){10}}
\put( 35, 35) {$\vec{\rho}$}
\put( 55, 37) {$\vec{\lambda}$}
\put( 63, 60) {$\vec{\eta}$}
\put( 20, 50) {$1$}
\put( 50, 20) {$2$}
\put( 85, 50) {$3$}
\put( 60, 95) {$4$}
\end{picture}
\caption[Jacobi coordinates for a tetrahedral configuration]
{Jacobi coordinates $\vec{\rho}$, $\vec{\lambda}$, $\vec{\eta}$ for a tetrahedral configuration.}
\label{Jacobi4}
\end{figure}
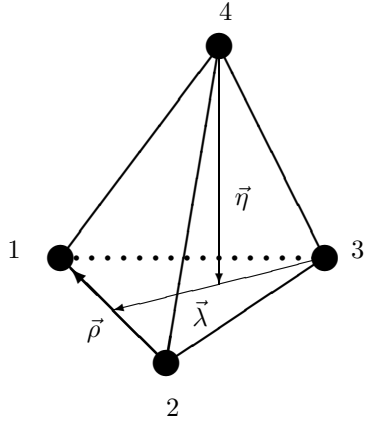

An algebraic description of this configuration is given by the algebra of $U(10)$ 
\cite{Bijker:2014tka,Bijker:2016bpb}. This algebra is constructed with boson creation operators  
$b^{\dagger}_{\rho,m}$, $b^{\dagger}_{\lambda,m}$, $b^{\dagger}_{\eta,m}$ with $m=0,\pm 1$ and 
$s^{\dagger}$, altogether denoted by $c^{\dagger}_{\alpha}$ with $\alpha=1,\ldots,10$, and 
annihilation operators $b_{\rho,m}$, $b_{\lambda,m}$, $b_{\eta,m}$, $s$. The three vector boson 
operators $b^{\dagger}_{\rho,m}$, $b^{\dagger}_{\lambda,m}$, $b^{\dagger}_{\eta,m}$ and $b_{\rho,m}$, 
$b_{\lambda,m}$, $b_{\eta,m}$ are the quantization of the three Jacobi variables
\ba
\vec{\rho} &=& \left( \vec{r}_{1} - \vec{r}_{2} \right) /\sqrt{2} ~,  
\nonumber\\
\vec{\lambda} &=& \left( \vec{r}_{1} + \vec{r}_{2} - 2\vec{r}_{3} \right)
/\sqrt{6} ~, 
\nonumber\\
\vec{\eta} &=& \left( \vec{r}_{1} + \vec{r}_{2} + \vec{r}_{3} - 3\vec{r}_{4} \right)
/\sqrt{12} ~,
\ea
shown in Fig.~\ref{Jacobi4}. The bilinear products $G_{\alpha\beta}=c^{\dagger}_{\alpha}c_{\beta}$ 
with $\alpha,\beta=1,\ldots,10$ generate the Lie algebra of $U(10)$. 
A specific form is given in \cite{Bijker:2016bpb}. 

\begin{figure}
\centering
\setlength{\unitlength}{1pt}
\begin{picture}(330,240)(0,0)
\thicklines
\put( 50,130) {\circle*{7}} 
\put( 30,150) {\circle*{7}}
\put( 80,150) {\circle*{7}}
\put( 60,190) {\circle*{7}}
\put( 50,130) {\vector( 0,-1){15}} 
\put( 30,150) {\vector(-1, 0){15}}
\put( 80,150) {\vector( 1, 0){15}}
\put( 60,190) {\vector( 0, 1){15}}
\put( 50,130) {\line(-1, 1){20}}
\put( 50,130) {\line( 3, 2){30}}
\put( 60,190) {\line(-3,-4){30}}
\put( 60,190) {\line( 1,-2){20}}
\put( 60,190) {\line(-1,-6){10}}
\put( 10,190) {$v_1(A)$}
\multiput( 30,150)(5,0){10}{\circle*{1}}
\put(150,130) {\circle*{7}} 
\put(130,150) {\circle*{7}}
\put(180,150) {\circle*{7}}
\put(160,190) {\circle*{7}}
\put(150,130) {\vector( 2,-1){15}} 
\put(130,150) {\vector(-1, 2){ 8}}
\put(180,150) {\vector( 1,-2){ 8}}
\put(160,190) {\vector(-2, 1){15}}
\put(150,130) {\line(-1, 1){20}}
\put(150,130) {\line( 3, 2){30}}
\put(160,190) {\line(-3,-4){30}}
\put(160,190) {\line( 1,-2){20}}
\put(160,190) {\line(-1,-6){10}}
\put(110,190) {$v_{2a}(E)$}
\multiput(130,150)(5,0){10}{\circle*{1}}
\put(250,130) {\circle*{7}} 
\put(230,150) {\circle*{7}}
\put(280,150) {\circle*{7}}
\put(260,190) {\circle*{7}}
\put(250,130) {\vector( 1, 2){10}} 
\put(230,150) {\vector( 3, 2){15}}
\put(280,150) {\vector(-3,-4){15}}
\put(260,190) {\vector(-3, 1){15}}
\put(250,130) {\line(-1, 1){20}}
\put(250,130) {\line( 3, 2){30}}
\put(260,190) {\line(-3,-4){30}}
\put(260,190) {\line( 1,-2){20}}
\put(260,190) {\line(-1,-6){10}}
\put(210,190) {$v_{2b}(E)$}
\multiput(230,150)(5,0){10}{\circle*{1}}
\put( 50, 30) {\circle*{7}} 
\put( 30, 50) {\circle*{7}}
\put( 80, 50) {\circle*{7}}
\put( 60, 90) {\circle*{7}}
\put( 50, 30) {\vector(-1, 1){10}} 
\put( 30, 50) {\vector( 1,-1){10}}
\put( 80, 50) {\vector( 1,-2){ 8}}
\put( 60, 90) {\vector(-1, 2){ 8}}
\put( 50, 30) {\line(-1, 1){20}}
\put( 50, 30) {\line( 3, 2){30}}
\put( 60, 90) {\line(-3,-4){30}}
\put( 60, 90) {\line( 1,-2){20}}
\put( 60, 90) {\line(-1,-6){10}}
\put( 10, 90) {$v_{3a}(F)$}
\multiput( 30, 50)(5,0){10}{\circle*{1}}
\put(150, 30) {\circle*{7}} 
\put(130, 50) {\circle*{7}}
\put(180, 50) {\circle*{7}}
\put(160, 90) {\circle*{7}}
\put(150, 30) {\vector( 0,-1){15}} 
\put(130, 50) {\vector( 1, 0){15}}
\put(180, 50) {\vector(-1, 0){15}}
\put(160, 90) {\vector( 0, 1){15}}
\put(150, 30) {\line(-1, 1){20}}
\put(150, 30) {\line( 3, 2){30}}
\put(160, 90) {\line(-3,-4){30}}
\put(160, 90) {\line( 1,-2){20}}
\put(160, 90) {\line(-1,-6){10}}
\put(110, 90) {$v_{3b}(F)$}
\multiput(130, 50)(5,0){10}{\circle*{1}}
\put(250, 30) {\circle*{7}} 
\put(230, 50) {\circle*{7}}
\put(280, 50) {\circle*{7}}
\put(260, 90) {\circle*{7}}
\put(250, 30) {\vector( 3, 2){15}} 
\put(230, 50) {\vector(-3,-4){10}}
\put(280, 50) {\vector(-3,-2){15}}
\put(260, 90) {\vector( 3, 4){10}}
\put(250, 30) {\line(-1, 1){20}}
\put(250, 30) {\line( 3, 2){30}}
\put(260, 90) {\line(-3,-4){30}}
\put(260, 90) {\line( 1,-2){20}}
\put(260, 90) {\line(-1,-6){10}}
\put(210, 90) {$v_{3c}(F)$}
\multiput(230, 50)(5,0){10}{\circle*{1}}
\end{picture}
\caption[Fundamental vibrations of a tetrahedral configuration]
{Fundamental vibrations of a tetrahedral configuration (point group $T_d$). 
The $A$ vibration is singly degenerate, $E$ is doubly degenerate with components  
$v_{2a}$, $v_{2b}$ and $F$ is triply degenerate with $v_{3a}$, $v_{3b}$, $v_{3c}$.}
\label{fundvib4}
\end{figure}
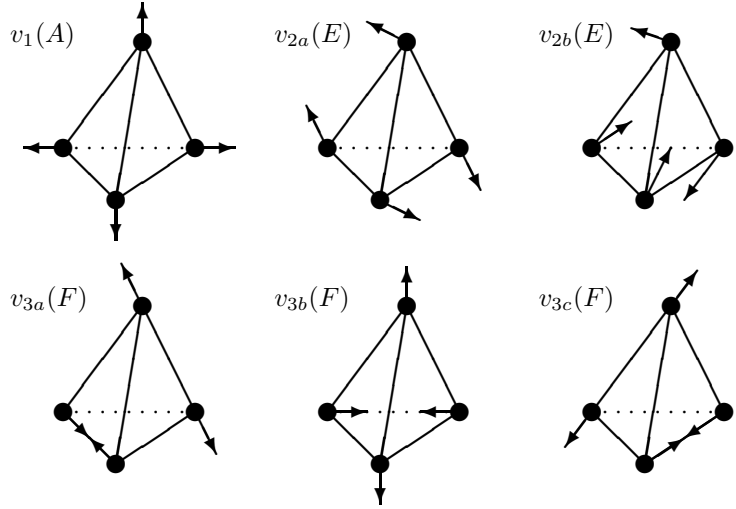

We consider here rotations and vibrations of a tetrahedral configuration. States can 
be classified as 
\ba
\left| (v_1,v_2^{\ell_2},v_3^{\ell_3});t,L^P,M \right> ~, 
\label{basis4}
\ea
where $(v_1,v_2^{\ell_2},v_3^{\ell_3})$ denote the vibrational quantum numbers and $t$ labels the 
representations of $T_{d}$. In the case in which the four constituents are identical (four 
$\alpha$-particles), the tetrahedral configuration of Fig.~\ref{Jacobi4} has $T_{d}$ symmetry. 
This imposes some conditions on the allowed values of $L$ which depend on $t$. A derivation 
of the allowed values is given in \cite{Bijker:2016bpb}. For the ground state, $t=A$, and for the 
fundamental vibrations with $t=A$, $E$ and $F$ of Fig.~\ref{fundvib4}, it can be summarized as follows
\ba
\begin{array}{lll}
t=A &\hspace{1cm}& L^P=0^+,3^-,4^+,6^{\pm},\ldots, \\ 
t=E && L^P=2^{\pm},4^{\pm},5^{\pm},6^{\pm},\ldots, \\
t=F && L^P=1^-,2^+,3^{\pm},4^{\pm},5^{-\pm},6^{+\pm},\ldots ~. 
\end{array}
\label{states4}
\ea

\subsection{Energy formulas}
\label{energies}

Energy levels in ACM can be obtained by diagonalizing the Hamiltonian. Computer programs have been written for all three cases, $U(4)$, $U(7)$ and $U(10)$ \cite{RB}. These programs can deal with all situations encountered in two-, three- and four-body problems, including both soft and rigid situations. For applications here, we consider only rigid situations and write down analytic formulas that can be used to analyze experimental data.

\subsubsection{Dumbbell configuration}

\begin{figure}
\centering
\setlength{\unitlength}{0.7pt} 
\begin{picture}(160,270)(0,0)
\thinlines
\put (  0,  0) {\line(1,0){160}}
\put (  0,270) {\line(1,0){160}}
\put (  0,  0) {\line(0,1){270}}
\put (160,  0) {\line(0,1){270}}
\thicklines
\put ( 30, 60) {\line(1,0){20}}
\put ( 30, 78) {\line(1,0){20}}
\put ( 30,120) {\line(1,0){20}}
\put ( 30,186) {\line(1,0){20}}
\multiput ( 80, 60)(5,0){9}{\circle*{0.1}}
\thinlines
\put ( 30, 25) {$(0)A$}
\put ( 55, 57) {$0^+$}
\put ( 55, 75) {$2^+$}
\put ( 55,117) {$4^+$}
\put ( 55,183) {$6^+$}
\thicklines
\put (100, 90) {\line(1,0){20}}
\put (100,108) {\line(1,0){20}}
\put (100,150) {\line(1,0){20}}
\put (100,216) {\line(1,0){20}}
\multiput (110, 60)(0,5){6}{\circle*{0.1}}
\thinlines
\put (100, 25) {$(1)A$}
\put (125, 87) {$0^+$}
\put (125,105) {$2^+$}
\put (125,147) {$4^+$}
\put (125,213) {$6^+$}
\thicklines
\put(30,230) {\circle*{10}} 
\put(60,230) {\circle*{10}}
\put(30,230) {\line( 1,0){30}}
\end{picture}
\caption[Schematic spectrum of a dumbbell configuration]
{Schematic spectrum of a dumbbell configuration. 
The rotational bands are labeled by $(v)$ and $t$ (bottom). 
All states are symmetric under $S_2 \sim Z_2$.} 
\label{dumbbell}
\end{figure}
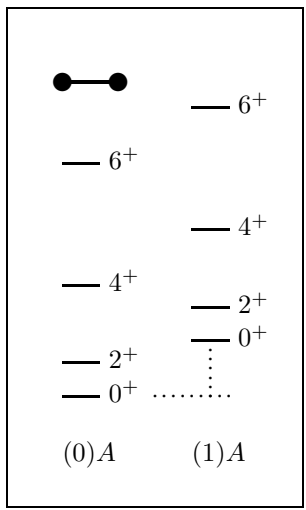

The algebraic Hamiltonian describing roto-vibrations of a dumbbell configuration (diatomic
molecule) is given in Eqs.~(2.108) and (2.112) of \cite{IachelloLevine}. Written explicitly 
in terms of boson operators, it has the form
\ba
H \;=\; E_0 + A ( \hat{D} \cdot \hat{D} + \hat{L} \cdot \hat{L} ) + B \hat{L} \cdot \hat{L} ~,
\ea
with
\ba
\hat{D}_m &=& (b^{\dagger}_{\rho} \times \tilde{s} + s^{\dagger} \times \tilde{b}_{\rho})^{(1)}_m ~,
\nonumber\\
\hat{L}_m &=& \sqrt{2} \, ( b^{\dagger}_{\rho} \times \tilde{b}_{\rho})^{(1)}_m ~,
\label{dipole}
\ea
with $m=0,\pm 1$. In this case, the Hamiltonian has a dynamic symmetry 
$U(4) \supset SO(4) \supset SO(3) \supset SO(2)$. 
The eigenvalues of $H$ can be written in explicit analytic form as
\ba
E(N,v,L) \;=\; E'_0 -4A(N+1) \left( v - \frac{v^2}{N+1} \right) + B L(L+1) ~.
\ea
Here $N$ is the so-called vibron number, that is the number of bosons that characterizes the irreducible
representations of $U(4)$. The vibrational quantum number $v$ takes the values
\ba
v \;=\; 0, 1, \ldots, \frac{N-1}{2} \mbox{ or } \frac{N}{2} ~,
\ea
for $N$ odd or even, and the rotational quantum number $L$ takes the integer values
\ba
L \;=\; 0, 1, \ldots, N-2v ~.
\ea
For identical constituents, {\it i.e.} $Z_2$ symmetry, only even values of $L$ are allowed. 
In the large $N$ limit, one obtains the semiclassical formula for the energy levels of a dumbbell 
configuration
\ba
E(N,v,L) \;=\; E''_0 + \omega \left( v + \frac{1}{2} \right) + B L(L+1) ~,
\label{ener2}
\ea
where $\omega$ is the vibrational energy and $B$ the inertial parameter $B=\hbar^2/2{\cal I}$.
A schematic spectrum of a rotating and vibrating dumbbell is shown in Fig.~\ref{dumbbell}.

\subsubsection{Equilateral triangle configuration}

\begin{figure}
\centering
\setlength{\unitlength}{0.7pt} 
\begin{picture}(240,270)(0,0)
\thinlines
\put (  0,  0) {\line(1,0){240}}
\put (  0,270) {\line(1,0){240}}
\put (  0,  0) {\line(0,1){270}}
\put (240,  0) {\line(0,1){270}}
\thicklines
\put ( 30, 60) {\line(1,0){20}}
\put ( 30, 78) {\line(1,0){20}}
\put ( 30, 96) {\line(1,0){20}}
\put ( 30,120) {\line(1,0){20}}
\put ( 30,150) {\line(1,0){20}}
\multiput ( 80, 60)(5,0){23}{\circle*{0.1}}
\thinlines
\put ( 25, 25) {$(00)A$}
\put ( 55, 57) {$0^+$}
\put ( 55, 75) {$2^+$}
\put ( 55, 93) {$3^-$}
\put ( 55,117) {$4^{\pm}$}
\put ( 55,147) {$5^-$}
\thicklines
\put (100, 90) {\line(1,0){20}}
\put (100,108) {\line(1,0){20}}
\put (100,126) {\line(1,0){20}}
\put (100,150) {\line(1,0){20}}
\put (100,180) {\line(1,0){20}}
\multiput (110, 60)(0,5){6}{\circle*{0.1}}
\thinlines
\put ( 95, 25) {$(10)A$}
\put (125, 87) {$0^+$}
\put (125,105) {$2^+$}
\put (125,123) {$3^-$}
\put (125,147) {$4^{\pm}$}
\put (125,177) {$5^-$}
\thicklines
\put (170,126) {\line(1,0){20}}
\put (170,138) {\line(1,0){20}}
\put (170,156) {\line(1,0){20}}
\put (170,180) {\line(1,0){20}}
\put (170,210) {\line(1,0){20}}
\multiput (180, 60)(0,5){13}{\circle*{0.1}}
\thinlines
\put (165, 25) {$(01)E$}
\put (195,123) {$1^-$}
\put (195,135) {$2^{\mp}$}
\put (195,153) {$3^{\mp}$}
\put (195,177) {$4^{\mp+}$}
\put (195,207) {$5^{\mp\pm}$}
\thicklines
\put(25,200) {\circle*{10}} 
\put(65,200) {\circle*{10}}
\put(45,240) {\circle*{10}}
\put(25,200) {\line( 1,0){40}}
\put(25,200) {\line( 1,2){20}}
\put(65,200) {\line(-1,2){20}}
\end{picture}
\caption[Schematic spectrum of a triangular configuration]
{Schematic spectrum of a triangular configuration. The rotational bands are labeled by 
$(v_1 v_2)$ and $t$ (bottom). All states are symmetric under $S_3 \sim D_3$.} 
\label{triangle}
\end{figure}
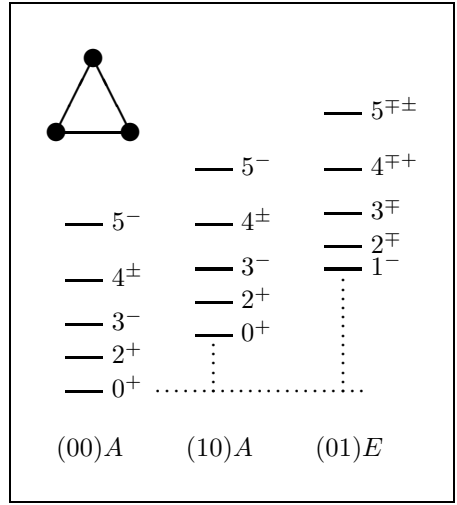

The situation here is more complicated than for the dumbbell configuration, since there is no dynamic
symmetry corresponding to the rotation and vibration of a rigid symmetric top. The explicit form
of the Hamiltonian is \cite{Bijker:2002ac}
\ba
H  &=& \xi_{1} \, ( s^{\dagger }s^{\dagger }-b_{\rho }^{\dagger
}\cdot b_{\rho }^{\dagger }-b_{\lambda }^{\dagger }\cdot b_{\lambda
}^{\dagger }) ( \tilde{s}\tilde{s}-\tilde{b}_{\rho }\cdot \tilde{b%
}_{\rho }-\tilde{b}_{\lambda }\cdot \tilde{b}_{\lambda })  \nonumber \\
&&+\xi _{2} \left[( b_{\rho }^{\dagger }\cdot b_{\rho }^{\dagger }-b_{\lambda
}^{\dagger }\cdot b_{\lambda }^{\dagger }) ( \tilde{b}_{\rho
}\cdot \tilde{b}_{\rho }-\tilde{b}_{\lambda }\cdot \tilde{b}_{\lambda
})  +4( b_{\rho }^{\dagger }\cdot b_{\lambda }^{\dagger }) 
( \tilde{b}_{\lambda }\cdot \tilde{b}_{\rho }) \right]  \nonumber \\
&&+2\kappa_1 \, (b_{\rho }^{\dagger} \times \tilde{b}_{\rho} 
+ b_{\lambda}^{\dagger} \times \tilde{b}_{\lambda})^{(1)} \cdot 
(b_{\rho}^{\dagger} \times \tilde{b}_{\rho}+b_{\lambda}^{\dagger} \times 
\tilde{b}_{\lambda})^{(1)}  \nonumber \\
&&+3\kappa_2 \, (b_{\rho}^{\dagger} \times \tilde{b}_{\lambda}
-b_{\lambda}^{\dagger} \times \tilde{b}_{\rho}) ^{(0)} \cdot 
(b_{\lambda}^{\dagger} \times \tilde{b}_{\rho}-b_{\rho}^{\dagger} \times 
\tilde{b}_{\lambda})^{(0)} ~.
\label{ham3}
\ea
In a generic situation, this Hamiltonian needs to be diagonalized in the space of given vibron
number $N$. However, in the limit $N \rightarrow \infty$, one can write down a semiclassical formula
\ba
E(v_{1},v_{2}^{\ell_{2}},K,L) &=& E_0 + \omega_{1} \left( v_{1}+\frac{1}{2} \right) 
+ \omega_{2}(v_{2}+1) 
\nonumber\\
&&+ \kappa_{1} \, L(L+1) + \kappa_{2} \, (K \mp 2\ell_{2})^{2} ~.
\label{ener3}
\ea
which describes the energy levels of a symmetric top \cite{Bijker:2002ac}. 
States are classified as in Eq.~(\ref{basis3}) and the values of $K$ and $L$ for 
$(v_1,v_2^{\ell_2=0,1})$ are given in Eqs.~(\ref{states3a}) and (\ref{states3b}).
In Fig.~\ref{triangle} we show a schematic rotational-vibrational spectrum of a 
triangular configuration. 

\subsubsection{Tetrahedral configuration}

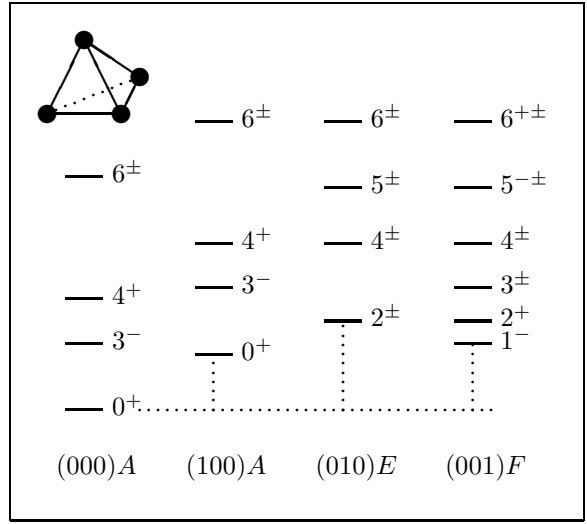
\begin{figure}
\centering
\setlength{\unitlength}{0.7pt} 
\begin{picture}(310,280)(0,0)
\thinlines
\put (  0,  0) {\line(1,0){310}}
\put (  0,280) {\line(1,0){310}}
\put (  0,  0) {\line(0,1){280}}
\put (310,  0) {\line(0,1){280}}
\thicklines
\put ( 30, 60) {\line(1,0){20}}
\put ( 30, 96) {\line(1,0){20}}
\put ( 30,120) {\line(1,0){20}}
\put ( 30,186) {\line(1,0){20}}
\multiput ( 70, 60)(5,0){39}{\circle*{0.1}}
\thinlines
\put ( 25, 25) {$(000)A$}
\put ( 55, 57) {$0^+$}
\put ( 55, 93) {$3^-$}
\put ( 55,117) {$4^+$}
\put ( 55,183) {$6^{\pm}$}
\thicklines
\put (100, 90) {\line(1,0){20}}
\put (100,126) {\line(1,0){20}}
\put (100,150) {\line(1,0){20}}
\put (100,216) {\line(1,0){20}}
\multiput (110, 60)(0,5){6}{\circle*{0.1}}
\thinlines
\put ( 95, 25) {$(100)A$}
\put (125, 87) {$0^+$}
\put (125,123) {$3^-$}
\put (125,147) {$4^+$}
\put (125,213) {$6^{\pm}$}
\thicklines
\put (170,108) {\line(1,0){20}}
\put (170,150) {\line(1,0){20}}
\put (170,180) {\line(1,0){20}}
\put (170,216) {\line(1,0){20}}
\multiput (180, 60)(0,5){10}{\circle*{0.1}}
\thinlines
\put (165, 25) {$(010)E$}
\put (195,105) {$2^{\pm}$}
\put (195,147) {$4^{\pm}$}
\put (195,177) {$5^{\pm}$}
\put (195,213) {$6^{\pm}$}
\thicklines
\put (240, 96) {\line(1,0){20}}
\put (240,108) {\line(1,0){20}}
\put (240,126) {\line(1,0){20}}
\put (240,150) {\line(1,0){20}}
\put (240,180) {\line(1,0){20}}
\put (240,216) {\line(1,0){20}}
\multiput (250, 60)(0,5){8}{\circle*{0.1}}
\thinlines
\put (235, 25) {$(001)F$}
\put (265, 93) {$1^-$}
\put (265,105) {$2^+$}
\put (265,123) {$3^{\pm}$}
\put (265,147) {$4^{\pm}$}
\put (265,177) {$5^{-\pm}$}
\put (265,213) {$6^{+\pm}$}
\thicklines
\put(20,220) {\circle*{10}} 
\put(60,220) {\circle*{10}}
\put(70,240) {\circle*{10}}
\put(40,260) {\circle*{10}}
\put(20,220) {\line( 1,0){40}}
\put(20,220) {\line( 1,2){20}}
\put(60,220) {\line(-1,2){20}}
\put(60,220) {\line( 1,2){10}}
\put(70,240) {\line(-3,2){30}}
\multiput(20,220)(5,2){11}{\circle*{2}}
\end{picture}
\caption[Schematic spectrum of a tetrahedral configuration]
{Schematic spectrum of a tetrahedral configuration. The rotational bands are labeled by 
$(v_1 v_2 v_3)$ and $t$ (bottom). All states are symmetric under $S_4 \sim T_d$.} 
\label{tetrahedron}
\end{figure}

Also in this case there is no dynamic symmetry corresponding to the rotation and vibration 
of a spherical top. The explicit form of the Hamiltonian describing the vibrations is 
\cite{Bijker:2014tka,Bijker:2016bpb}
\ba
H_{\rm vib} &=& \xi_{1} \, (s^{\dagger} s^{\dagger} 
- b_{\rho}^{\dagger} \cdot b_{\rho}^{\dagger} 
- b_{\lambda }^{\dagger} \cdot b_{\lambda}^{\dagger} 
- b_{\eta}^{\dagger} \cdot b_{\eta}^{\dagger}) \, ( {\rm h.c.} )  
\nonumber\\
&& + \xi_2 \, \left[ ( -2\sqrt{2} \, b_{\rho}^{\dagger} \cdot b_{\eta}^{\dagger} 
+ 2 b_{\rho}^{\dagger} \cdot b_{\lambda}^{\dagger} ) \, ( {\rm h.c.} ) \right.
\nonumber\\
&& \hspace{2cm} + \left. ( -2\sqrt{2} \, b_{\lambda}^{\dagger} \cdot b_{\eta}^{\dagger} 
+ ( b_{\rho}^{\dagger} \cdot b_{\rho}^{\dagger} 
- b_{\lambda}^{\dagger} \cdot b_{\lambda}^{\dagger} )) \, ( {\rm h.c.} ) \right]
\nonumber\\
&& + \xi_3 \, \left[ ( 2 b_{\rho}^{\dagger} \cdot b_{\eta}^{\dagger}
+ 2\sqrt{2} \, b_{\rho}^{\dagger} \cdot b_{\lambda}^{\dagger} ) \, ( {\rm h.c.} ) \right.
\nonumber\\
&& \hspace{2cm} + ( 2 b_{\lambda}^{\dagger} \cdot b_{\eta}^{\dagger} 
+ \sqrt{2} \, ( b_{\rho}^{\dagger} \cdot b^{\dagger}_{\rho}  
- b_{\lambda}^{\dagger} \cdot b_{\lambda}^{\dagger} )) \, ( {\rm h.c.} )
\nonumber\\
&& \hspace{2cm} \left. + ( b_{\rho}^{\dagger} \cdot b_{\rho}^{\dagger} 
     + b_{\lambda}^{\dagger} \cdot b_{\lambda}^{\dagger} 
   - 2 b_{\eta}^{\dagger} \cdot b_{\eta}^{\dagger} ) \, ( {\rm h.c.} ) \right] ~. 
\label{h3vib}
\ea
The Hamiltonian describing rotations can be written as
\ba
H_{3,\rm rot} &=& \kappa_1 \, \vec{L} \cdot \vec{L} 
+ \kappa_2 \, (\vec{L} \cdot \vec{L} - \vec{I} \cdot \vec{I})^2 ~,
\label{h3rot}
\ea
where $\vec{L}$ and $\vec{I}$ denote the angular momentum in coordinate space and 
index space, respectively, the explicit form of which is  
\ba
L_m &=& \sqrt{2} \, ( b^{\dagger}_{\rho} \tilde{b}_{\rho}
+ b^{\dagger}_{\lambda} \tilde{b}_{\lambda} 
+ b^{\dagger}_{\eta} \tilde{b}_{\eta})^{(1)}_m ~,
\nonumber\\
I_{\rho} &=& -i \sqrt{3} \, 
( b^{\dagger}_{\lambda} \tilde{b}_{\eta} 
- b^{\dagger}_{\eta} \tilde{b}_{\lambda} )^{(0)} ~,
\nonumber\\
I_{\lambda} &=& -i \sqrt{3} \, 
( b^{\dagger}_{\eta} \tilde{b}_{\rho} 
- b^{\dagger}_{\rho} \tilde{b}_{\eta} )^{(0)} ~,
\nonumber\\
I_{\eta} &=& -i \sqrt{3} \, 
( b^{\dagger}_{\rho} \tilde{b}_{\lambda} 
- b^{\dagger}_{\lambda} \tilde{b}_{\rho} )^{(0)} ~.
\ea
Again, in a generic situation, this Hamiltonian needs to be diagonalized in the space of given
vibron number $N$. For $N \rightarrow \infty$, one can write down a semiclassical formula 
\cite{Bijker:2016bpb}
\ba
E(v_1,v_2^{\ell_2},v_3^{\ell_3}),L) &=& E_0 + \omega_{1} \left( v_{1}+\frac{1}{2} \right) 
+ \omega_{2}(v_{2}+1) + \omega_{3} \left( v_{3}+\frac{3}{2} \right) 
\nonumber\\
&& + \kappa_1 \, L(L+1) ~.
\label{ener4}
\ea
which describes the energy levels of a spherical top. States are classified as in Eq.~(\ref{basis4}) and the
values of $t$ and $L^P$ for the ground state band and the fundamental vibrations are given in Eq.~(\ref{states4}).
In Fig.~\ref{tetrahedron} we show a schematic rotational-vibrational spectrum of a 
tetrahedral configuration. 

\subsection{Form factors and transition probabilities}
\label{emff}

The transition form factors are the matrix elements of 
$\sum_{i=1}^k \exp(i \vec{q} \cdot \vec{r}_i)$, where 
$\vec{q}$ is the momentum transfer and $\vec{r}_i$ is the location of the $\alpha$-particles. 
To do this calculation in the ACM, one first converts the transition operator to algebraic form 
and then calculates the form factors
\ba
{\cal F}(i \rightarrow f;q) \;=\; \langle \gamma_f, L_f, M \, 
| \, \hat T(q) \, | \, \gamma_i, L_i, M \rangle ~.
\label{ff}
\ea
The transition probabilities $B(EL)$ can be extracted from the form factors 
in the long wavelength limit
\ba
B(EL;i \rightarrow f) \;=\; (Ze)^2 \, \frac{[(2L+1)!!]^{2}}{4\pi (2L_{i}+1)} 
\, \lim_{q\rightarrow 0} \sum_M \frac{
\left| {\cal F}(i \rightarrow f;q) \right|^{2}}{q^{2L}} ~,
\label{belif}
\ea
where $Ze$ is the total electric charge of the cluster. 

\subsubsection{Dumbbell configuration}

Choosing the $z$-axis along the direction of the momentum transfer and using the 
fact that the two particles are identical, it is sufficient to consider the matrix 
elements of $\exp(i q r_{2z})$. After converting to Jacobi coordinates and integrating 
over the center-of-mass coordinate one has $\exp(-i q \rho_{z})$. The matrix elements of 
this operator can be obtained algebraically by making the replacement  
\ba
\rho_z \;\rightarrow\; \beta \hat D_{z}/X_D ~, 
\label{map2} 
\ea
where $\beta$ represents the scale of the coordinate and $X_D$ is given 
by the reduced matrix element of the dipole operator of Eq.~(\ref{dipole}). 
Explicit evaluation in the large $N$ limit gives 
\ba
{\cal F}(0^+ \rightarrow L^P;q) \;\rightarrow\; c_L \, j_L(q \beta) ~, 
\ea
with
\ba
c_L^2 \;=\; \frac{2L+1}{4} \left[ 2+2P_{L}(-1) \right] \;=\; (2L+1) \frac{1+(-1)^L}{2} ~. 
\label{cl2}
\ea
where $j_L$ is the spherical Bessel function, and $P_L$ the Legendre polynomial. From these, 
one can obtain the $B(EL)$ value 
\ba
B(EL;0^+ \rightarrow L^P) &=& \frac{(Ze \beta^L c_L)^2}{4\pi} 
\nonumber\\
&=& \left(\frac{Ze\beta^{L}}{2}\right)^{2} 
\frac{2L+1}{4\pi} \left[ 2+2P_{L}(-1) \right] ~. 
\label{BEL2}
\ea

\subsubsection{Equilateral triangle configuration}

Choosing again the $z$-axis along the direction of the momentum transfer and using the 
fact that the three particles are identical, it is sufficient to consider the matrix 
elements of $\exp(-i q \sqrt{2/3} \lambda_{z})$. By making the replacement  
\ba
\sqrt{\frac{2}{3}} \, \lambda_z \;\rightarrow\; \beta \hat D_{\lambda,z}/X_D ~, 
\label{map3} 
\ea
one can obtain the form factors for $N \rightarrow \infty$ in explicit form as 
\ba
{\cal F}(0^+ \rightarrow L^P;q) \;\rightarrow\; c_L \, j_L(q \beta) ~, 
\label{ff3}
\ea
with
\ba
c_L^2 &=& \frac{2L+1}{9} \left[ 3+6P_{L}(-\frac{1}{2}) \right] ~,
\label{cl3}
\ea
which gives $c_0^2=1$, $c_2^2=5/4$, $c_3^2=35/8$, $c_4^2=81/64$ and $c_5^2=385/128$. 
For a triangular configuration there is no dipole radiation $c_1^2=0$. 
The corresponding $B(EL)$ values are 
\ba
B(EL;0^+ \rightarrow L^P) &=& \frac{(Ze \beta^L c_L)^2}{4\pi} 
\nonumber\\
&=& \left(\frac{Ze\beta^{L}}{3}\right)^{2} 
\frac{2L+1}{4\pi} \left[ 3+6P_{L}(-\frac{1}{2}) \right] ~. 
\label{BEL3}
\ea

\subsubsection{Tetrahedral configuration}

The operator here is $\exp(-i q \sqrt{3/4} \, \eta_{z})$, and the replacement is  
\ba
\sqrt{3/4} \, \eta_z \;\rightarrow\; \beta \hat D_{\eta,z}/X_D ~, 
\label{map4}
\ea
One obtains
\ba
{\cal F}(0^+ \rightarrow L^P;q) \;\rightarrow\; c_L \, j_L(q \beta) ~, 
\label{ff4}
\ea
with
\ba
c_L^2 &=& \frac{2L+1}{16} \left[ 4+12P_{L}(-\frac{1}{3}) \right] ~,
\label{cl4}
\ea
which gives $c_0^2=1$, $c_3^2=35/9$, $c_4^2=7/3$ and $c_6^2=416/81$. 
For a tetrahedral configuration one has $c_1^2=c_2^2=c_5^2=0$.  
The corresponding $B(EL)$ values are give by 
\ba
B(EL;0^+ \rightarrow L^P) &=& \frac{(Ze \beta^L c_L)^2}{4\pi} 
\nonumber\\
&=& \left(\frac{Ze\beta^{L}}{4}\right)^{2} 
\frac{2L+1}{4\pi} \left[ 4+12P_{L}(-\frac{1}{3}) \right] ~. 
\label{BEL4}
\ea

\subsection{Cluster densities}
\label{cluster}

All results in Sect.~\ref{emff} are for point-like constituents, with density
\ba
\rho (\vec{r}) &=& \sum_{i=1}^{k}\delta (\vec{r}-\vec{r}_{i}) ~.
\label{charge}
\ea
This situation is not realistic, since the constituent $\alpha$-particles are not point-like. Assuming a
Gaussian form of the density of the $\alpha$-particle, one has the more realistic cluster density
\ba
\rho (\vec{r}) &=& \left( \frac{\alpha }{\pi }\right)^{3/2} 
\sum_{i=1}^{3}\exp \left[ -\alpha \left( \vec{r}-\vec{r}_{i}\right)^{2}\right] ~.
\label{rhor}
\ea
Here $\alpha=0.56$ fm$^{-2}$ describes the form factor of the $\alpha$-particle \cite{Sick:1970ma}.
For the density of Eq.~(\ref{rhor}), the form factors become
\ba
F(0^+ \rightarrow L^P;\vec{q}) &=& c_L j_L(q\beta) \, \mbox{e}^{-q^{2}/4\alpha} ~.  
\label{fq}
\ea
which represents the convolution of the form factor of the cluster with that of the $\alpha$-particle.
$B(EL)$ values, however, remain the same as in Sect.~\ref{emff} since in the long-wavelength limit 
$q \rightarrow 0$, the exponential factor $\mbox{exp}(-q^{2}/4\alpha) \rightarrow 1$. 
The density of Eq.~(\ref{rhor}) can be visualized by making an expansion into multipoles. By placing
the particles at a distance $\beta$ from the center of mass with spherical coordinates 
$(\beta,\theta_i,\phi_i)$ we then have \cite{DellaRocca:2017qkx}
\ba
\rho (\vec{r}) &=& \left( \frac{\alpha }{\pi }\right)^{3/2} 
\sum_{i=1}^{k}\exp \left[ -\alpha \left( \vec{r}-\vec{r}_{i}\right)^{2}\right] 
\nonumber \\
&=& \left( \frac{\alpha }{\pi }\right)^{3/2} \mbox{e}^{-\alpha(r^{2}+\beta ^{2})} 
\, 4\pi \,\sum_{\lambda\mu} i_{\lambda}(2\alpha \beta r) Y_{\lambda\mu}(\theta,\phi) 
\sum_{i=1}^{k} Y_{\lambda\mu}^{\ast}(\theta_{i},\phi_{i}) ~,  
\label{rhor2}
\ea
where $i_{\lambda}(x)=j_{\lambda}(ix)/i^{\lambda}$ is the modified spherical Bessel function. 
The matter and charge density for each configuration can be obtained from Eq.~(\ref{rhor2}) 
by multiplying by $Am/k$ and $Ze/k$, respectively. One should note that all results in 
Sect.~\ref{emff} can also be obtained from Eq.~(\ref{rhor2}) without making use of the algebraic 
approach, by taking the Fourier transform of the density.

\subsubsection{Dumbbell configuration}

\begin{figure}
\centering
\includegraphics[width=4.5in]{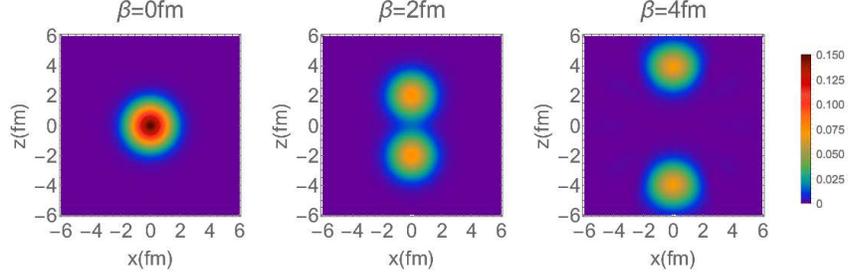} 
\caption[Densities of a $k=2$ $\alpha$-cluster]
{Densities of a $k=2$ $\alpha$-cluster as given in Eq.~(\ref{rhor2}). 
The value of $\alpha=0.56$ fm$^{-2}$. The color scale is in fm$^{-3}$. 
Reproduced from \cite{DellaRocca:2017qkx} with permission.}
\label{density2}
\end{figure}

For $Z_2$ symmetry, the origin is chosen in the center of mass, and the angles of the two particles 
are given by $(\theta_1,\phi_1)=(0,-)$ and $(\theta_2,\phi_2)=(\pi,-)$, and 
\ba
\sum_{i=1}^{2} Y^{\ast}_{\lambda \mu}(\theta_i,\phi_i) &=& \sqrt{\frac{2\lambda+1}{4\pi}} 
\left[ \delta_{\mu,0} + \sqrt{\frac{(\lambda+\mu)!}{(\lambda-\mu)!}} \, P_{\lambda}^{-\mu} (-1) \right]
\nonumber\\
&=& \delta_{\mu,0} \sqrt{\frac{2\lambda+1}{4\pi}} \left[ 1 + P_{\lambda}(-1) \right] ~.
\label{geometry2}
\ea
This configuration has axial symmetry. In the multipole expansion, only $\mu=0$ and
$\lambda=\mbox{even}=0,2,\ldots,$ remain. The charge and matter densities of the dumbbell 
configuration are shown in Fig.~\ref{density2} for $\beta=0$, $2$ and $4$ fm. The density 
describes the entire range from united constituent particles ($\beta=0$) to separated 
constituent particles ($\beta \rightarrow \infty$). Note that the density describes also 
break-up into two fragments, as shown in the panel on the right-hand side of Fig.~\ref{density2}. 

\subsubsection{Equilateral triangle configuration}

\begin{figure}
\centering
\includegraphics[width=4.5in]{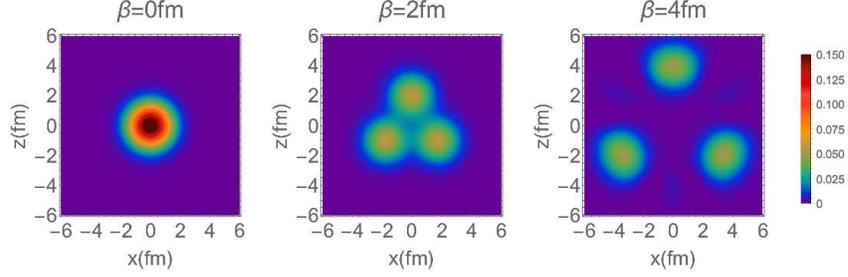} 
\caption[Densities of a $k=3$ $\alpha$-cluster]
{Densities of a $k=3$ $\alpha$-cluster as given in Eq.~(\ref{rhor2}). 
The value of $\alpha=0.56$ fm$^{-2}$. The color scale is in fm$^{-3}$. 
Reproduced from \cite{DellaRocca:2017qkx} with permission.}
\label{density3}
\end{figure}

For the $D_{3h}$ symmetry of an equilateral triangle, the angles of the particles 
are given by $(\theta_1,\phi_1)=(0,-)$, $(\theta_2,\phi_2)=(2\pi/3,0)$ and $(\theta_3,\phi_3)=(2\pi/3,\pi)$, and 
\ba
\sum_{i=1}^{3} Y^{\ast}_{\lambda \mu}(\theta_i,\phi_i) &=& \sqrt{\frac{2\lambda+1}{4\pi}} 
\left[ \delta_{\mu,0} + \sqrt{\frac{(\lambda+\mu)!}{(\lambda-\mu)!}} \, 
P_{\lambda}^{-\mu} (-\tfrac{1}{2}) (1 + (-1)^{\mu}) \right]
\nonumber\\
&=& \left\{ \begin{array}{lll} \sqrt{\frac{2\lambda+1}{4\pi}} \left[ 1 + 2P_{\lambda}(-\frac{1}{2}) \right] 
&& \mu=0 \\ && \\
\sqrt{\frac{2\lambda+1}{4\pi}} \sqrt{\frac{(\lambda+\mu)!}{(\lambda-\mu)!}} \, 2P_{\lambda}^{-\mu}(-\frac{1}{2}) 
&& \mu=2\kappa \neq 0 
\end{array} \right.
\label{geometry3}
\ea
where $\kappa=1,2,\ldots,$ and $\mu \leq \lambda$. 
For this configuration, the remaining multipoles are $\lambda=0,2,3,4,\ldots,$ corresponding to the fact
that the density is invariant under $D_{3h}$ transformations and thus belongs to the symmetric representation of
$D_{3h}$ \cite{Bijker:2002ac}. The charge (and matter) densities of a triangular configuration are shown 
in Fig.~\ref{density3} for different values of $\beta$. 

\subsubsection{Tetrahedral configuration}

\begin{figure}
\centering
\includegraphics[width=4.5in]{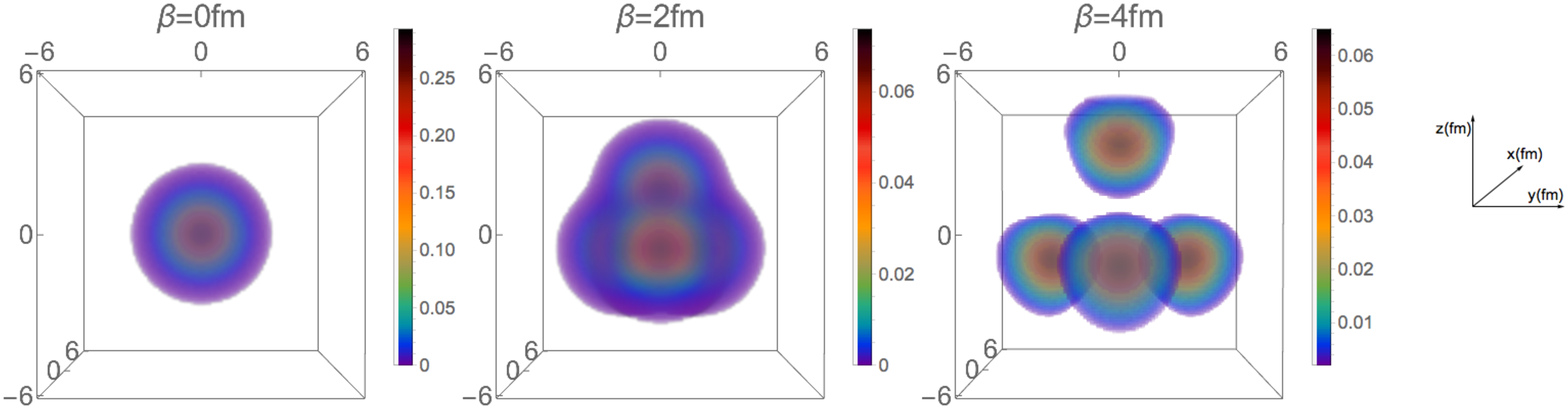} 
\caption[Densities of a $k=4$ $\alpha$-cluster]
{Densities of a $k=4$ $\alpha$-cluster as given in Eq.~(\ref{rhor2}). 
The value of $\alpha=0.56$ fm$^{-2}$. The color scale is in fm$^{-3}$. 
Reproduced from \cite{DellaRocca:2017qkx} with permission.}
\label{density4}
\end{figure}

For the $T_d$ symmetry of the regular tetrahedron, the angles of the particles are given by $(\theta_1,\phi_1)=(0,-)$, $(\theta_2,\phi_2)=(\gamma,0)$, $(\theta_3,\phi_3)=(\gamma,2\pi/3)$ 
and $(\theta_4,\phi_4)=(\gamma,4\pi/3)$ with $\cos \gamma=-1/3$, and 
\ba
\sum_{i=1}^{4} Y^{\ast}_{\lambda \mu}(\theta_i,\phi_i) &=& \sqrt{\frac{2\lambda+1}{4\pi}} 
\left[ \delta_{\mu,0} + \sqrt{\frac{(\lambda+\mu)!}{(\lambda-\mu)!}} \, 
P_{\lambda}^{-\mu} (-\tfrac{1}{3}) (1 + 2\cos \tfrac{2\mu \pi}{3}) \right]
\nonumber\\
&=& \left\{ \begin{array}{lll} \sqrt{\frac{2\lambda+1}{4\pi}} \left[ 1 + 3P_{\lambda}(-\frac{1}{3}) \right] 
&& \mu=0 \\ && \\
\sqrt{\frac{2\lambda+1}{4\pi}} \sqrt{\frac{(\lambda+\mu)!}{(\lambda-\mu)!}} \, 3P_{\lambda}^{-\mu}(-\frac{1}{3}) 
&& \mu=3\kappa \neq 0 
\end{array} \right. 
\label{geometry4}
\ea
where $\kappa=1,2,\ldots,$ and $\mu \leq \lambda$. 
For this configuration, the remaining multipoles are $\lambda=0,3,4,6,\ldots,$ corresponding to the $A_1$
representation of the tetrahedral group, $T_d$ \cite{Bijker:2016bpb}. The charge and matter densities of a tetrahedral
configuration are shown in Fig.~\ref{density4}.

\subsection{Moments of inertia and radii}

From the density Eq.~(\ref{rhor2}) one can calculate the moments of inertia and radii. The three
components of the moment of inertia are given by
\ba
{\cal I}_{x} &=& \int (y^{2}+z^{2}) \rho(\vec{r}) \, d^3 r ~,  
\nonumber \\
{\cal I}_{y} &=& \int (z^{2}+x^{2}) \rho(\vec{r}) \, d^3 r ~,  
\nonumber \\
{\cal I}_{z} &=& \int (x^{2}+y^{2}) \rho(\vec{r}) \, d^3 r ~,
\label{inertia}
\ea
and radii by
\ba
\left< r^2 \right> &=& \int r^2 \rho(\vec{r}) \, d^3 r ~.
\label{radius}
\ea

\subsubsection{Dumbbell configuration}

Introducing the appropriate normalization, one has
\ba
{\cal I}_{x} \;=\; {\cal I}_{y} &=& Am\beta^{2} \left( 1 + \frac{1}{\alpha\beta^{2}} \right) ~,  
\nonumber \\
{\cal I}_{z} &=& \frac{Am}{\alpha} ~. 
\label{inertia2}
\ea
where $A=4k=8$, corresponding to a prolate top, and 
\ba
\left< r^2 \right>^{1/2} &=& \sqrt{\beta^2+\frac{3}{2\alpha}} ~.
\ea

\subsubsection{Equilateral triangle configuration}

In this case, one has \cite{Bijker:2002ac}
\ba
{\cal I}_{x} \;=\; {\cal I}_{z} &=& \frac{1}{2} Am \beta^{2} 
\left( 1+\frac{2}{\alpha \beta^{2}} \right) ~,  
\nonumber \\
{\cal I}_{y} &=& Am \beta^{2} \left( 1+\frac{1}{\alpha \beta^{2}} \right) ~. 
\label{inertia3}
\ea
where $A=4k=12$, corresponding to an oblate top, and 
\ba
\left< r^2 \right>^{1/2} &=& \sqrt{\beta^2+\frac{3}{2\alpha}} ~.
\ea

\subsubsection{Tetrahedral configuration}

For the tetrahedral configuration, all three moments of inertia are the same 
\ba
{\cal I}_{x} \;=\; {\cal I}_{y} \;=\; {\cal I}_{z} \;=\; \frac{2}{3} Am \beta^{2} 
\left( 1+\frac{3}{2\alpha \beta^{2}} \right) ~,  
\label{inertia4}
\ea
where $A=4k=16$, corresponding to a spherical top, and 
\ba
\left< r^2 \right>^{1/2} &=& \sqrt{\beta^2+\frac{3}{2\alpha}} ~.
\ea

\section{Evidence for cluster structures}
\label{evidence}

The ACM provides a simple way to analyze experimental data, thus determining whether or not
the symmetries $Z_2$, $D_{3h}$, and $T_d$ appear in the spectra of $^8$Be, $^{12}$C and $^{16}$O.

\subsection{Energies}

\subsubsection{Dumbbell configuration}
\label{sec311}

\begin{figure}
\centering
\includegraphics[width=4.5in]{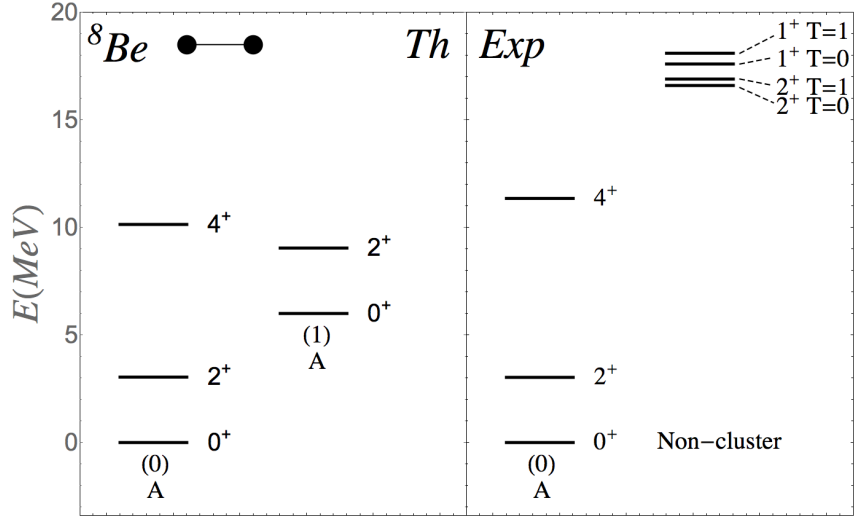} 
\caption[Energy spectrum of $^{8}$Be]
{Comparison between the cluster spectrum and the experimental spectrum \cite{Tilley:2004zz} of $^{8}$Be. 
The theoretical spectrum is calculated using Eq.~(\ref{ener2}) with $D=6$ MeV and $B=0.507$ MeV. 
Figure adapted from \cite{DellaRocca:2018mrt}.}
\label{Be8}
\end{figure}

\begin{figure}
\centering
\includegraphics[width=2.5in]{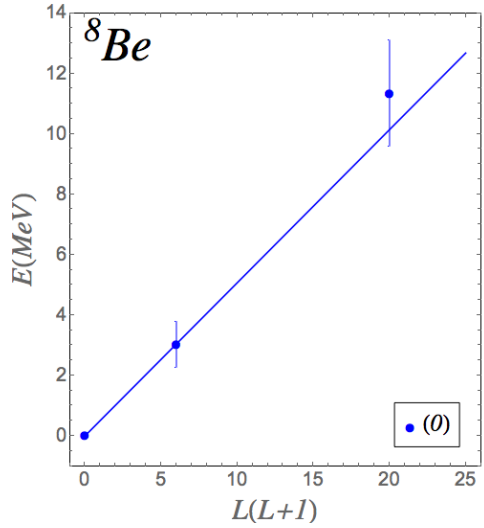} 
\caption[Rotational band in $^{8}$Be]
{Observed cluster rotational band in $^{8}$Be, $v=0$. 
The experimental bar is the width $\Gamma$. Figure adapted from \cite{DellaRocca:2018mrt}.}
\label{bands2}
\end{figure}

Energy levels for this configuration can be analyzed with Eq.~(\ref{ener2}). A comparison with data in
$^8$Be \cite{Tilley:2004zz} is shown in Fig.~\ref{Be8}. The occurrence of a rotational band in the 
experimental spectrum is clearly seen in Fig.~\ref{bands2}, where the energy of the states is shown 
as a function of $L(L+1)$. No evidence for the vibrational bands is reported in \cite{Tilley:2004zz}, 
although Barker \cite{Barker_1968,Barker_1969} suggested in the 1960's one such a band 
at $E \sim 6$ MeV, in accordance to similar vibrational bands observed in $^{12}$C and $^{16}$O. 
The non-observation of the vibrational band in $^{8}$Be may be due to its expected large width. 

From the value of $B=\hbar^2/2{\cal I}$ extracted from the experimental energy difference, 
$E_{2^+}-E_{0^+}$, one can determine the moment of inertia ${\cal I}={\cal I}_x={\cal I}_y$ and from 
Eq.~(\ref{inertia2}) the value of $\beta=1.82 \pm 0.04$ fm \cite{DellaRocca:2018mrt}.

\subsubsection{Equilateral triangle configuration}

\begin{figure}
\centering
\includegraphics[width=4.5in]{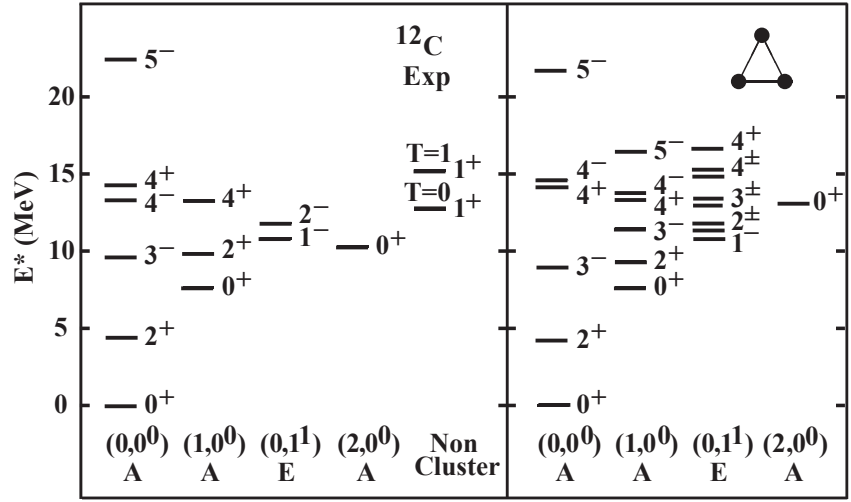} 
\caption[Energy spectrum of $^{12}$C]
{Comparison between the cluster spectrum and the experimental spectrum of $^{12}$C. 
Reproduced from \cite{Marin-Lambarri:2014zxa} with permission.}
\label{C12}
\end{figure}

\begin{figure}
\centering
\includegraphics[width=4in]{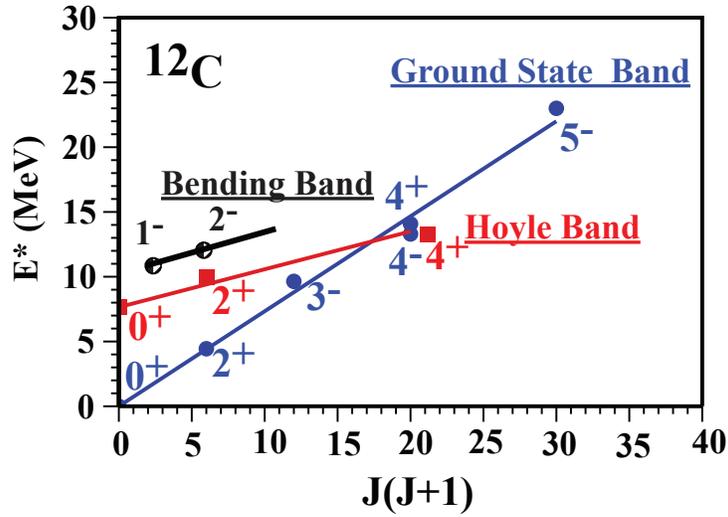} 
\caption[Rotational bands in $^{12}$C]
{Rotational bands in $^{12}$C. Reproduced from \cite{Marin-Lambarri:2014zxa} with permission.}
\label{bands3}
\end{figure}

Recent experiments \cite{Marin-Lambarri:2014zxa} have confirmed the occurrence of $D_{3h}$ symmetry 
in $^{12}$C. Energy levels have been analyzed with a variation of Eq.~(\ref{ener3}) which includes 
anharmonic terms. The results are shown in Figs.~\ref{C12} and \ref{bands3}. One can see here the 
occurrence of not only rotational bands with angular momentum content expected from $D_{3h}$ symmetry, 
but also the occurrence of the fundamental vibrations of the triangle of Fig.~\ref{triangle} 
$(v_1,v_2^{\ell_2})=(1,0^0)$ and $(0,1^1)$ with symmetry $A$ and $E$, respectively. 

\subsubsection{Tetrahedral configuration}

\begin{figure}
\centering
\vspace{15pt}
\setlength{\unitlength}{0.6pt}
\begin{picture}(560,335)(-50,0)
\normalsize
\thinlines
\put (  0,  0) {\line(0,1){335}}
\put (  0,  0) {\line(1,0){510}}
\put (  0,335) {\line(1,0){510}}
\put (280,  0) {\line(0,1){335}}
\put (510,  0) {\line(0,1){335}}
\thicklines
\put (  0, 60) {\line(1,0){5}}
\put (  0,110) {\line(1,0){5}}
\put (  0,160) {\line(1,0){5}}
\put (  0,210) {\line(1,0){5}}
\put (  0,260) {\line(1,0){5}}
\put (  0,310) {\line(1,0){5}}
\put (275, 60) {\line(1,0){10}}
\put (275,110) {\line(1,0){10}}
\put (275,160) {\line(1,0){10}}
\put (275,210) {\line(1,0){10}}
\put (275,260) {\line(1,0){10}}
\put (275,310) {\line(1,0){10}}
\put (505, 60) {\line(1,0){5}}
\put (505,110) {\line(1,0){5}}
\put (505,160) {\line(1,0){5}}
\put (505,210) {\line(1,0){5}}
\put (505,260) {\line(1,0){5}}
\put (505,310) {\line(1,0){5}}
\put (-20, 55) { 0}
\put (-20,155) {10}
\put (-20,255) {20}
\put (-30,305) {E}
\put (-50,290) {(MeV)}
\put ( 20, 60.00) {\line(1,0){20}}
\put ( 20,121.30) {\line(1,0){20}}
\put ( 20,163.56) {\line(1,0){20}}
\put ( 20,270.52) {\line(1,0){20}}
\put ( 20, 35) {$(000)$}
\put ( 30, 20) {$A$}
\put ( 45, 55.00) {$0^+$}
\put ( 45,116.30) {$3^-$}
\put ( 45,158.56) {$4^+$}
\put ( 45,265.52) {$6^+$}
\put ( 70,120.49) {\line(1,0){20}}
\put ( 70,176.00) {\line(1,0){20}}
\put ( 70,206.20) {\line(1,0){20}}
\put ( 70,298.79) {\line(1,0){20}}
\put ( 70, 35) {$(100)$}
\put ( 80, 20) {$A$}
\put ( 95,115.49) {$0^+$}
\put ( 95,171.00) {$3^-$}
\put ( 95,201.20) {$4^{(+)}$}
\put ( 95,293.79) {$6^+$}
\put (120,129.17) {\line(1,0){20}}
\put (120,148.72) {\line(1,0){20}}
\put (120,170.97) {\line(1,0){20}}
\put (120, 35) {$(010)$}
\put (130, 20) {$E$}
\put (145,124.17) {$2^+$}
\put (145,143.72) {$2^-$}
\put (145,165.97) {$4^+$}
\put (170,131.16) {\line(1,0){20}}
\put (170,158.44) {\line(1,0){20}}
\put (170,170.80) {\line(1,0){20}}
\put (170, 35) {$(001)$}
\put (180, 20) {$F$}
\put (195,126.16) {$1^-$}
\put (195,153.44) {$2^+$}
\put (195,165.80) {$3^+$}
\put (220,155.85) {\line(1,0){20}}
\put (220,169.57) {\line(1,0){20}}
\put (220,187.96) {\line(1,0){20}}
\put (220,190.90) {\line(1,0){20}}
\put (230, 35) {Non}
\put (220, 20) {Cluster}
\put (245,150.85) {$1^-$}
\put (245,164.57) {$0^-$}
\put (220,130) {$T=0$}
\put (245,177.96) {$0^-$}
\put (245,190.90) {$1^-$}
\put (220,210) {$T=1$}
\Large
%\put (210,270) {Exp}
\put (210,290) {$^{16}$O}
\normalsize
\put (300, 60.0) {\line(1,0){20}}
\put (300,121.3) {\line(1,0){20}}
\put (300,162.2) {\line(1,0){20}}
\put (300,274.6) {\line(1,0){20}}
\put (300, 35) {$(000)$}
\put (310, 20) {$A$}
\put (325, 55.00) {$0^+$}
\put (325,116.30) {$3^-$}
\put (325,157.2) {$4^+$}
\put (325,269.6) {$6^{\pm}$}
\put (350,120.5) {\line(1,0){20}}
\put (350,169.7) {\line(1,0){20}}
\put (350,202.5) {\line(1,0){20}}
\put (350,292.7) {\line(1,0){20}}
\put (350, 35) {$(100)$}
\put (360, 20) {$A$}
\put (375,115.5) {$0^+$}
\put (375,164.7) {$3^-$}
\put (375,197.5) {$4^+$}
\put (375,287.7) {$6^{\pm}$}
\put (400,137.4) {\line(1,0){20}}
\put (400,176.9) {\line(1,0){20}}
\put (400,205.1) {\line(1,0){20}}
\put (400, 35) {$(010)$}
\put (410, 20) {$E$}
\put (425,132.4) {$2^{\pm}$}
\put (425,171.9) {$4^{\pm}$}
\put (425,200.1) {$5^{\pm}$}
\put (450,128.5) {\line(1,0){20}}
\put (450,144.6) {\line(1,0){20}}
\put (450,168.7) {\line(1,0){20}}
\put (450,200.9) {\line(1,0){20}} 
\put (450, 35) {$(001)$}
\put (460, 20) {$F$}
\put (475,123.5) {$1^-$}
\put (475,139.6) {$2^+$}
\put (475,163.7) {$3^{\pm}$}
\put (475,195.9) {$4^{\pm}$}
\put(430,270) {\circle*{10}} 
\put(470,270) {\circle*{10}}
\put(480,290) {\circle*{10}}
\put(450,310) {\circle*{10}}
\put(430,270) {\line( 1,0){40}}
\put(430,270) {\line( 1,2){20}}
\put(470,270) {\line(-1,2){20}}
\put(470,270) {\line( 1,2){10}}
\put(480,290) {\line(-3,2){30}}
\multiput(430,270)(5,2){11}{\circle*{2}}
\end{picture}
\caption[Energy spectrum of $^{16}$O]
{Comparison between the cluster spectrum and the experimental spectrum of $^{16}$O. 
Reproduced from \cite{Bijker:2016bpb} with permission.}
\label{O16}
\end{figure}
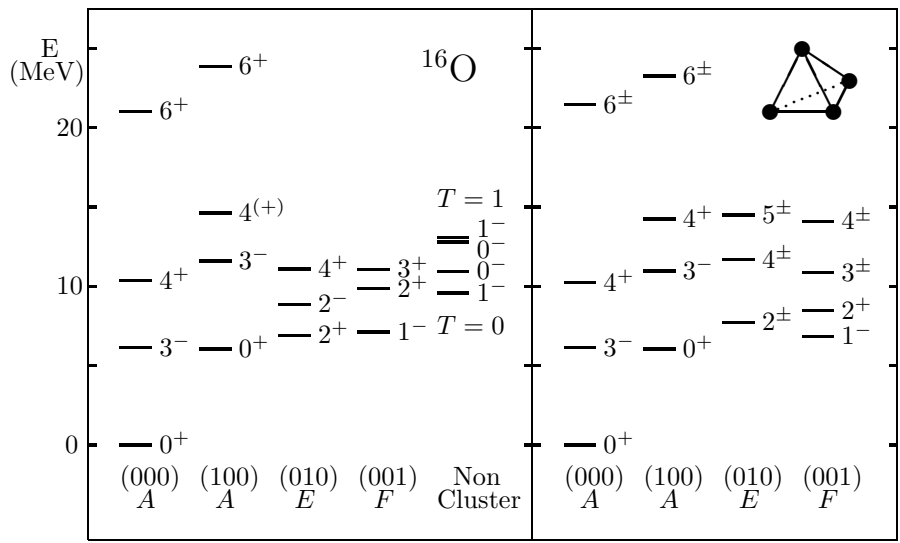

\begin{figure}
\centering
\includegraphics[width=4in]{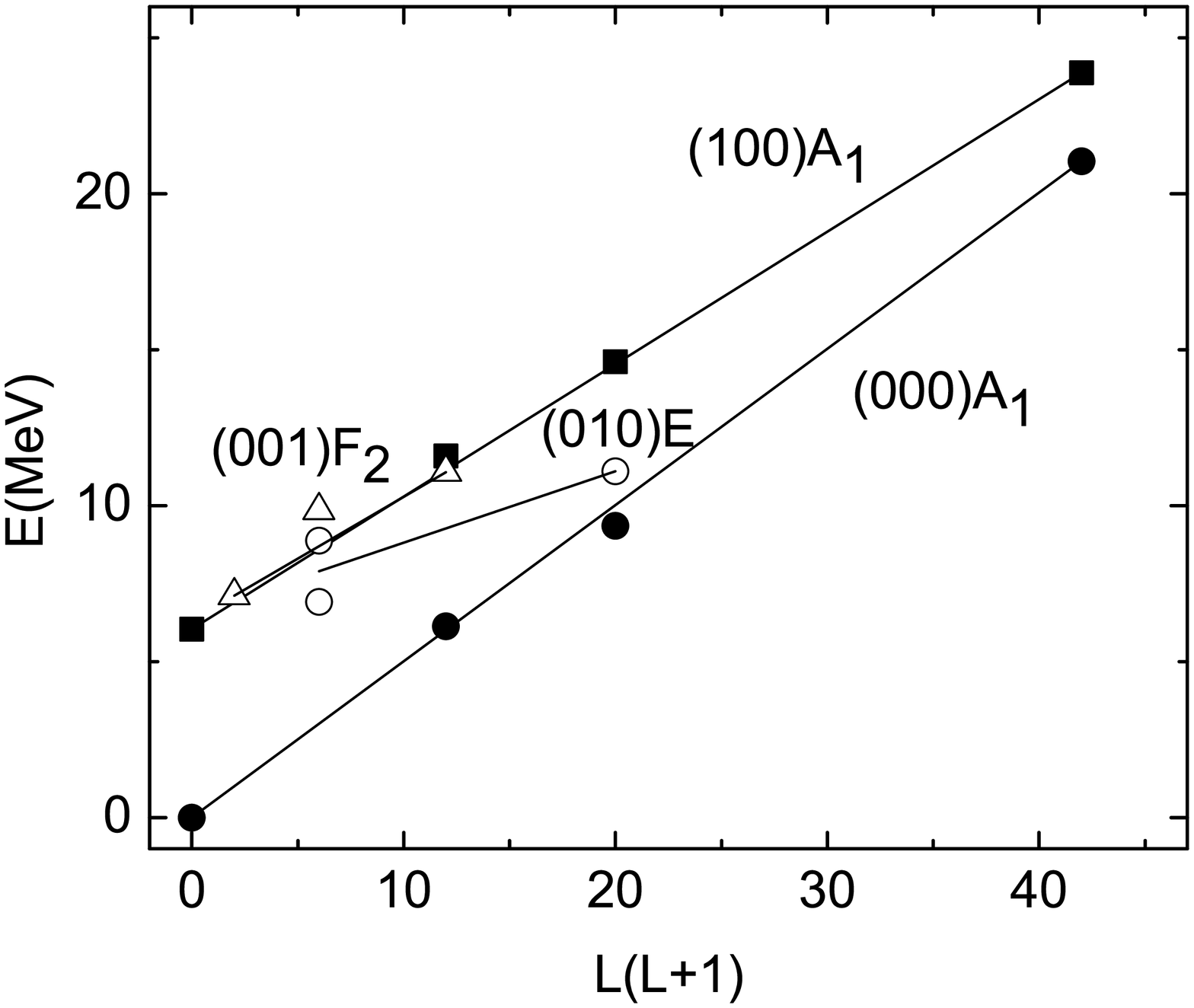}
\vspace{-0.5cm}
\caption[Rotational bands in $^{16}$O]
{Rotational bands in $^{16}$O. Reproduced from \cite{Bijker:2016bpb} with permission.}
\label{bands4}
\end{figure}

The occurrence of $T_d$ symmetry in $^{16}$O was emphasized by Robson \cite{Robson:1978vh,ROBSON1982257} 
in the 1970’s and more recently revisited in \cite{Bijker:2014tka,Bijker:2016bpb}. Energy levels have 
been analyzed with Eq.~(\ref{ener4}). A comparison with data is shown in Figs.~\ref{O16} and \ref{bands4}.

\subsection{Form factors}
\label{formfactor}

Form factors in electron scattering can be simply derived by making use of the formulas given in
Sect.~\ref{emff} in the rigid case, or, in the more general situation, by evaluating the matrix elements of
the operator $\hat{T}(q)$ of Eq.~(\ref{ff}) in the wave functions obtained by diagonalizing the Hamiltonian
of Sect.~\ref{energies}.

\subsubsection{Dumbbell configuration}

The nucleus $^{8}$Be is unstable and therefore form factors in electron scattering cannot be measured. 
The value of $\beta$ for this configuration is estimated from the moment of inertia to be 
$\beta=1.82 \pm 0.04$ fm, as given in Section~\ref{sec311}.

\subsubsection{Equilateral triangle configuration}

\begin{figure}
\vfill 
\begin{minipage}{.5\linewidth}
\centerline{\includegraphics[width=\linewidth]{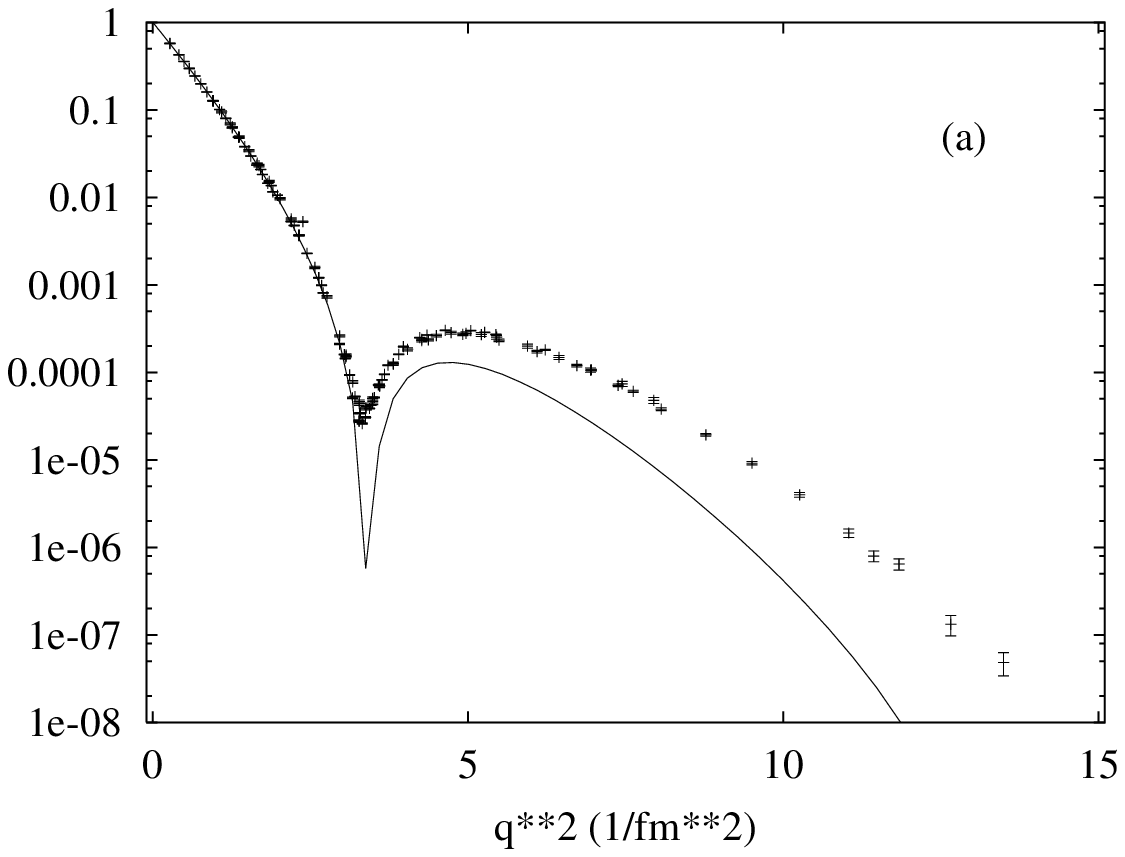}}
\end{minipage}\hfill
\begin{minipage}{.5\linewidth}
\centerline{\includegraphics[width=\linewidth]{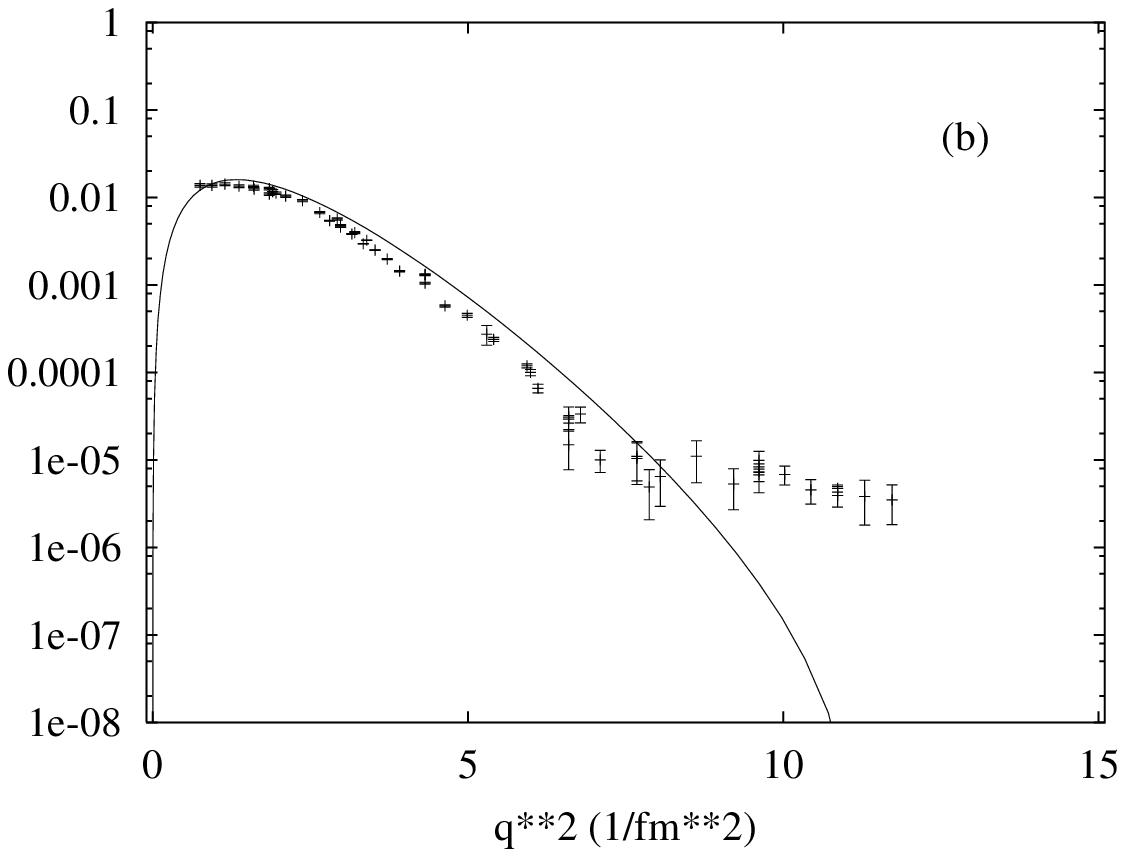}}
\end{minipage}
\begin{minipage}{.5\linewidth}
\centerline{\includegraphics[width=\linewidth]{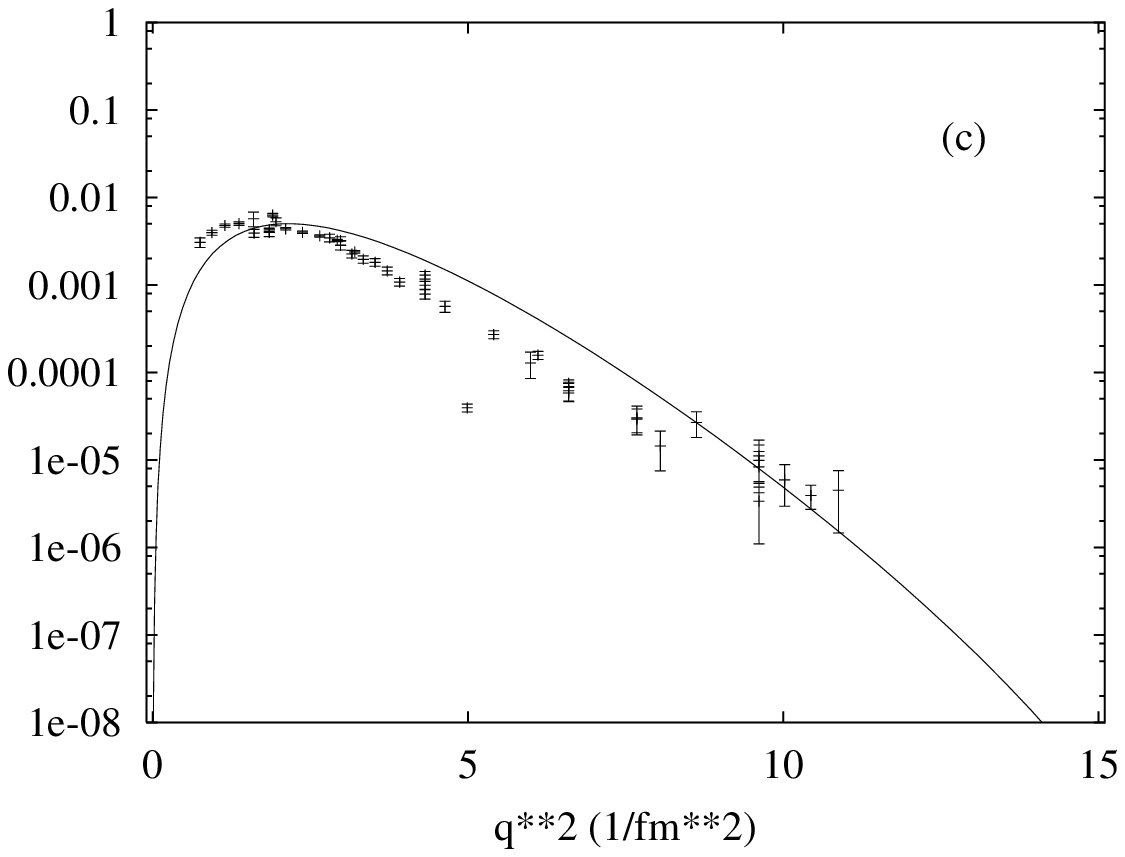}}
\end{minipage}\hfill
\begin{minipage}{.5\linewidth}
\centerline{\includegraphics[width=\linewidth]{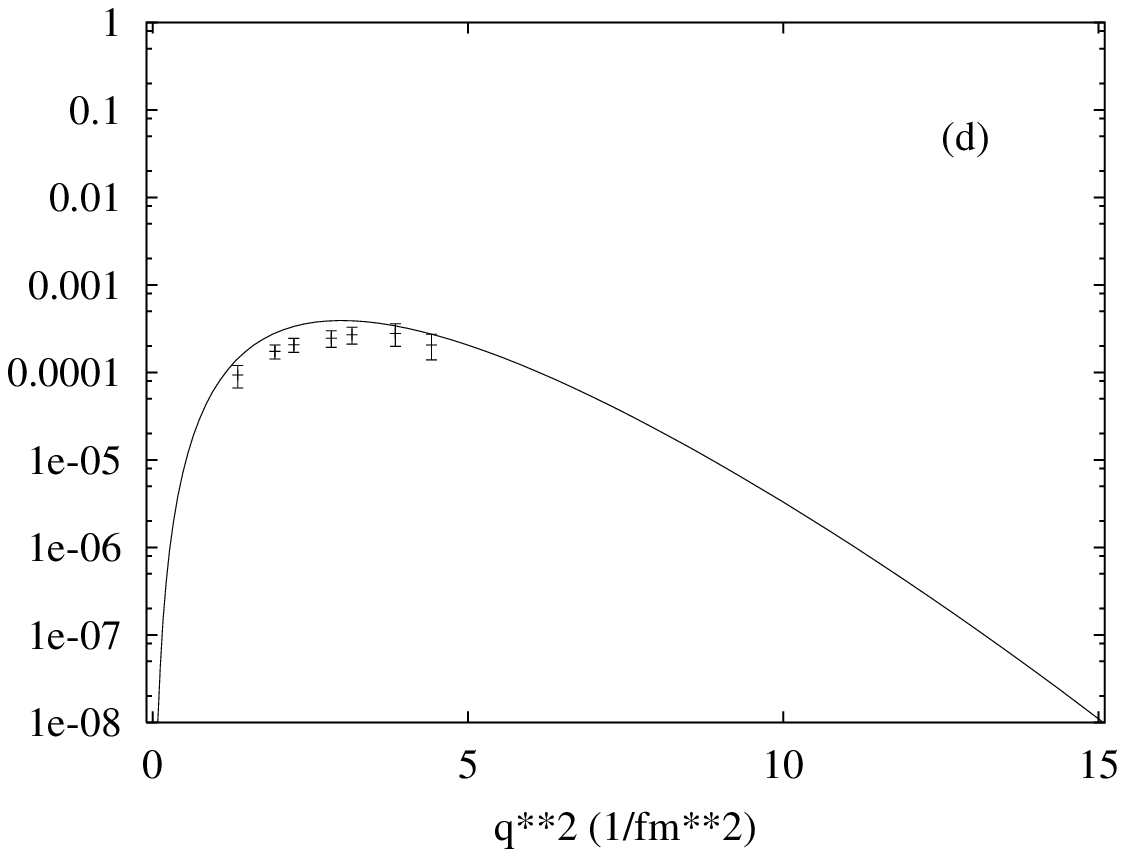}}
\end{minipage}
\begin{minipage}{.5\linewidth}
\centerline{\includegraphics[width=\linewidth]{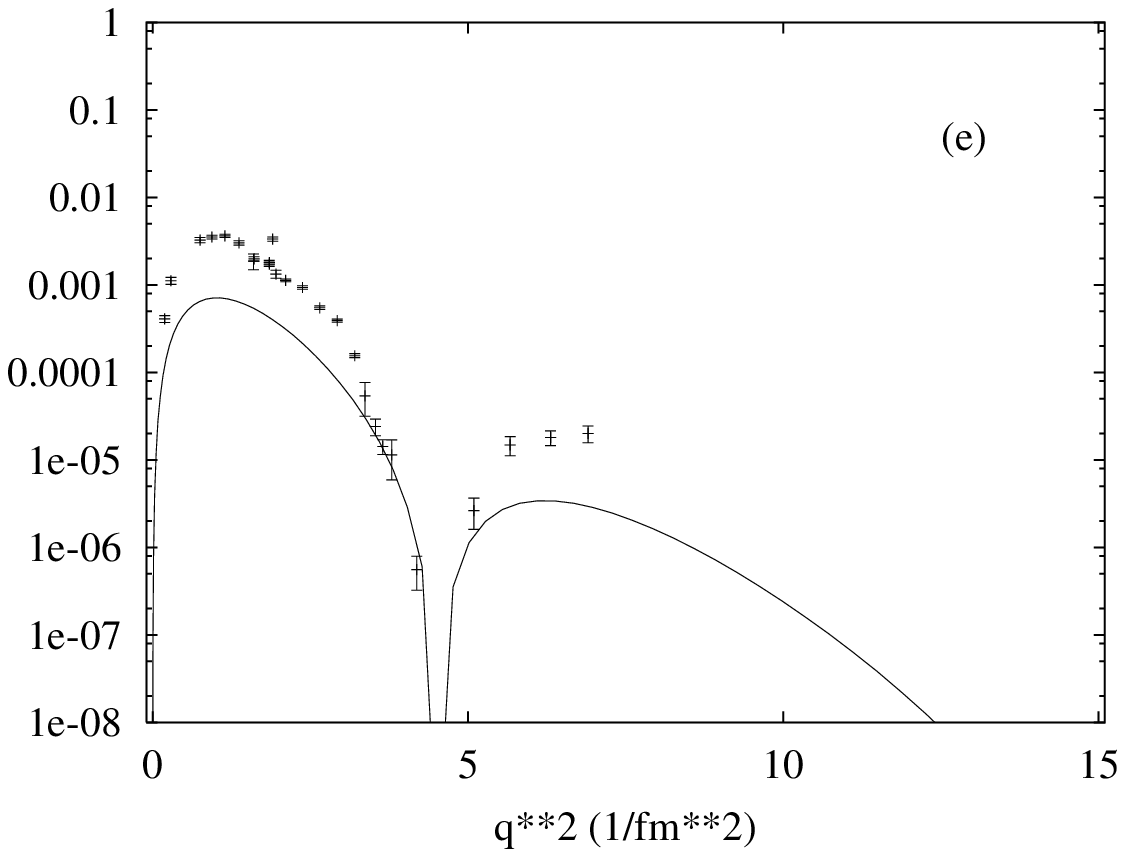}}
\end{minipage}\hfill
\begin{minipage}{.5\linewidth}
\centerline{\includegraphics[width=\linewidth]{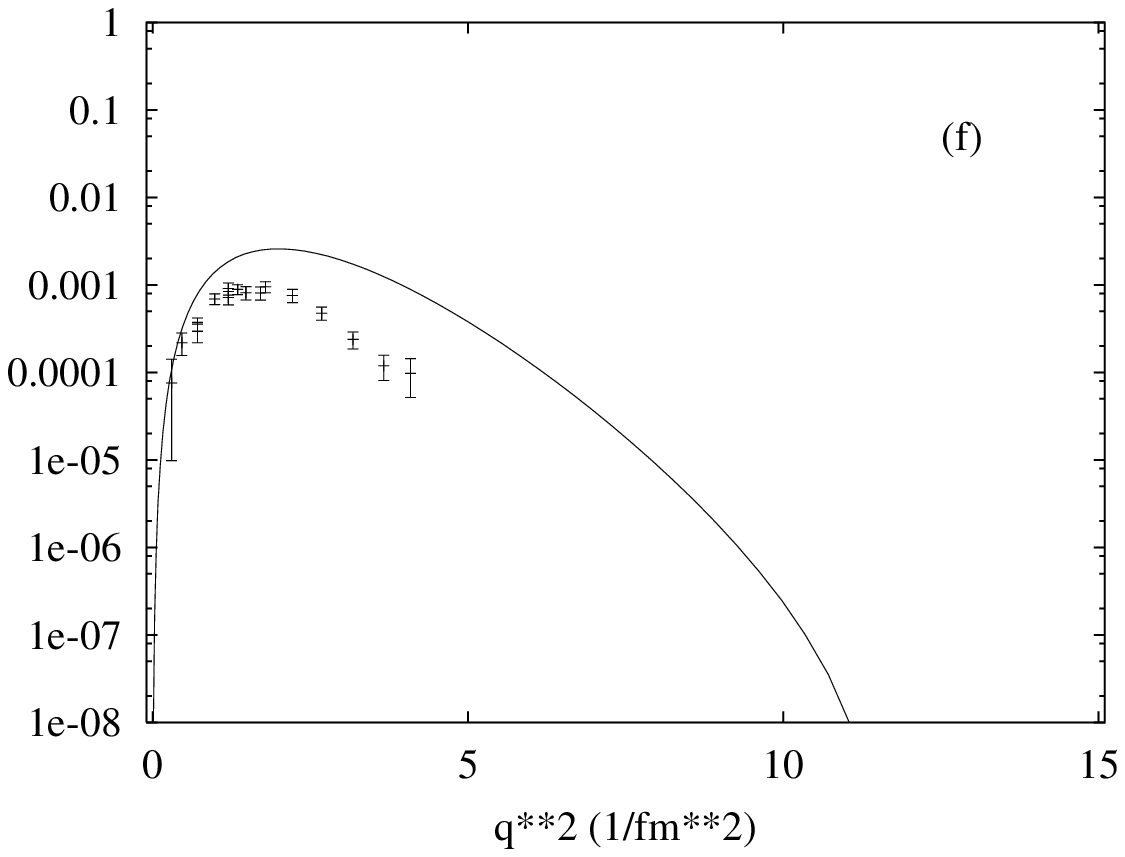}}
\end{minipage}
\caption[Form factors for $^{12}$C]
{Comparison between the experimental form factors 
$|{\cal F}(0^+_1 \rightarrow L^P_i;q)|^2$ of $^{12}$C 
for the final states (a) $L^P_i=0^+_1$ (elastic), (b) $L^P_i=2^+_1$, 
(c) $L^P_i=3^-_1$, (d) $L^P_i=4^+_1$, (e) $L^P_i=0^+_2$, and 
(f) $L^P_i=1^-_1$ and those obtained for the oblate top with $N=10$. 
Reproduced from \cite{Bijker:2002ac} with permission.}
\label{ffc12}
\end{figure}

Form factors in $^{12}$C have been extensively investigated. In the rigid case, only states in the
ground state band are excited with form factors given by Eq.~(\ref{ff3}) and no excitation of the
vibrational bands occurs. Since experimentally excitation of these bands occurs, although with a
small strength, one needs in this case to do a calculation with the general algebraic Hamiltonian,
Eq.~(\ref{ham3}) \cite{Bijker:2002ac}. The resulting form factors are shown in Fig.~\ref{ffc12}, 
where they are compared with experimental data. The value of $\beta$ is determined from the first 
minimum of the elastic form factor to be $\beta=1.74 \pm 0.04$ fm \cite{Bijker:2002ac} 
with an estimated error of 2 \%. The experimental form factors in Fig.~\ref{ffc12} compare well with 
the theoretical form factors, except for the transition form factor $|{\cal F}(0^+_1 \rightarrow 0^+_2)|^2$ 
whose shape is correctly given but whose magnitude is smaller by a factor of 10. This discrepancy is the 
subject of current investigations \cite{Bijker_2017} and seems to indicate that the structure of the 
$0^+_2$ Hoyle state may be somewhat softer than calculated by the rigid oblate configuration. 

\subsubsection{Tetrahedral configuration}

\begin{figure}
\begin{minipage}{.5\linewidth}
\centerline{\includegraphics[width=\linewidth]{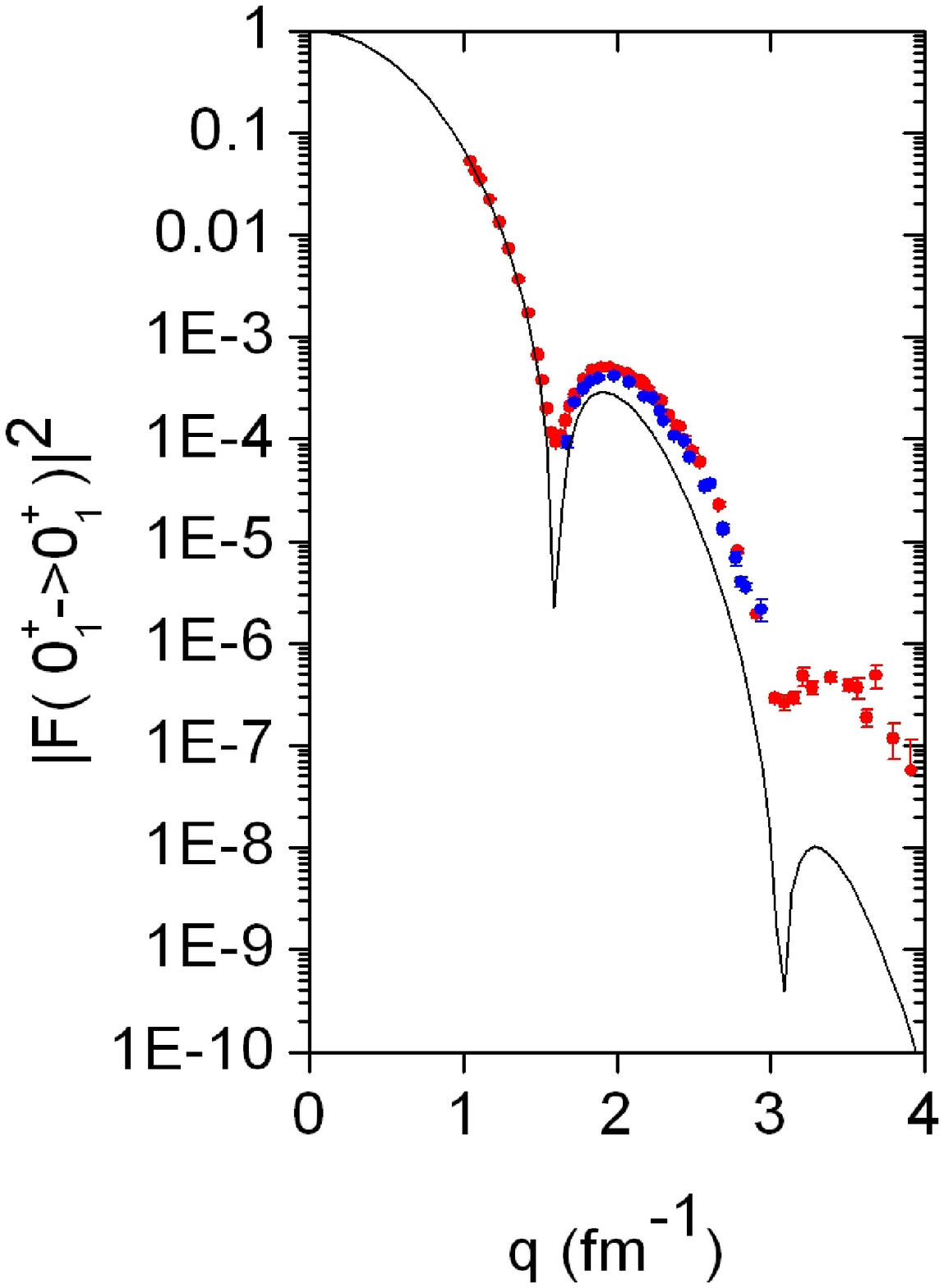}}
\end{minipage}\hfill
\begin{minipage}{.5\linewidth}
\centerline{\includegraphics[width=\linewidth]{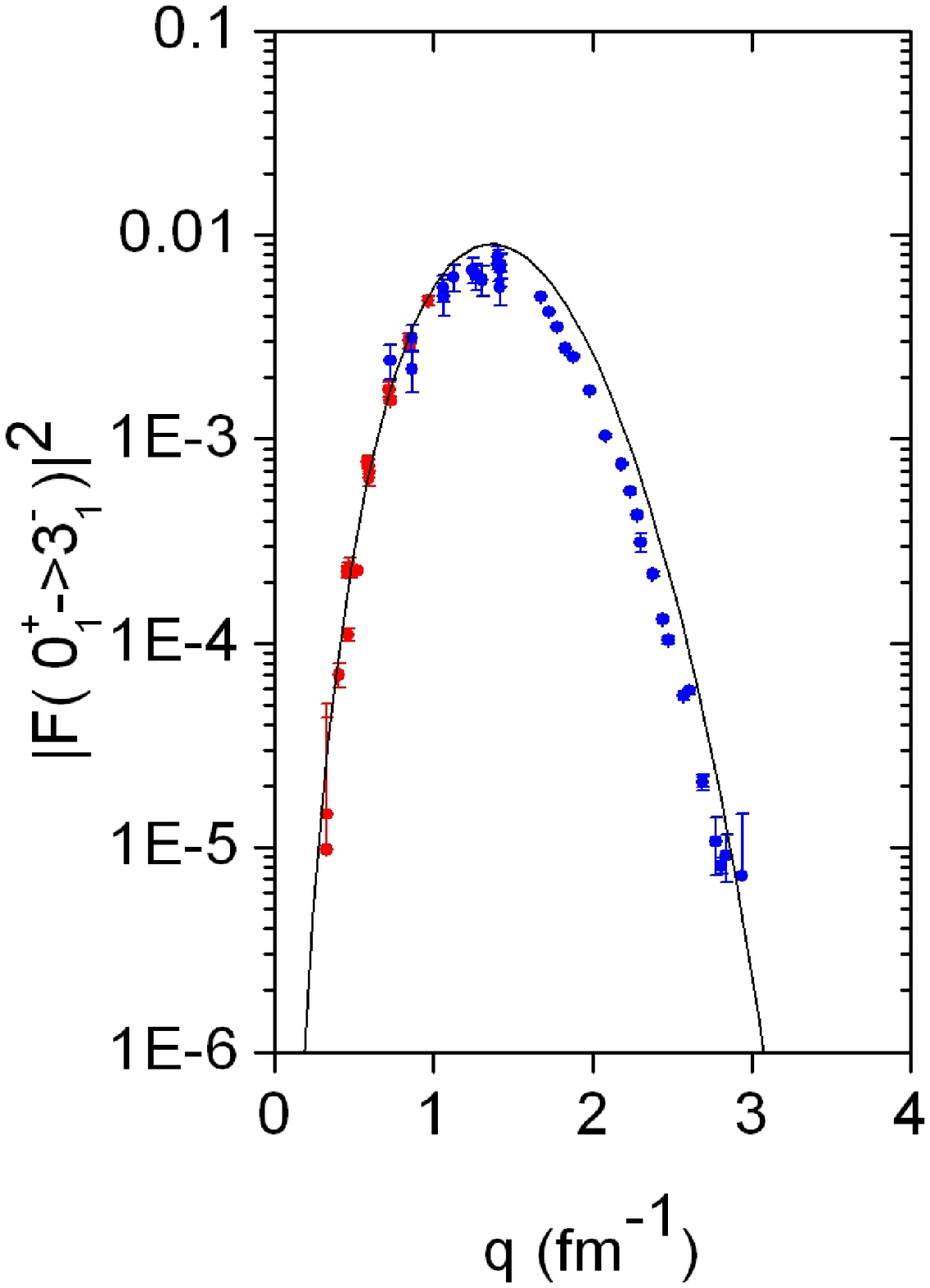}}
\end{minipage}
\vfill
\begin{minipage}{.5\linewidth}
\centerline{\includegraphics[width=\linewidth]{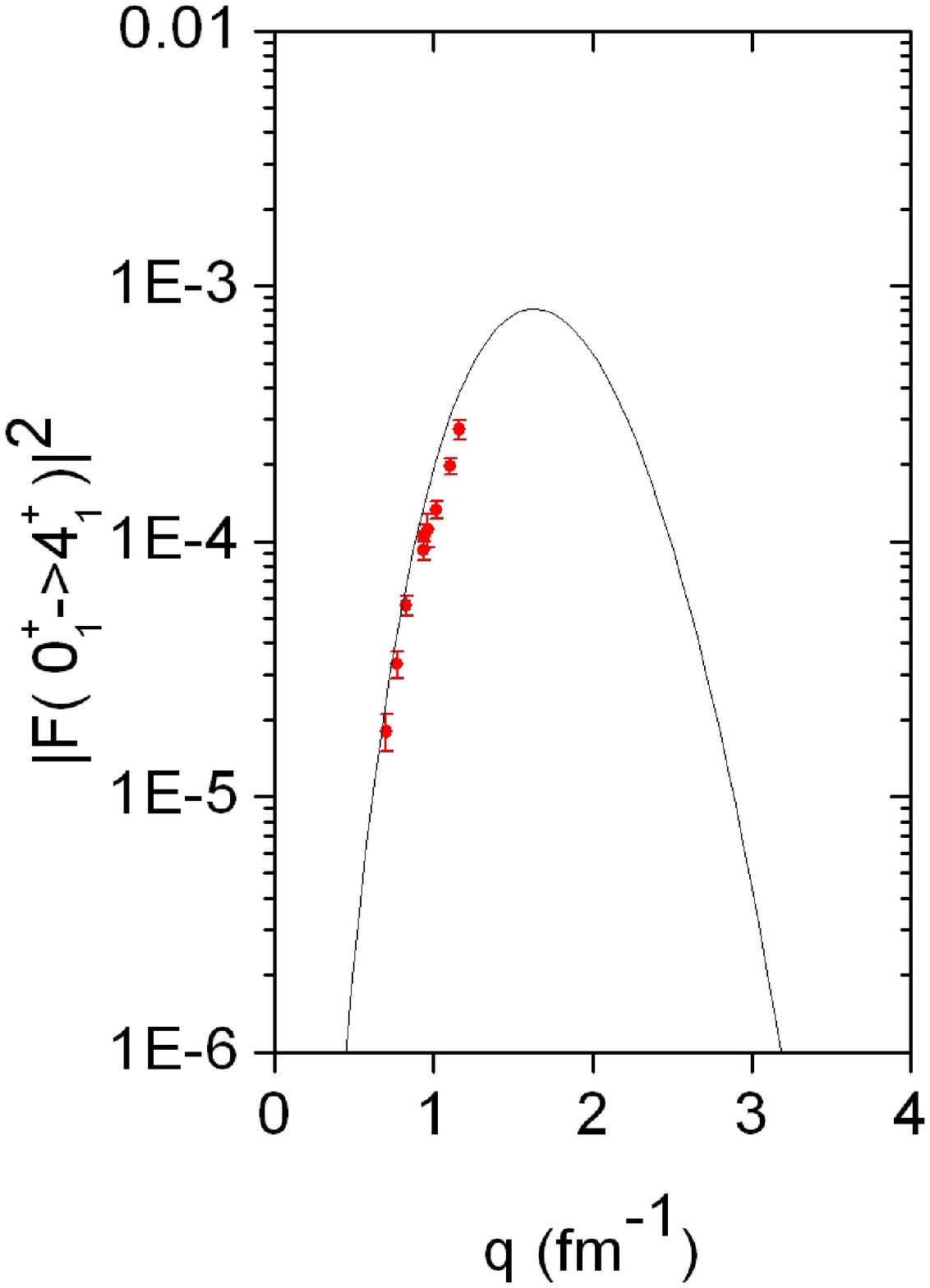}}
\end{minipage}\hfill
\begin{minipage}{.5\linewidth}
\centerline{\includegraphics[width=\linewidth]{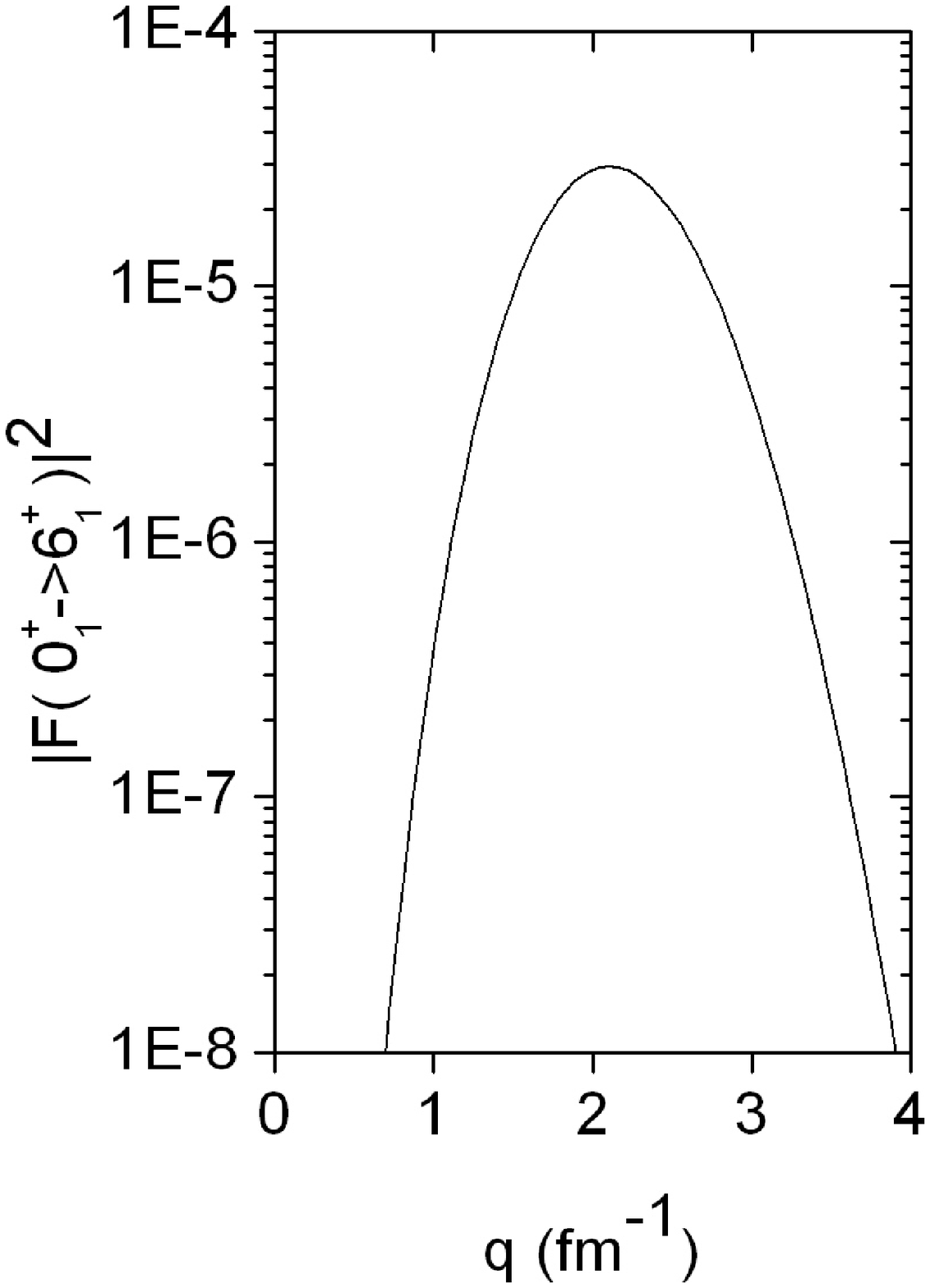}}
\end{minipage}
\caption[Form factors of $^{16}$O. I]
{Comparison between the experimental form factors $|{\cal F}(0_1^+ \rightarrow L^P_i)|^2$ 
of $^{16}$O for the final states with $L^P_i=0^+_1$, $3^-_1$, $4^+_1$ and $6^+_1$ and those obtained 
for the spherical top. Reproduced from \cite{Bijker:2016bpb} with permission.}
\label{ffgsb}
\end{figure}

\begin{figure}
\centering
\begin{minipage}{.5\linewidth}
\centerline{\includegraphics[width=\linewidth]{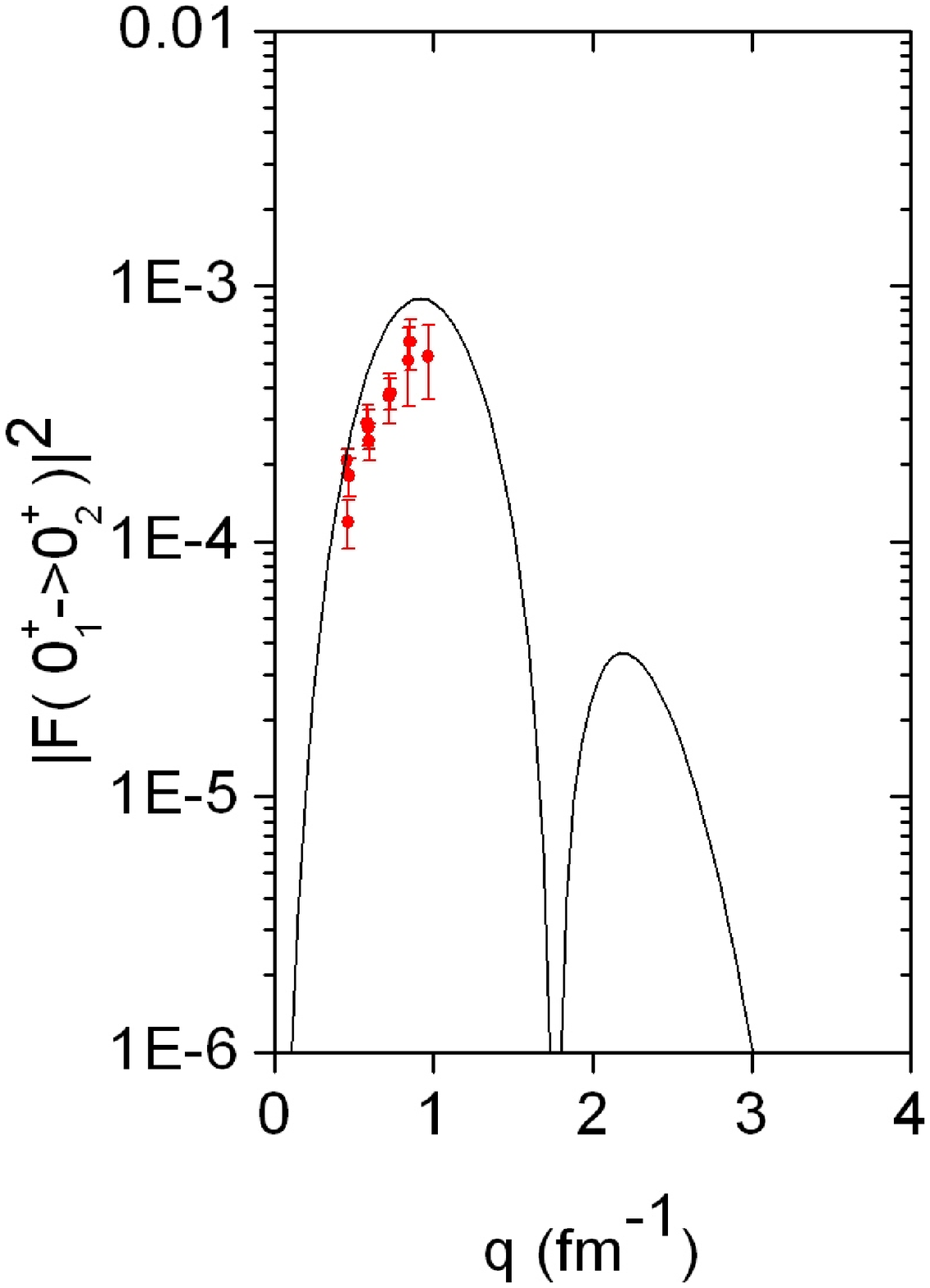}}
\end{minipage}\hfill
\begin{minipage}{.5\linewidth}
\centerline{\includegraphics[width=\linewidth]{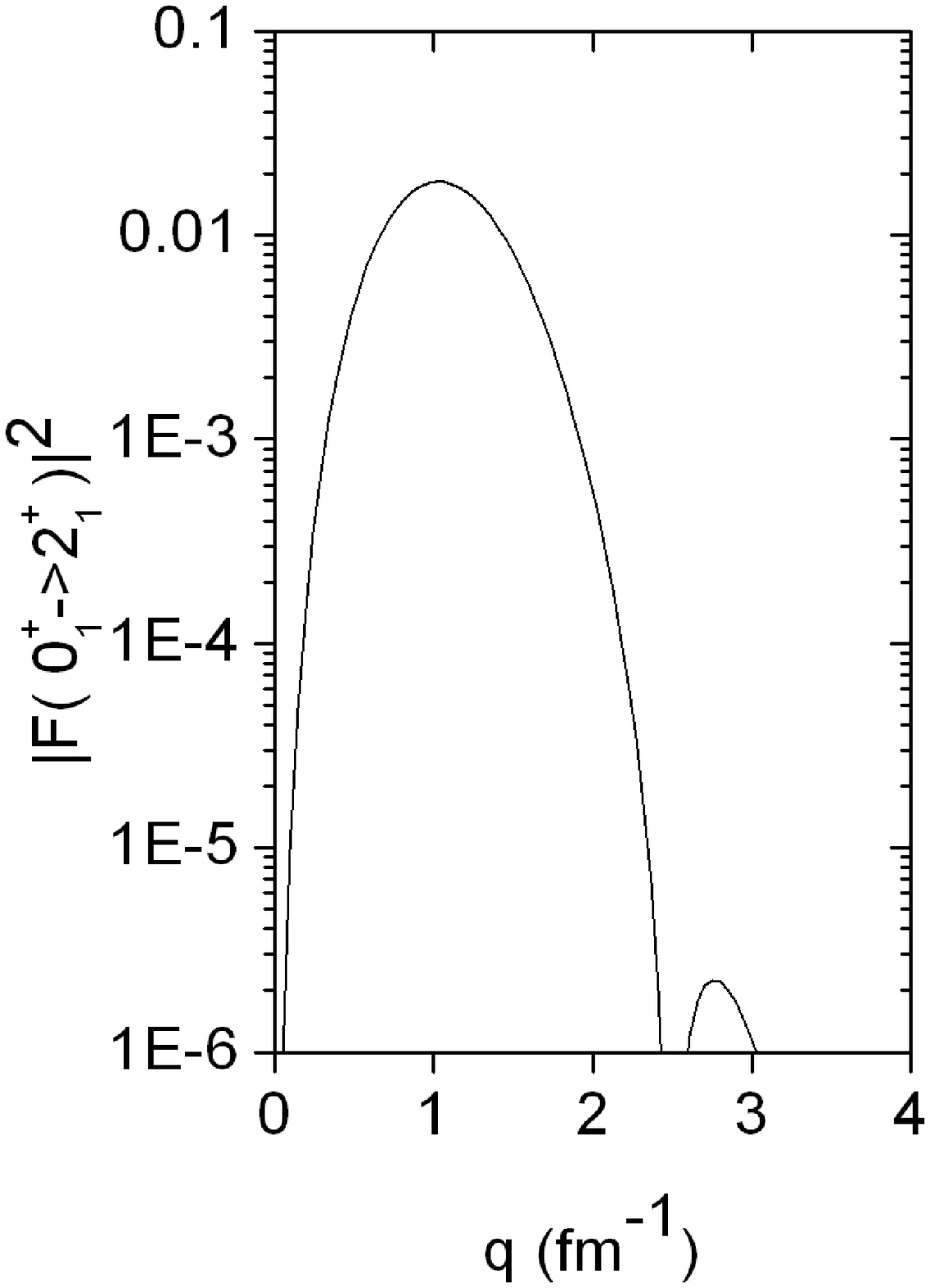}}
\end{minipage}
\vfill
\begin{minipage}{.5\linewidth}
\centerline{\includegraphics[width=\linewidth]{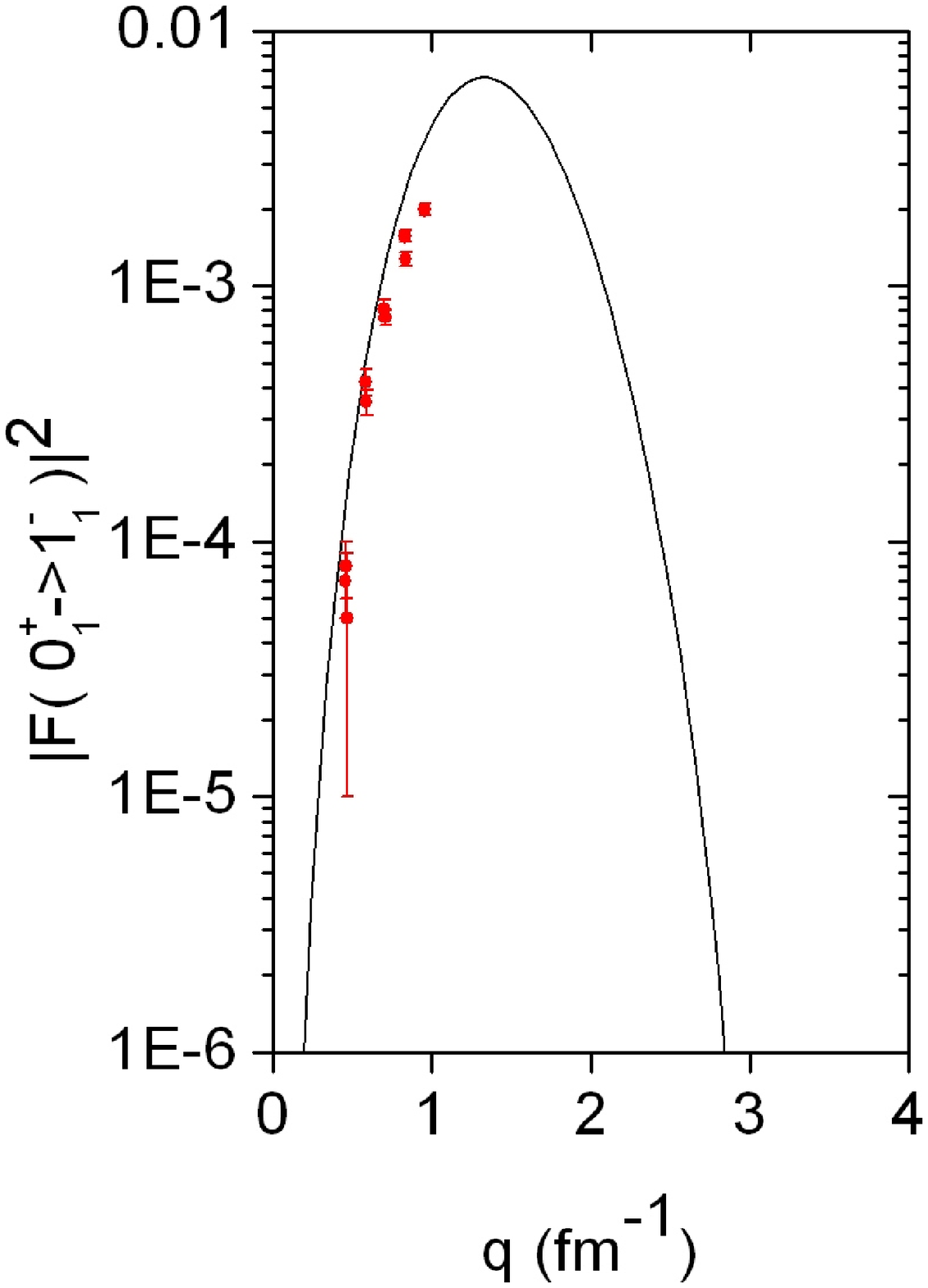}}
\end{minipage}
\caption[Form factors of $^{16}$O. II]
{Comparison between the experimental form factors $|{\cal F}(0_1^+ \rightarrow L^P_i)|^2$ 
of $^{16}$O for the final states with $L^P_i=0^+_2$, $2^+_1$ and $1^-_1$ and those obtained 
for the spherical top. Reproduced from \cite{Bijker:2016bpb} with permission.}
\label{ffvib}
\end{figure}

Form factors in $^{16}$O have also been extensively investigated. In the rigid case, the form factors
are given by Eq.~(\ref{ff4}). Since also here excitation of the vibrational bands occurs, one needs to
do a full calculation \cite{Bijker:2016bpb}. The resulting form factors are shown in Figs.~\ref{ffgsb} 
and \ref{ffvib}, where they are compared with experimental data. The value of $\beta$ is determined 
from the first minimum of the elastic form factor to be $\beta=2.07 \pm 0.04$ fm \cite{Bijker:2016bpb}, 
where again we have estimated an error of 2 \%. The experimental form factors in Figs.~\ref{ffgsb} and \ref{ffvib} 
compare very well with the theoretical form factors, which is an astonishing result since the theoretical form 
factors contain no free parameters, exept from $\beta$ which is determined from the first minimum of the elastic 
form factor.

We remark at this stage that the values of $\beta$ extracted from the moment of inertia in $^{8}$Be, and from 
the first minimum in the elastic form factors in $^{12}$C and $^{16}$O are consistent with each other, $\sim 2$ fm, 
corresponding to a close-packing of $\alpha$-particles, as shown in Figs.~\ref{density2}, \ref{density3} and \ref{density4} \cite{DellaRocca:2017qkx}.

\subsection{Electromagnetic transition rates}

Electromagnetic transition rates and $B(EL)$ values can be analyzed by making use of Eqs.~(\ref{BEL2}), 
(\ref{BEL3}) and (\ref{BEL4}). A comparison with data for $^{8}$Be, $^{12}$C and $^{16}$O is shown in 
Table~\ref{BEL}. In this table, the experimental value for $^{8}$Be was estimated using the Green’s 
function Monte Carlo method (GFMC) \cite{Datar:2013pbd}. The value of $\beta=1.82$ fm for $^{8}$Be is 
obtained from the moment of inertia, and for $^{12}$C and $^{16}$O from the first minimum in the 
elastic form factor, see Sect.~\ref{formfactor}. 

\begin{table}
\centering
\caption[$B(EL)$ values]
{$B(EL)$ values in $^8$Be, $^{12}$C and $^{16}$O in e$^2$fm$^{2L}$.
Experimental data are taken from \cite{Datar:2013pbd,KELLEY201771,TILLEY19931}, 
and theoretical ACM results from \cite{DellaRocca:2018mrt,Bijker:2002ac,Bijker:2016bpb}.}
\label{BEL} 
\vspace{10pt}
\begin{tabular}{lcccccc}
\hline
\noalign{\smallskip}
& $B(EL;L^P \rightarrow 0^+)$ & Exp & ACM & $E(L^P)$ & Exp & ACM \\
\noalign{\smallskip}
\hline
\noalign{\smallskip}
$^{8}$Be & $B(E2;2^+ \rightarrow 0^+)$ & $21.0 \pm 2.3$ &  14.0 & $E(2^+)$ &  3030 &  3060 \\ 
         & $B(E4;4^+ \rightarrow 0^+)$ &                & 153.3 & $E(4^+)$ & 11350 & 10200 \\ 
\noalign{\smallskip}
\hline
\noalign{\smallskip}
$^{12}$C & $B(E2;2^+ \rightarrow 0^+)$ & $7.61 \pm 0.42$ &  8.4 & $E(2^+)$ &  4439 &  4400 \\ 
         & $B(E3;3^- \rightarrow 0^+)$ & $104 \pm 14$    & 73.0 & $E(3^-)$ &  9641 &  9640 \\ 
         & $B(E4;4^+ \rightarrow 0^+)$ &                 &      & $E(4^+)$ & 14080 & 14670 \\ 
\noalign{\smallskip}
\hline
\noalign{\smallskip}
$^{16}$O & $B(E3;3^- \rightarrow 0^+)$ & $205 \pm 11$  &  215 & $E(3^-)$ &  6130 &  6132 \\ 
         & $B(E4;4^+ \rightarrow 0^+)$ & $378 \pm 133$ &  425 & $E(4^+)$ & 10356 & 10220 \\ 
         & $B(E2;6^+ \rightarrow 0^+)$ &               & 9626 & $E(6^+)$ & 21052 & 21462 \\ 
\noalign{\smallskip}
\hline
\end{tabular}
\end{table}

\section{Non-cluster states}

The cluster model assumes that $k\alpha$ nuclei are composed of $\alpha$-particles. However, these in turn
are composed of two protons and two neutrons. At some excitation energy in the nucleus, the $\alpha$-particle
structure may break. In some cases non-cluster states can be clearly identified, since
some states are forbidden by the discrete symmetry. Specifically, for $Z_2$ symmetry ($^8$Be) $L^P=1^+$ 
states cannot be formed, for $D_{3h}$ symmetry ($^{12}$C) $L^P=1^+$ states cannot be formed, and for $T_d$ 
symmetry ($^{16}$O) $L^P=0^-$ states cannot be formed. These are signatures of non-cluster states. In
addition, since $\alpha$-particles have isospin $T=0$, states with $T=1$ cannot be formed. This is
another signature of non-cluster states. The energy at which non-cluster states occur is shown in
Figs.~\ref{Be8}, \ref{C12} and \ref{O16}: at $\sim 15$ MeV in $^8$Be, at $\sim 13$ MeV in $^{12}$C 
and at $\sim 10$ MeV in $^{16}$O. Above these energies, cluster states co-exist with non-cluster states. 

\section{Softness and higher-order corrections}

The situations described by the energy formulas in Eqs.~(\ref{ener2}), (\ref{ener3}) and (\ref{ener4}) 
correspond to rigid configurations. As mentioned in previous sections, 
soft (non-rigid) situations can be described by diagonalizing the full algebraic Hamiltonian.
However, one can also write, for comparison with experimental data, simpler analytic
expressions for non-rigid situations. 

\subsection{Dumbbell configuration}

An analytic formula including anharmonic corrections and vibration-rotation interaction is
\ba
E &=& E_0 + \omega v + x v^2  
\nonumber\\
&& + B L(L+1) + B' [L(L+1)]^2 
\nonumber\\
&& + \lambda \, vL(L+1) ~.
\label{soft2}
\ea

\subsection{Equilateral triangle configuration}

In this case an analytic expression is \cite{Marin-Lambarri:2014zxa}
\ba
E &=& E_0 + \sum_{i=1}^2 \omega_{i} v_{i} + \sum_{i,j=1}^2 x_{ij} v_i v_j 
\nonumber\\
&& + B L(L+1) + B' [L(L+1)]^2 + C (K \mp 2\ell_{2})^{2}
\nonumber\\
&& + \left( \sum_{i=1}^2 \lambda_i v_i \right)L(L+1) ~.
\label{soft3}
\ea

\subsection{Tetrahedral configuration}

The effect of anharmonicities here can be written as
\ba
E &=& E_0 + \sum_{i=1}^3 \omega_{i} v_{i} + \sum_{i,j=1}^3 x_{ij} v_i v_j 
\nonumber\\
&& + B L(L+1) + B' [L(L+1)]^2 
\nonumber\\
&& + \left( \sum_{i=1}^3 \lambda_i v_i \right)L(L+1) ~.
\label{soft4}
\ea

\section{Other geometric configurations}

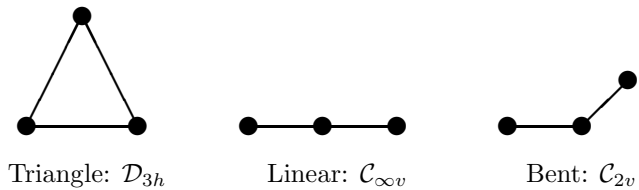
\begin{figure}
\centering
\vspace{15pt}
\setlength{\unitlength}{0.7pt}
\begin{picture}(400,130)(0,0)
\thicklines
\put( 60, 50) {\circle*{10}} 
\put(120, 50) {\circle*{10}}
\put( 90,110) {\circle*{10}}
\put( 60, 50) {\line( 1,0){60}}
\put( 60, 50) {\line( 1,2){30}}
\put(120, 50) {\line(-1,2){30}}
\put( 50, 20) {Triangle: \bf ${\cal D}_{3h}$}

\put(180, 50) {\circle*{10}} 
\put(220, 50) {\circle*{10}}
\put(260, 50) {\circle*{10}}
\put(180, 50) {\line( 1,0){80}}
\put(190, 20) {Linear: \bf ${\cal C}_{\infty v}$}
\put(320, 50) {\circle*{10}} 
\put(360, 50) {\circle*{10}}
\put(385, 75) {\circle*{10}}
\put(320, 50) {\line( 1,0){40}}
\put(360, 50) {\line( 1,1){25}}
\put(330, 20) {Bent: \bf ${\cal C}_{2v}$}
\end{picture}
\vspace{15pt}
\caption[Three-body cluster configurations]{Three-body cluster configurations.}
\label{shapes3}
\end{figure}

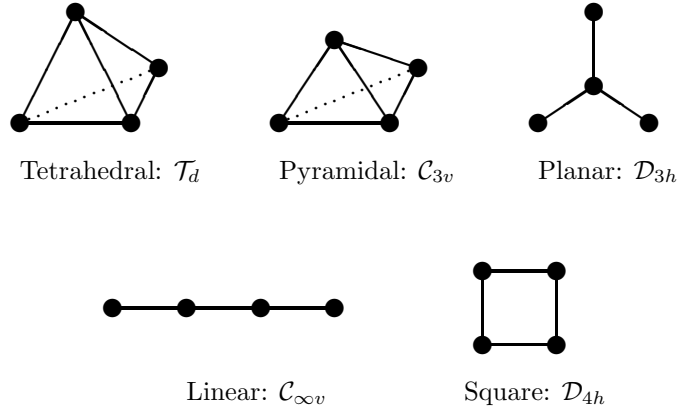
\begin{figure}
\centering
\vspace{15pt}
\setlength{\unitlength}{0.7pt}
\begin{picture}(460,220)(0,0)
\thicklines
\put(110, 50) {\circle*{10}}
\put(150, 50) {\circle*{10}}
\put(190, 50) {\circle*{10}}
\put(230, 50) {\circle*{10}}
\put(110, 50) {\line( 1,0){120}}
\put(150,  0) {Linear: \bf ${\cal C}_{\infty v}$}
\put(310, 30) {\circle*{10}}
\put(350, 30) {\circle*{10}}
\put(310, 70) {\circle*{10}}
\put(350, 70) {\circle*{10}}
\put(310, 30) {\line( 1,0){40}}
\put(310, 70) {\line( 1,0){40}}
\put(310, 30) {\line( 0,1){40}}
\put(350, 30) {\line( 0,1){40}}
\put(300,  0) {Square: \bf ${\cal D}_{4h}$}
\put( 60,150) {\circle*{10}} 
\put(120,150) {\circle*{10}}
\put(135,180) {\circle*{10}}
\put( 90,210) {\circle*{10}}
\put( 60,150) {\line( 1,0){60}}
\put( 60,150) {\line( 1,2){30}}
\put(120,150) {\line(-1,2){30}}
\put(120,150) {\line( 1,2){15}}
\put(135,180) {\line(-3,2){45}}
\multiput( 60,150)(5,2){16}{\circle*{2}}
\put( 60,120) {Tetrahedral: \bf ${\cal T}_d$}
\put(200,150) {\circle*{10}} 
\put(260,150) {\circle*{10}}
\put(275,180) {\circle*{10}}
\put(230,195) {\circle*{10}}
\put(200,150) {\line( 1,0){60}}
\put(200,150) {\line( 2,3){30}}
\put(260,150) {\line(-2,3){30}}
\put(260,150) {\line( 1,2){15}}
\put(275,180) {\line(-3,1){45}}
\multiput(200,150)(5,2){16}{\circle*{2}}
\put(200,120) {Pyramidal: \bf ${\cal C}_{3v}$}
\put(340,150) {\circle*{10}} 
\put(400,150) {\circle*{10}}
\put(370,170) {\circle*{10}}
\put(370,210) {\circle*{10}}
\put(340,150) {\line( 3,2){30}}
\put(400,150) {\line(-3,2){30}}
\put(370,170) {\line( 0,1){40}}
\put(340,120) {Planar: \bf ${\cal D}_{3h}$}
\end{picture}
\vspace{15pt}
\caption[Four-body cluster configurations]{Four-body cluster configurations.}
\label{shapes4}
\end{figure}

Within the ACM it is possible to provide analytic formulas for energies and electromagnetic
transition rates for all possible geometric configurations and, most importantly, by diagonalizing
the full algebraic Hamiltonian, it is possible to study non-rigid situation intermediate between
two or more geometric situations and thus study the transitions between these, called ground-state 
phase transitions \cite{FI}. Some possible geometric configurations for three and four $\alpha$-particles
are shown in Figs.~\ref{shapes3} and \ref{shapes4}.
Another situation that can be studied with ACM is that in which the ground state has a different
geometric configuration than the excited states. In the interpretation of the previous Sect.~\ref{evidence}, 
the excited states are (large amplitude) vibrations of the ground-state configuration. An alternative
interpretation was given by Brink \cite{Brink,Brink:1970ufk}, in which the excited states have a different 
geometric configuration as the ground state. Specifically, in $^{12}$C the ground state was suggested to be an
equilateral triangle ($D_{3h}$ symmetry) and the excited state to be linear ($C_{\infty v}$  
symmetry) \cite{PhysRev.101.254} or bent ($C_{2v}$ symmetry) as obtained in lattice EFT calculations 
\cite{Epelbaum:2012qn}. 
Similarly in $^{16}$O, the ground state was suggested to be a regular tetrahedron ($T_d$ symmetry) 
and the excited state to be a square ($D_{4h}$ symmetry) \cite{PhysRevLett.112.102501}. 
This situation, in which the symmetry of the state changes as a function of excitation energy is called 
an excited-state quantum-phase transitions (ESQPT) \cite{CAPRIO20081106}.

Work is currently underway to see whether experimental data support Brink's hypothesis
\cite{Brink,Brink:1970ufk} or lattice EFT calculations \cite{Epelbaum:2011md} or rather the oblate structure 
of the previous sections for the excited $0^+_2$ state (Hoyle state) of $^{12}$C. Within the ACM, the transition from 
bent to linear can be studied by adding to the Hamilonian of Eq.~(\ref{ham3}) which describes triangular configurations, 
a term $\epsilon \hat{n}$ where $\hat{n}=\sum_m (b_{\rho,m}^{\dagger} b_{\rho,m} + b_{\lambda,m}^{\dagger} b_{\lambda,m})$ 
\cite{Bijker:2002ac,Bijker_2017}. Bent to linear transitions have been extensively investigated in molecular physics
by making use of the algebraic approach described here \cite{Larese_2011,Larese_2013}.

\section{The cluster shell model}
\label{CSM}

The cluster shell model has been introduced recently \cite{DellaRocca:2017qkx,DellaRocca:2018mrt} to describe 
nuclei composed of $k$ $\alpha$-particles plus additional nucleons, simply denoted as $k \alpha + x$ nuclei. 
For each of the three configurations with $Z_2$, $D_{3h}$ and $T_d$ symmetry, it is possible to determine the 
cluster densities $\rho(\vec{r})$ given in Sect.~\ref{cluster}, and study the motion of a single-particle in 
the potential, $V(\vec{r})$, obtained by convoluting the density with the nucleon-alpha interaction 
$v(\vec{r}-\vec{r}^{\, \prime})$,
\begin{equation}
\label{convolution}
V(\vec{r}) \;=\; \int\rho(\vec{r}^{\, \prime}) v(\vec{r}-\vec{r}^{\, \prime})d^3\vec{r}^{\, \prime}.
\end{equation}
Several forms of the nucleon-alpha interaction have been considered. By taking a Volkov-type
Gaussian interaction \cite{Volkov:1965zz}, one obtains a potential $V(\vec{r})$ with the same dependence 
on $r$, $\theta$, $\phi$ as in the density of Eq.~(\ref{rhor2}), but with a different value of the parameter 
$\alpha$. The basic form of the potential has been assumed to be
\begin{equation}
\label{defpotential}
V(\vec{r}) \;=\; -V_0 \sum_{\lambda\mu} f_\lambda(r) Y_{\lambda\mu}(\theta,\phi) 
\sum_{i=1}^k Y_{\lambda\mu}^*(\theta_i,\phi_i) ~,
\end{equation}
where
\begin{equation}
\label{flambda}
f_\lambda(r) \;=\; e^{-\alpha(r^2+\beta^2)}4\pi i_\lambda(2\alpha\beta r) ~.
\end{equation}
In addition, the odd-particle experiences also a spin-orbit interaction. Since $V(\vec{r})$ is not
spherically symmetric, one must take for $V_{\rm so}(\vec{r})$ the symmetrized form
\begin{equation}
\label{spinorbitpot}
V_{so}(\vec{r}) \;=\; V_{0,\rm so} \frac{1}{2} \left[ 
\left(\frac{\hat{r}}{r} \cdot \vec{\nabla} V(\vec{r}) \right) \left(\vec{s} \cdot \vec{l}\right) 
+ \left(\vec{s} \cdot \vec{l}\right) \left(\frac{\hat{r}}{r} \cdot \vec{\nabla} V(\vec{r}) \right) \right]
\end{equation}
From Eq.~(\ref{defpotential}), one has
\begin{equation}
\label{eq:spinorbit}
\frac{\hat{r}}{r} \cdot \vec{\nabla} V(\vec{r}) \;=\; \sum_{\lambda\mu} \left( -2\alpha f_{\lambda} 
+ \frac{\lambda}{r^2} f_{\lambda} + \frac{2\alpha\beta}{r} f_{\lambda+1} \right) Y_{\lambda\mu}(\theta,\phi) 
\sum_{i=1}^k Y_{\lambda\mu}^*(\theta_i,\phi_i) ~.
\end{equation}
Finally, if the odd particle is a proton one must add the Coulomb interaction between the odd particle
and the cluster, given by
\ba
V_C(\vec{r}) &=& \frac{Ze^2}{k} \left(\frac{\alpha}{\pi}\right)^{3/2} \sum_{\lambda\mu} \frac{4\pi}{2\lambda+1} Y_{\lambda\mu}(\theta,\phi) \sum_{i=1}^k Y^*_{\lambda\mu}(\theta_i,\phi_i) 
\nonumber\\
&& \times \left[ \frac{1}{r^{\lambda+1}} \int_0^r f_\lambda(r') r'^{\lambda} r'^2 dr' 
+ r^\lambda \int_r^{\infty} f_\lambda(r') \frac{1}{r'^{\lambda+1}} r'^2 dr'\right] ~. 
\ea
The total single-particle Hamiltonian is then the sum of kinetic, nuclear spin-independent, nuclear spin-orbit, 
and Coulomb terms
\begin{equation}
H \;=\; \frac{\vec{p}^{\, 2}}{2m} + V(\vec{r}) + V_{\rm so}(\vec{r}) + \frac{1}{2}(1+\tau_3) V_C(\vec{r}) ~.
\label{hcsm}
\end{equation}
The single-particle energies, $\epsilon_{\Omega}$, and the intrinsic wave functions, $\chi_{\Omega}$, are then 
obtained by diagonalizing the Hamiltonian of Eq.~(\ref{hcsm}) in a harmonic oscillator basis
\begin{equation}
|\chi_\Omega\rangle \;=\; \sum_{nljm} C_{nljm}^{\Omega} \left| n\frac{1}{2}ljm \right> ~, 
\label{wfint}
\end{equation}

\subsection{Dumbbell configuration}

\begin{figure}
\centering
\includegraphics[width=4in]{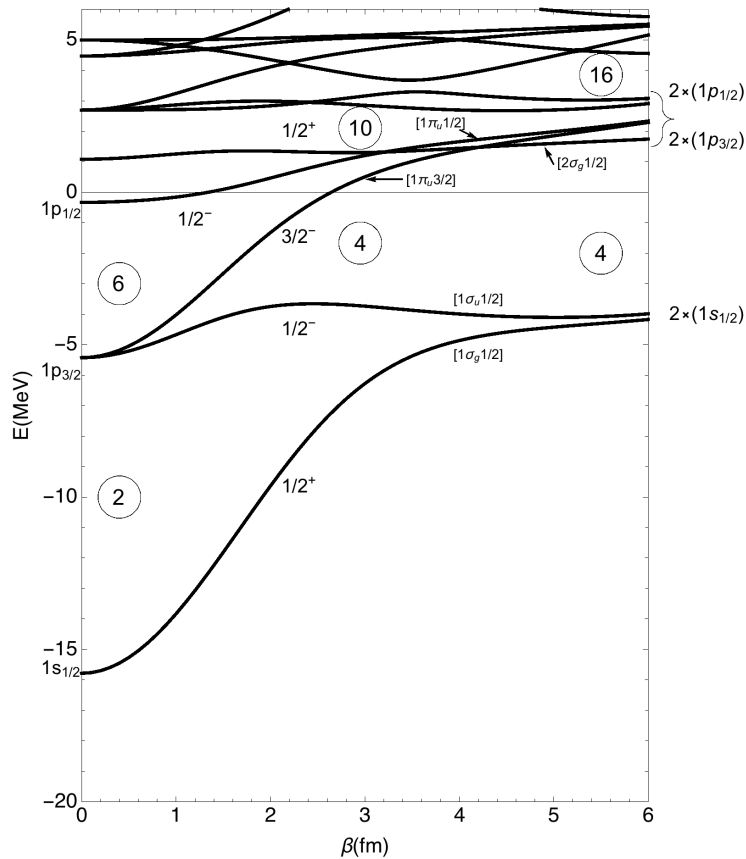} 
\caption[Energy levels in a two-body cluster potential]
{Single-particle energies in a cluster potential with $Z_2$ symmetry 
calculated with $V_0=20$ MeV, $V_{0,\rm so}=22$ MeV fm$^{2}$, and $\alpha=0.1115$ fm$^{-2}$. 
Reproduced from \cite{DellaRocca:2017qkx} with permission.}
\label{csm2}
\end{figure}

The energy levels of a neutron in a potential with $Z_2$ symmetry are given in Fig.~\ref{csm2}. 
At $\beta=0$ the single-particle levels are those of the spherical shell model and can be labelled accordingly. As $\beta$ increases the spherical levels split. However, since the potential has axial symmetry, the projection of the angular momentum on the symmetry $z$-axis, $K$, is a good quantum number. All levels are doubly degenerate with $\pm K$. The values of $K$ contained in each $j$ level are $K=j,j-1,\ldots,1/2$, Table~\ref{resZ2}. In Fig.~\ref{csm2}, levels are labelled by $K$ and the parity $P=(-)^l$ of the spherical level $nlj$ from which they originate. These quantum numbers are conserved in the correlation diagram of Fig.~\ref{csm2}. Alternatively the energy levels can be labelled by the molecular notation $n\sigma K$, $n\pi K$, $n\delta K$, $\ldots$ and a $g$ (gerade), $u$ (ungerade) label. Here $n=1,2,\dots$ denotes the $1^{st}, 2^{nd}, \dots$, state, $\sigma,\pi,\delta,\dots$ denotes the projection of the orbital angular momentum $L$ on the $z$-axis in spectroscopic notation $0 \equiv \sigma$, $1 \equiv \pi$, $2 \equiv \delta$, $\ldots$, $K$ the total projection including spin, and $g$, $u$ the parity, $g \equiv +$, $u \equiv -$.

\begin{table}
\centering
\caption[$K$ values]
{Values of $K$ for each $j$ level.}
\label{resZ2}
\vspace{10pt}
\begin{tabular}{cl}
\hline
\noalign{\smallskip}
$j$ & $K$ \\
\noalign{\smallskip}
\hline
\noalign{\smallskip}
$1/2$ & $1/2$ \\
$3/2$ & $1/2$, $3/2$ \\ 
$5/2$ & $1/2$, $3/2$, $5/2$ \\
$7/2$ & $1/2$, $3/2$, $5/2$, $7/2$ \\
\noalign{\smallskip}
\hline
\end{tabular}
\end{table}

\begin{figure}
\centering
\centerline{\includegraphics[width=\linewidth]{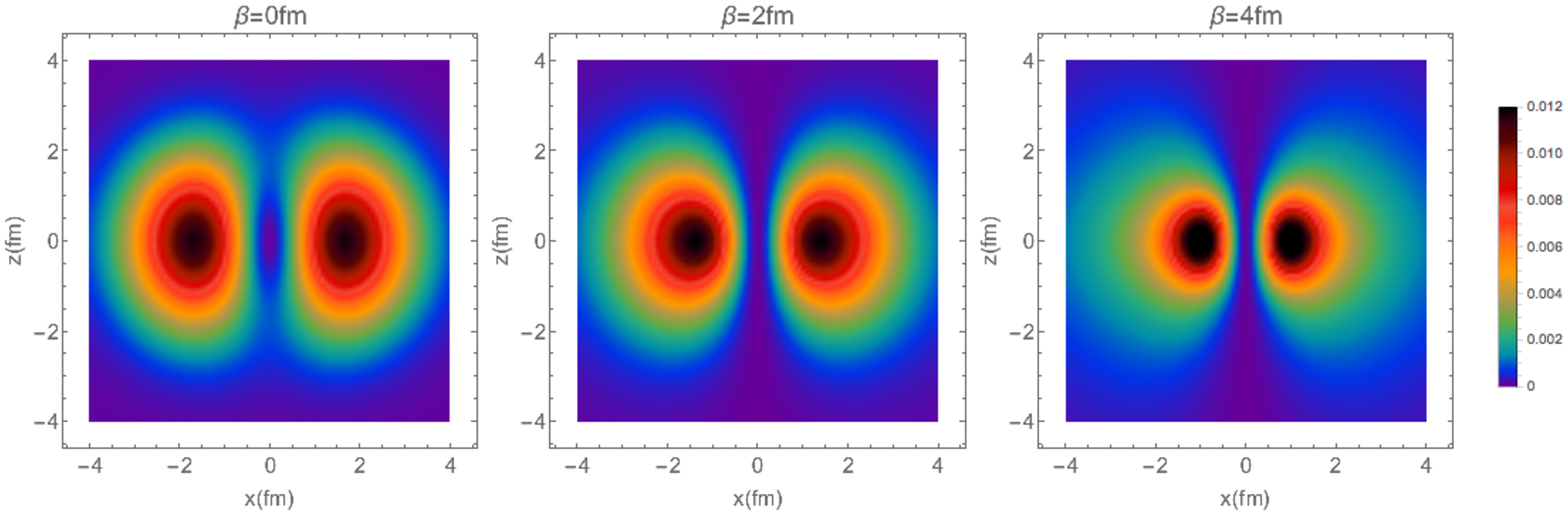}}
\centerline{\includegraphics[width=\linewidth]{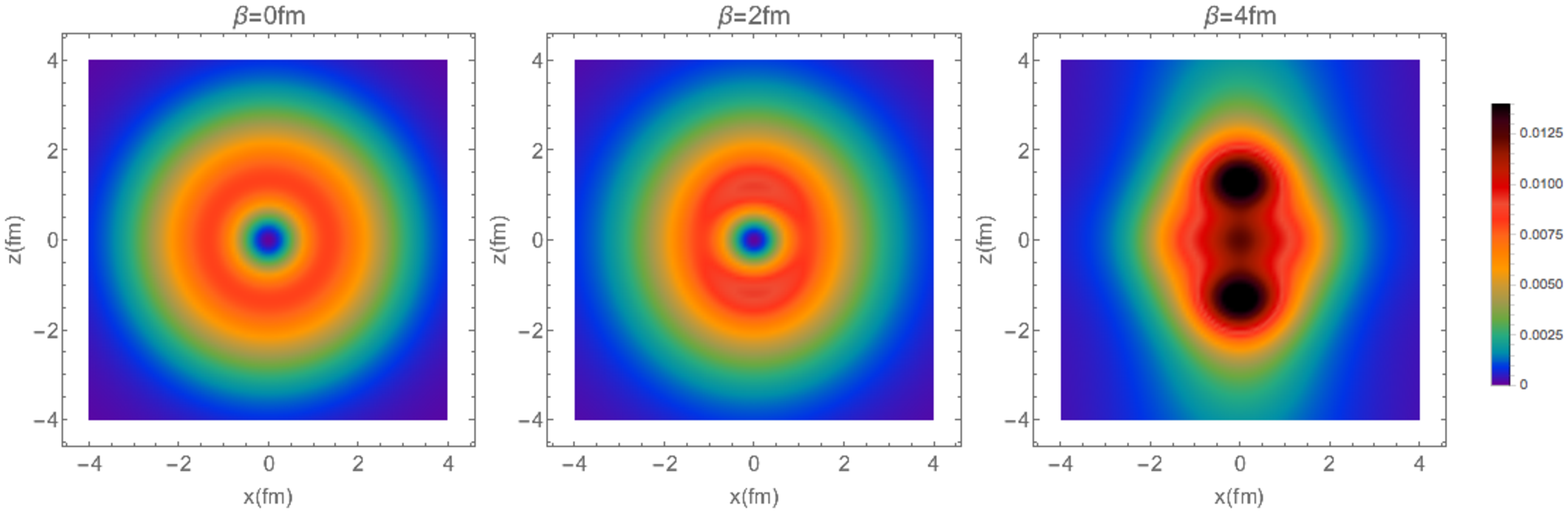}}
\centerline{\includegraphics[width=\linewidth]{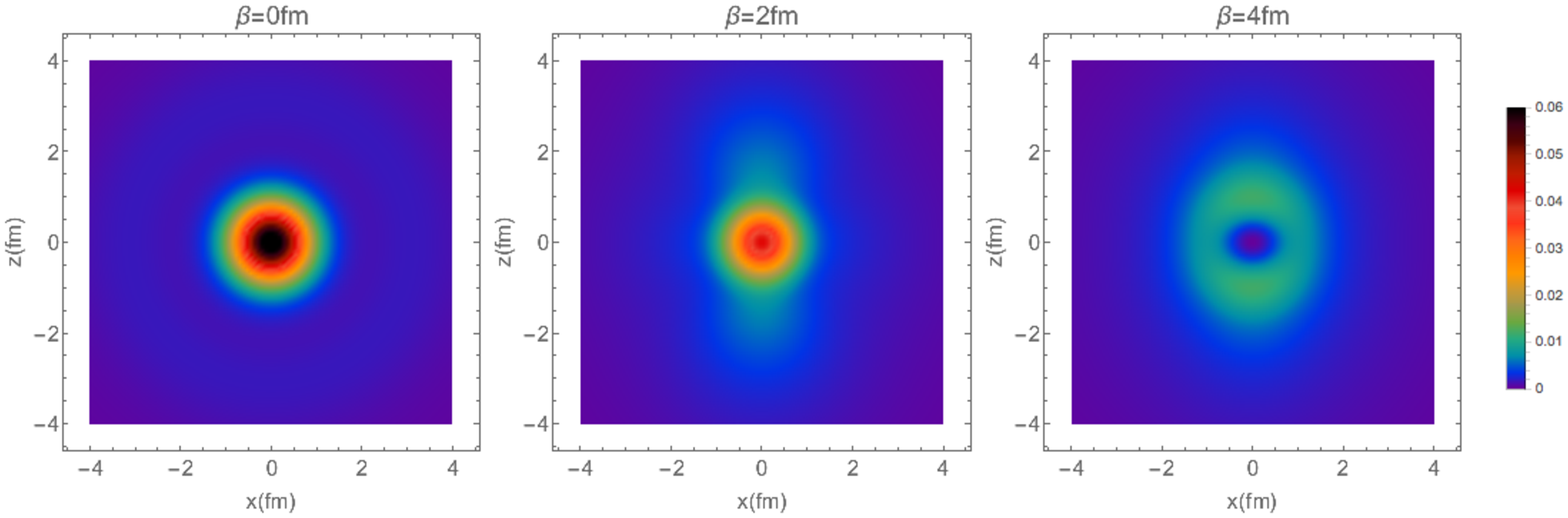}}
\caption[Single-particle densities]
{Single-particle densities of the states $K^P=3/2^-$, $1/2^-$ and $1/2^+$ 
in a $Z_2$ symmetric potential with $V_0=20$ MeV, $V_{0,\rm so}=22$ MeV fm$^{2}$ and $\alpha=0.1115$ fm$^{-2}$. 
The color code is in fm$^{-3}$. Reproduced from \cite{DellaRocca:2017qkx} with permission.}
\label{spdensities}
\end{figure}

The single-particle densities of the three levels $K^P=3/2^-,[1\pi_u3/2]$, $K^P=1/2^-,[1\pi_u1/2]$ 
and $K^P=1/2^+,[2\sigma_g1/2]$ are shown in Fig.~\ref{spdensities} as a function of $\beta$. It should be 
noted that in this case the energy levels and wave functions are identical to those of the two-center 
shell model \cite{Andersen:1970tov,Scharnweber:1971qpn}. 
The idea of molecular wave functions was introduced by von Oertzen in the 1970's, and applied 
to $^{9}$Be and $^{9}$B in 1996 \cite{VONOERTZEN19751,IMANISHI198729,Oertzen}. Fig.~\ref{csm2} is similar 
to Fig.~1 of \cite{Oertzen}, except for the fact that in constructing Fig.~\ref{csm2} a realistic Gaussian 
potential is used in Eq.~(\ref{defpotential}) which goes to zero at $r \rightarrow \infty$, while in \cite{Oertzen} 
a harmonic oscillator potential is used that goes to infinity at $r \rightarrow \infty$.

\subsection{Equilateral triangle configuration}

\begin{figure}
\centering
\includegraphics[width=4in]{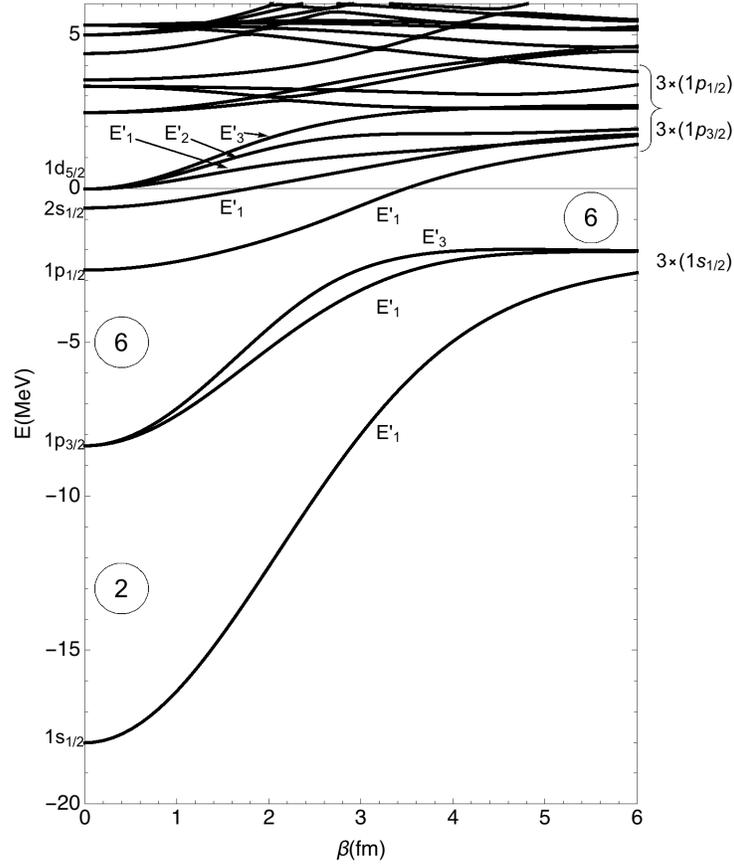} 
\caption[Energy levels in a three-body cluster potential]
{Single-particle energies in a cluster potential with $D_{3h}$ symmetry 
calculated with $V_0= 13.3$ MeV, $V_{0,\rm so}=16.9$ MeV fm$^{2}$, $\alpha=0.0872$ fm$^{-2}$. 
Reproduced from \cite{DellaRocca:2017qkx} with permission.}
\label{csm3}
\end{figure}

The energy levels of a neutron in a potential with $D_{3h}$ symmetry are shown in Fig.~\ref{csm3}. 
At $\beta=0$ the levels are again those of the spherical shell-model. As $\beta$ increases, the levels split. 
The resolution of the representations $D_j^P$ of $SU(2)$ with angular momentum $j$ and parity $P$ into 
representations of the double group $D'_{3h}$ which describes fermions is a complicated group-theoretical problem. 
It was solved by Koster \cite{koster1963properties} for applications to crystal physics and by Herzberg \cite{herzberg1989molecular} for applications to molecular physics. The solution is given in Table~\ref{d3hj}, 
where we have used a notation more appropriate to nuclear physics \cite{BI}. Various notations are used for
representations of $D'_{3h}$ , whose conversion is $E_{1/2}^{(+)} \equiv E_{1/2} \equiv \Gamma_7 \equiv E'_1$, 
$E_{1/2}^{(-)} \equiv E_{5/2} \equiv \Gamma_8 \equiv E'_2$ and $E_{3/2} \equiv E_{3/2} \equiv \Gamma_9 \equiv E'_3$,  
where the first notation is the nuclear physics notation \cite{BI}, the second is that of Herzberg 
\cite{herzberg1989molecular}, the third is that of Koster \cite{koster1963properties} and the fourth is 
that of Hamermesh \cite{hamermesh1964group}.

\begin{table}
\centering
\caption[Resolution into $D'_{3h}$]
{Resolution of rotational states into irreducible representations of $D'_{3h}$.}
\label{d3hj}
\vspace{10pt}
\begin{tabular}{c|ccc|c|cccc}
\hline
\noalign{\smallskip}
& $\Gamma_7$ & $\Gamma_8$ & $\Gamma_9$ &
& $\Gamma_7$ & $\Gamma_8$ & $\Gamma_9$ & \cite{koster1963properties} \\
$D_j^P$ & $E_{1/2}^{(+)}$ & $E_{1/2}^{(-)}$ & $E_{3/2}$ & 
$D_j^P$ & $E_{1/2}^{(+)}$ & $E_{1/2}^{(-)}$ & $E_{3/2}$ & \cite{BI} \\ 
\noalign{\smallskip}
\hline
\noalign{\smallskip}
$1/2^+$ & 1 & 0 & 0 & $1/2^-$ & 0 & 1 & 0 & \\
$3/2^+$ & 1 & 0 & 1 & $3/2^-$ & 0 & 1 & 1 & \\
$5/2^+$ & 1 & 1 & 1 & $5/2^-$ & 1 & 1 & 1 & \\
$7/2^+$ & 1 & 2 & 1 & $7/2^-$ & 2 & 1 & 1 & \\
$9/2^+$ & 1 & 2 & 2 & $9/2^-$ & 2 & 1 & 2 & \\
$11/2^+$ & 2 & 2 & 2 & $11/2^-$ & 2 & 2 & 2 & \\
$13/2^+$ & 3 & 2 & 2 & $13/2^-$ & 2 & 3 & 2 & \\
\noalign{\smallskip}
\hline
\end{tabular}
\end{table}

The representations $E_{1/2}^{(+)}$, $E_{1/2}^{(-)}$ and $E_{3/2}$ can be further 
decomposed into values of $K$ \cite{BI} 
\ba
\Omega = E_{1/2}^{(+)} &:& K^P=1/2^+ 
\nonumber\\
&& K=3n \pm 1/2 \hspace{1cm} P=(-)^{n} 
\nonumber\\ 
\Omega = E_{1/2}^{(-)} &:& K^P=1/2^- 
\nonumber\\
&& K=3n \pm 1/2 \hspace{1cm} P=(-)^{n+1} 
\nonumber\\
\Omega = E_{3/2} &:& K^P=(3n - 3/2)^{\pm}  
\label{omega3a}
\ea
with $n=1,2,3,\ldots,$ and $K > 0$. The angular momenta are given by $J=K,K+1,K+2,\ldots$. 
As a result, the rotational sequences for each one of the irreps of $D'_{3h}$ 
are given by (see also Table~\ref{d3hj})
\ba
\Omega = E_{1/2}^{(+)} &:& J^P = \frac{1}{2}^+, \frac{3}{2}^+, \frac{5}{2}^{\pm}, 
\frac{7}{2}^+, \left(\frac{7}{2}^-\right)^2, \frac{9}{2}^+, \left(\frac{9}{2}^-\right)^2, \ldots
\nonumber\\
\Omega = E_{1/2}^{(-)} &:& J^P = \frac{1}{2}^-, \frac{3}{2}^-, \frac{5}{2}^{\pm}, 
\left(\frac{7}{2}^+\right)^2, \frac{7}{2}^-, \left(\frac{9}{2}^+\right)^2, \frac{9}{2}^-, \ldots
\nonumber\\
\Omega = E_{3/2} &:& J^P = \frac{3}{2}^{\pm}, \frac{5}{2}^{\pm}, \frac{7}{2}^{\pm}, 
\left(\frac{9}{2}^{\pm}\right)^2, \ldots
\label{omega3b}
\ea

\subsection{Tetrahedral configuration}

\begin{figure}
\centering
\includegraphics[width=4in]{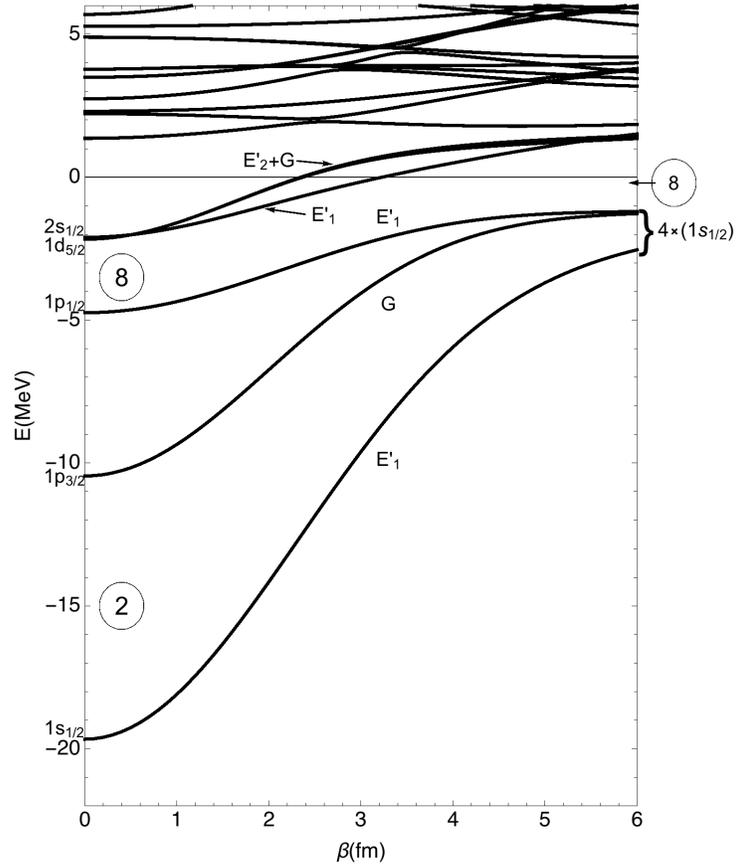} 
\caption[Energy levels in a four-body cluster potential]
{Single-particle energies in a cluster potential with $T_d$ symmetry 
calculated with $V_0=10$ MeV, $V_{0,\rm so}=13.4$ MeV fm$^{2}$, $\alpha=0.0729$ fm$^{-2}$. 
Reproduced from \cite{DellaRocca:2017qkx} with permission.}
\label{csm4}
\end{figure}

The energy levels of a neutron in a potential with $T_d$ symmetry are shown in Fig.~\ref{csm4}. 
The resolution of single-particles levels $j^P$ into representations of $T'_d$ is given in Table~\ref{tdj}. 
In this table, the notation of \cite{koster1963properties} is used as well as the notation appropriate 
to nuclear physics. The conversion between our notation and that of others is
$E_{1/2}^{(+)} \equiv E_{1/2} \equiv \Gamma_6 \equiv E'_1$, 
$E_{1/2}^{(-)} \equiv E_{5/2} \equiv \Gamma_7 \equiv E'_2$ and $G_{3/2} \equiv G_{3/2} \equiv \Gamma_8 \equiv G$,  
where the second notation is that of Herzberg \cite{herzberg1989molecular}, the third is that of Koster 
\cite{koster1963properties} and the fourth is that of Hamermesh \cite{hamermesh1964group}.

In this case the projection of the angular momentum $K$ is not a good quantum number. From
Table~\ref{tdj} one obtains directly the values of the angular momentum contained in each
representation $E_{1/2}^{(+)}$, $E_{1/2}^{(-)}$ and $G_{3/2}$ 
\ba
\Omega = E_{1/2}^{(+)} &:& J^P = \frac{1}{2}^+, \frac{5}{2}^-, \frac{7}{2}^{\pm}, 
\frac{9}{2}^+, \frac{11}{2}^{\pm}, \ldots
\nonumber\\
\Omega = E_{1/2}^{(-)} &:& J^P = \frac{1}{2}^-, \frac{5}{2}^+, \frac{7}{2}^{\pm}, 
\frac{9}{2}^-, \frac{11}{2}^{\pm}, \ldots
\nonumber\\
\Omega = G_{3/2} &:& J^P = \frac{3}{2}^{\pm}, \frac{5}{2}^{\pm}, \frac{7}{2}^{\pm}, 
\left(\frac{9}{2}^{\pm}\right)^2, \left(\frac{11}{2}^{\pm}\right)^2, \ldots
\ea

\begin{table}
\centering
\caption[Resolution into $T'_d$]
{Resolution of rotational states into irreducible representations of $T'_d$.}
\label{tdj}
\vspace{10pt}
\begin{tabular}{c|ccc|c|ccccc}
\hline
\noalign{\smallskip}
& $\Gamma_6$ & $\Gamma_7$ & $\Gamma_8$ &
& $\Gamma_6$ & $\Gamma_7$ & $\Gamma_8$ & \cite{koster1963properties} \\
$D_j^P$ & $E_{1/2}^{(+)}$ & $E_{1/2}^{(-)}$ & $G_{3/2}$ & 
$D_j^P$ & $E_{1/2}^{(+)}$ & $E_{1/2}^{(-)}$ & $G_{3/2}$ & \cite{BI} \\ 
\noalign{\smallskip}
\hline
\noalign{\smallskip}
$1/2^+$ & 1 & 0 & 0 & $1/2^-$ & 0 & 1 & 0 & \\
$3/2^+$ & 0 & 0 & 1 & $3/2^-$ & 0 & 0 & 1 & \\
$5/2^+$ & 0 & 1 & 1 & $5/2^-$ & 1 & 0 & 1 & \\
$7/2^+$ & 1 & 1 & 1 & $7/2^-$ & 1 & 1 & 1 & \\
$9/2^+$ & 1 & 0 & 2 & $9/2^-$ & 0 & 1 & 2 & \\
$11/2^+$ & 1 & 1 & 2 & $11/2^-$ & 1 & 1 & 2 & \\
$13/2^+$ & 1 & 2 & 2 & $13/2^-$ & 2 & 1 & 2 & \\
$15/2^+$ & 1 & 1 & 3 & $15/2^-$ & 1 & 1 & 3 & \\
$17/2^+$ & 2 & 1 & 3 & $17/2^-$ & 1 & 2 & 3 & \\
$19/2^+$ & 2 & 2 & 3 & $19/2^-$ & 2 & 2 & 3 & \\
$21/2^+$ & 1 & 2 & 4 & $21/2^-$ & 2 & 1 & 4 & \\
$23/2^+$ & 2 & 2 & 4 & $23/2^-$ & 2 & 2 & 4 & \\
$25/2^+$ & 3 & 2 & 4 & $25/2^-$ & 2 & 3 & 4 & \\
\noalign{\smallskip}
\hline
\end{tabular}
\end{table}

\subsection{Energy formulas}

We consider here the rotational and vibrational spectra of rigid configurations. Only the
dumbbell and equilateral triangle configurations have been analyzed so far.

\subsubsection{Dumbbell configuration}

The rotational spectra of a dumbbell configuration can be analyzed with the energy formula 
\cite{DellaRocca:2018mrt}
\ba
E_{\rm rot}(\Omega,K,J) &=& \varepsilon_{\Omega} + A_{\Omega} \left[ J(J+1)-K^{2} \right.
\nonumber\\
&& \hspace{1.5cm} \left. + a_{\Omega} (-1)^{J+1/2}(J+1/2) \delta_{K,1/2} \right] ~, 
\label{erot2}
\ea
where $J=K,K+1,K+2,\ldots$. The energy levels depend on the inertial parameter $A_{\Omega}=\hbar^2/2{\cal I}$, 
where ${\cal I}$ is the moment of inertia, and on the so-called decoupling parameter $a_{\Omega}$ 
\cite{nla.cat-vn298793}
\ba
a_{\Omega} \;=\; -\sum_{nlj} (-1)^{j+1/2}(j+1/2) \left| C_{nlj1/2}^{\Omega} \right|^2 ~,
\ea
where the expansion coefficients are given by Eq.~(\ref{wfint}) and $\Omega$ is restricted to states 
with $K=1/2$.
Eq.~(\ref{erot2}) is identical to that used in the collective model which describes the rigid motion of an
ellipsoidal shape \cite{nla.cat-vn298793}. The moment of inertia ${\cal I}$ in odd nuclei can be obtained 
by adding the contribution of the odd particles ${\cal I}^n$ to that of the cluster ${\cal I}^c$
\ba
{\cal I} \;=\; {\cal I}^c + {\cal I}^n ~,
\ea
where ${\cal I}^c$ is given in Eq.~(\ref{inertia}). The assumption here is that the odd particle is dragged along in a
rigid fashion. The odd particle contribution to the three components of the moment of inertia can be calculated as
\ba
{\cal I}_{x}^n &=& m \int (y^{2}+z^{2}) \left| \chi_{\Omega} \right|^2 d^3 r ~,  
\nonumber \\
{\cal I}_{y}^n &=& m \int (z^{2}+x^{2}) \left| \chi_{\Omega} \right|^2 d^3 r ~,  
\nonumber \\
{\cal I}_{z}^n &=& m \int (x^{2}+y^{2}) \left| \chi_{\Omega} \right|^2 d^3 r ~,
\ea
where $m$ is the nucleon mass, and $\chi_{\Omega}$ is the intrinsic wave function of Eq.~(\ref{wfint}). 

The vibrational spectra of a dumbbell configuration plus additional particles can be analyzed
with the formula
\ba
E_{\rm vib}(\Omega,v_{\Omega}) \;=\; \omega_{\Omega} v_{\Omega} ~,
\ea
where the zero-point energy has been removed. There is in this case only one vibrational
quantum number $v_{\Omega}=0,1,\ldots,$ as in Eq.~(\ref{ener2}).

\subsubsection{Equilateral triangle configuration}

\begin{figure}
\centering
\includegraphics[width=4in]{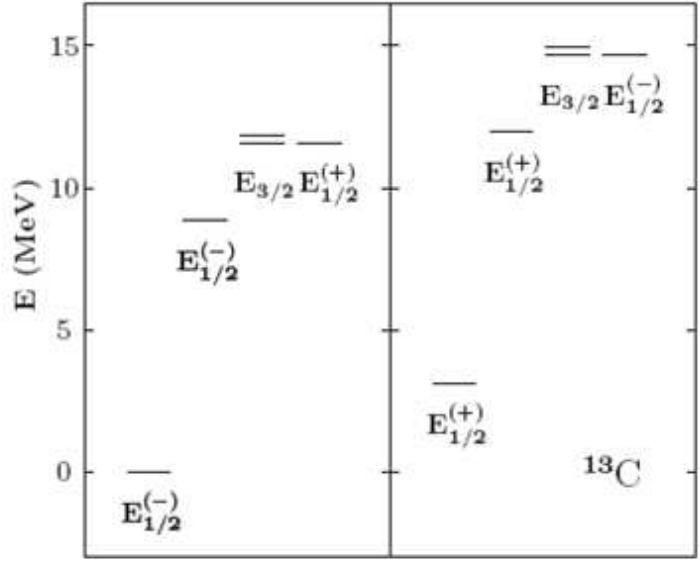} 
\caption[Vibrational spectra of a triangular configuration]
{Vibrational spectra of a triangular configuration.}
\label{vib3}
\end{figure}

An expression similar to Eq.~(\ref{erot2}) applies to the rotational energy levels of an equilateral triangle
configuration. The rotational formula is
\ba
E_{\rm rot}(\Omega,K,J) &=& \varepsilon_{\Omega} + A_{\Omega} 
\left[ J(J+1) + b_{\Omega} K^{2} \right. 
\nonumber\\
&& \hspace{1.5cm} \left. + a_{\Omega} (-1)^{J+1/2}(J+1/2) \delta_{K,1/2} \right] ~,
\label{erot3}
\ea
where $\varepsilon_{\Omega}$ is the intrinsic energy, $A_{\Omega} = \hbar^{2}/2{\cal I}$  
the inertial parameter, $b_{\Omega}$ a Coriolis term, and $a_{\Omega}$ the decoupling parameter. 
The latter term applies only to representations $\Omega = E_{1/2}^{(\pm)}$ and $K^P=1/2^{\pm}$.
$A_{\Omega}$, $b_{\Omega}$ and $a_{\Omega}$ can be calculated as in the previous
subsection. For each intrinsic state $\Omega$, the values of $K$ are given by Eq.~(\ref{omega3a}) and
$J=K,K+1,\ldots$. We note here that the rotational bands for a triangular configuration are
different from those of a dumbbell and of the collective model \cite{Nilsson:1955fn} since they include several
values of $K$ as given in Eq.~(\ref{omega3a}).
The vibrational spectra of a triangular configuration can be analyzed with the formula \cite{BI}
\begin{equation}
E_{\rm vib}(\Omega;v_{1\Omega},v_{2\Omega},v_{3\Omega}) \;=\; \omega_{1\Omega} v_{1\Omega} 
+ \omega_{2\Omega} v_{2\Omega} + \omega_{3\Omega} v_{3\Omega} ~,
\end{equation}
where again the zero-point energy has been removed. There are here three vibrational quantum
numbers and therefore the situation is more complex than in the case of a dumbbell
configuration. In Fig.~\ref{vib3}, the expected vibrational levels of a triangular configuration are shown.

\subsection{Electromagnetic transition probabilities}

Electromagnetic transition probabilities and moments can be calculated in the same way as in the
collective model. The wave functions are factorized as a product of the intrinsic wave functions,
$\chi_{\Omega}$, obtained as in Sect.~\ref{CSM}, the vibrational functions of the cluster $\psi_{\rm vib}$ 
which depend on the vibrational quantum numbers $v_i$ and a rotational part 
which is that of the symmetric top,
\ba
\left| \Omega;JMK \right> \;=\; \sqrt{\frac{2J+1}{4\pi^2}} \psi_{\rm vib} \left[ 
\chi_{\Omega,K} D^{(J)}_{M,K}(\Theta_i) + (-1)^{J+K} \chi_{\Omega,-K} D^{(J)}_{M,-K}(\Theta_i) \right] ~,
\label{wfk}
\ea
where $\Omega,K$ labels the intrinsic state, $J$ the angular momentum, and $M$ and $K$ its projection 
on the $z$-axis and the symmetry axis, respectively. Eq.~(\ref{wfk}) is valid when $K$ is 
a good quantum number, as is the case for the dumbbell and equilateral triangle configuration. 

The electric and magnetic multipole operators in the laboratory frame are written as a sum of single-particle 
and cluster contributions 
\ba
M_{\rm el}(\lambda,\mu) &=& T_{\lambda\mu}^{\rm el,sp} + T_{\lambda\mu}^{\rm el,c} ~, 
\nonumber\\ 
M_{\rm mag}(\lambda,\mu) &=& T_{\lambda\mu}^{\rm mag,sp} + T_{\lambda\mu}^{\rm mag,c} ~. 
\label{emlambda}
\ea
The matrix elements of the operators in Eq.~(\ref{emlambda}) can be calculated in the standard way and the
transition probabilities, defined as
\ba
B(\lambda; \Omega';K',J' \rightarrow \Omega;K,J) \;=\; 
\sum_{M,\mu} \left| \left< \Omega;K,J,M \left| M(\lambda,\mu) \right| \Omega';K',J',M' \right> \right|^2 ~, 
\ea
thus obtained as 
\ba
&& B(\lambda; \Omega';K',J' \rightarrow \Omega;K,J) 
\nonumber\\
&& \hspace {1cm} \;=\; \Big| \left<J',K',\lambda,K-K' | J,K \right> 
\left( \delta_{v,v'} G_{\lambda}^{\rm sp}(\Omega,\Omega') 
+ \delta_{\Omega,\Omega'} G_{\lambda}^{\rm c} \right)  
\nonumber\\
&& \hspace{2cm} +(-1)^{J+K} \left<J',K',\lambda,-K-K' | J,-K \right> 
\nonumber\\
&& \hspace{3cm}
\left( \delta_{v,v'} \tilde{G}_{\lambda}^{\rm sp}(\Omega,\Omega') 
+ \delta_{\Omega,\Omega'} G_{\lambda}^{\rm c} \right) \Big|^2 ~.
\label{Blambda}
\ea
The two terms in Eq.~(\ref{Blambda}) come from the symmetrization of the wave function in Eq.~(\ref{wfk}), 
and
\ba
G_{\lambda,K-K'}^{\rm sp}(\Omega,\Omega') &=& \left< \chi_{\Omega,K} \left| 
M_{\lambda,\mu=K-K'} \right| \chi_{\Omega',K'} \right> ~,
\nonumber\\
\tilde{G}_{\lambda,-K-K'}^{\rm sp}(\Omega,\Omega') &=& \left< \chi_{\Omega,-K} \left| 
M_{\lambda,\mu=-K-K'} \right| \chi_{\Omega',K'} \right> ~.
\label{Glambda}
\ea
The second term in Eq.~(\ref{Blambda}) contributes only in the case $\lambda \geq K+K'$. 

Similarly, the electric multipole moments are defined in the usual fashion as 
\ba
Q^{(\lambda)}(K,J) &=& \sqrt{\frac{16\pi}{2\lambda+1}} 
\left< \Omega;K,J,M=J \left| M_{\rm el}(\lambda,0) \right| \Omega;K,J,M=J \right> 
\nonumber\\
&=& \sqrt{\frac{16\pi}{2\lambda+1}} \left< J,K,\lambda,0 | J,K \right> 
\left< J,J,\lambda,0 | J,J \right> 
\nonumber\\
&& \hspace{2cm} \times \left( G_{\lambda,0}^{\rm el,sp}(\Omega,\Omega) 
+ G_{\lambda,0}^{\rm el,c} \right) ~,
\ea
and the magnetic multipoles as 
\ba
\mu^{(\lambda)}(K,J) &=& \sqrt{\frac{4\pi}{2\lambda+1}} 
\left< \Omega;K,J,M=J \left| M_{\rm mag}(\lambda,0) \right| \Omega;K,J,M=J \right> 
\nonumber\\
&=& \sqrt{\frac{4\pi}{2\lambda+1}} \left< J,K,\lambda,0 | J,K \right> 
\left< J,J,\lambda,0 | J,J \right> 
\nonumber\\
&& \hspace{2cm} \times \left( G_{\lambda,0}^{\rm mag,sp}(\Omega,\Omega) 
+ G_{\lambda,0}^{\rm mag,c} \right) ~.
\ea
Electromagnetic transition rates and moments in odd nuclei have contributions from both the
single particle and the cluster, Eq.~(\ref{emlambda}). The single-particle contribution is written 
in the standard form \cite{shalit1963nuclear,brussaard1977shell}
\ba
T_{\lambda,\mu}^{\rm el,sp} &=& e_{\rm eff} r^{\lambda} Y_{\lambda\mu}(\theta,\phi) ~,
\nonumber\\
T_{\lambda,\mu}^{\rm mag,sp} &=& \frac{\hbar c}{2mc^2} \left( g_s \vec{s} + \frac{2}{\lambda+1} g_l \vec{l} \right) 
\cdot \vec{\nabla} \left[ r^{\lambda} Y_{\lambda\mu}(\theta,\phi) \right] ~,
\ea
where $e_{\rm eff}$ is the effective charge center-of-mass corrected \cite{brussaard1977shell}
\ba
e_{\rm eff}^p &=& e + (-1)^{\lambda} \frac{Ze}{A^{\lambda}} ~,
\nonumber\\
e_{\rm eff}^n &=& (-1)^{\lambda} \frac{Ze}{A^{\lambda}} ~,
\ea
and the $g$-factors are given by 
\ba
g_s^p \;=\; +5.5855 ~, \qquad g_l^p \;=\; 1 ~,
\nonumber\\
g_s^n \;=\; -3.8256 ~, \qquad g_l^n \;=\; 0 ~.
\ea
The cluster contribution depends on the vibrational quantum numbers $v_i$  and on the charge
and magnetization distribution. The electric cluster contribution can be evaluated using the 
algebraic cluster model (ACM) described in Sect.~\ref{ACM}, and it depends on the configuration.

\subsubsection{Dumbbell configuration}

For the dumbbell configuration 
\ba
G_{\lambda}^{\rm el,c} \;=\; \frac{Z \beta^{\lambda} c_\lambda}{\sqrt{4\pi}} ~, 
\ea
with $c_{\lambda}=\sqrt{2\lambda+1}$ with $\lambda=\mbox{even}=0,2,4,\ldots$ 
(see Eq.~(\ref{cl2})..

\subsubsection{Equilateral triangle configuration}

For the equilateral triangle configuration 
\ba
G_{\lambda}^{\rm el,c} \;=\; \frac{Z \beta^{\lambda} c_\lambda}{\sqrt{4\pi}} ~, 
\ea
where $c_{\lambda}$ is given by Eq.~(\ref{cl3}). 

The magnetic cluster contribution is rather difficult to evaluate. Since the cluster is composed of
spin-less $\alpha$-particles 
\ba
G_{\lambda}^{\rm mag,c} \;=\; 0 ~, 
\ea
has been taken in all calculations performed so far.

\subsection{Form factors in electron scatterng}

Form factors in electron scattering can also be split into a single-particle 
and collective cluster contribution, 
\ba
{\cal F}(i \rightarrow f;q) \;=\; 
{\cal F}^{\rm sp}(i \rightarrow f;q) + {\cal F}^{\rm c}(i \rightarrow f;q) ~.
\ea
The single-particle contribution ${\cal F}^{\rm sp}$ gives rise to longitudinal electric, transverse magnetic 
and transverse electric form factors. These contributions were derived in the laboratory frame by 
De Forest and Walecka \cite{DeForest:1966ycn}. They were converted to the intrinsic frame in 
\cite{DellaRocca:2018mrt}, where explicit expressions are given. Since the cluster contribution is composed 
of spin-less $\alpha$-particles, it is assumed that the cluster contribution ${\cal F}^{\rm c}$ applies only 
to the longitudinal form factors. 

\subsubsection{Dumbbell configuration}

For the dumbbell configuration
\ba
{\cal F}_{\lambda}^{\rm c}(J,K \rightarrow J',K';q) \;=\; \delta_{K,K'} 
\left< J,K,\lambda,0 \mid J',K' \right> \, c_{\lambda} j_{\lambda}(q\beta) \mbox{e}^{-q^2/4\alpha} ~,
\label{cff2}
\ea
where $\lambda=\mbox{even}=0,2,\ldots$. Here $\alpha$ and $\beta$ are the parameters of the cluster density of 
Eq.~(\ref{geometry2}). For odd multipolarities it has a more complicated dependence on $\beta$, as discussed in 
\cite{DellaRocca:2018mrt}.

\subsubsection{Equilateral triangle configuration}

In this case, the cluster contribution is given by
\ba
{\cal F}_{\lambda}^{\rm c}(J,K \rightarrow J',K';q) \;=\; \delta_{K,K'} 
\left< J,K,\lambda,0 \mid J',K' \right> \, c_{\lambda} j_{\lambda}(q\beta) \mbox{e}^{-q^2/4\alpha} ~,
\label{cff3}
\ea
where $\lambda=0,2,3,4,\ldots$, as in Eq.~(\ref{geometry3}).

\section{Evidence for cluster structure in odd nuclei}
\label{sec8}

The CSM provides a simple way to analyze cluster structures in $k\alpha+x$ nuclei, in particular odd nuclei ($x=1$), consistent with the Pauli principle. To this end, one places nucleons in the single-particle orbitals of Figs.~\ref{csm2}, \ref{csm3} and \ref{csm4} but with no two particles in the same level. For example, the ground state of $^{9}$Be is the configuration $[1\sigma_g 1/2]^2$ $[1\sigma_u 1/2]^2$ $[1\pi_u 3/2]$, where the molecular notation has been used. For $x>1$, one can use the same approach as in the spherical shell model or in the deformed shell model by introducing an effective interaction between the valence $x$-particles and diagonalizing this interaction in the cluster shell model basis. Preliminary results have been obtained for $^{10}$Be, $^{10}$B ($x=2$). In the following subsections, however, we analyze only 
cluster structures in nuclei with $Z_2$ symmetry plus one particle, $^{9}$Be and $^{9}$B, 
and in nuclei with $D_{3h}$ symmetry plus one particle, $^{13}$C. The study of nuclei with 
$T_d$ symmetry plus one particle, $^{17}$O and $^{17}$F, and of nuclei with $Z_2$, $D_{3h}$ and $T_d$ symmetry plus two particles, $^{10}$Be, $^{14}$C and $^{18}$O, is planned for future investigations. We note here that while the case of the two-center shell model ($Z_2$ symmetry) has been extensively investigated \cite{Andersen:1970tov,Scharnweber:1971qpn}, the three- and four-center shell model has not been studied within the context of nuclear physics.

\subsection{Energy levels}

\subsubsection{Dumbbell configuration}

\begin{figure}
\centering
\includegraphics[width=4in]{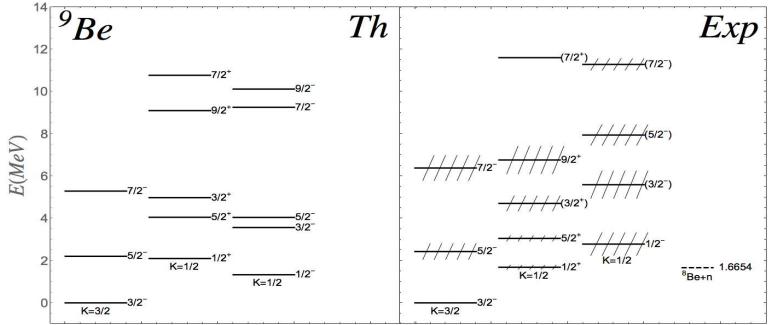} 
\caption[Energy spectrum of $^9$Be]
{Comparison between the theoretical and experimental [63] spectrum of $^9$Be. 
The dashed region is given by the width of the states. 
Reproduced from \cite{DellaRocca:2018mrt} with permission.}
\label{Be9}
\end{figure}

\begin{figure}
\centering
\includegraphics[width=4in]{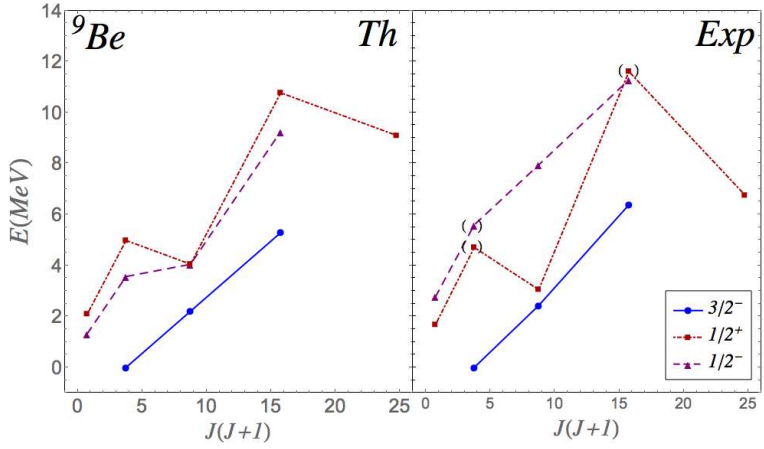} 
\caption[Rotational bands in $^9$Be]
{Rotational bands in $^9$Be. Reproduced from \cite{DellaRocca:2018mrt} with permission.}
\label{bandsBe9}
\end{figure}

The energy spectrum of $^9$Be is shown in Fig.~\ref{Be9} where it is compared with the experimental
spectrum. Three rotational bands have been observed with $K^P=3/2^-$, $1/2^-$ and $1/2^+$ which can 
be assigned to the representations $\Omega=[1\pi_u3/2]$, $[1\pi_u1/2]$, $[2\sigma_g1/2]$, respectively. 
It is convenient to visualize the three bands by plotting the energies of each level as a function of
$J(J+1)$ as shown in Fig.~\ref{bandsBe9}. It is seen that the band with $K^P=1/2^+$ has a large 
decoupling parameter. The ACM appears to describe the energy levels well, including the large decoupling
of the $K^P=1/2^+$. 

Table~\ref{inertiapar} shows a comparison between the experimental inertia and decoupling parameters, 
$A_{\Omega}$ and $a_{\Omega}$. The agreement is remarkable in view of the fact that there are no free 
parameters that have been adjusted, the value of $\beta$ having been fixed from the moment of 
inertia of $^{8}$Be.

\begin{table}
\centering
\caption[Inertia and decoupling parameters]{Inertia parameters and decoupling parameters in $^{9}$Be 
\cite{DellaRocca:2018mrt}.}
\label{inertiapar}
\vspace{10pt}
\begin{tabular}{ccccccc}
\hline
\noalign{\smallskip}
$^{9}$Be && \multicolumn{2}{c}{$A_{\Omega}$ (MeV)} && \multicolumn{2}{c}{$a_{\Omega}$} \\
\cline{3-4} \cline{6-7}
&& Exp & Calc && Exp & Calc \\
\noalign{\smallskip}
\hline
\noalign{\smallskip}
$K^P=3/2^-$ && $0.486 \pm 0.024$ & $0.441$ && & \\
$K^P=1/2^+$ && $0.385 \pm 0.019$ & $0.387$ && $1.61 \pm 0.08$ & $1.48$ \\
$K^P=1/2^-$ && $0.542 \pm 0.054$ & $0.420$ && $0.89 \pm 0.09$ & $0.77$ \\
\noalign{\smallskip}
\hline
\end{tabular}
\end{table}

\begin{figure}
\centering
\includegraphics[width=4in]{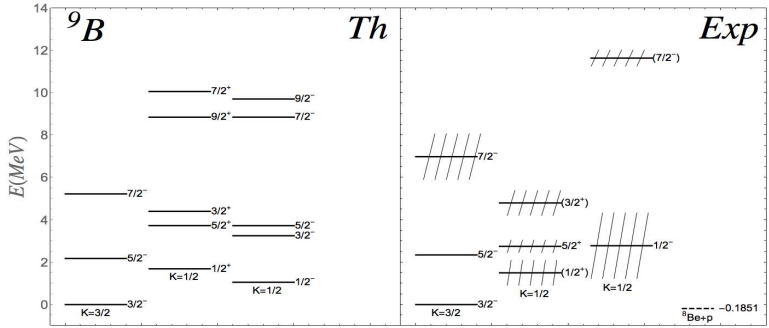} 
\caption[Energy spectrum of $^9$B]
{Comparison between the theoretical and experimental [63] spectrum of $^9$B. 
The observed level at 1.5 MeV is tentatively assigned as $J^P=1/2^+$ and 
that at 4.8 MeV to $3/2^+$ in analogy with $^9$Be. 
Reproduced from \cite{DellaRocca:2018mrt} with permission.}
\label{B9}
\end{figure}

\begin{figure}
\centering
\includegraphics[width=4in]{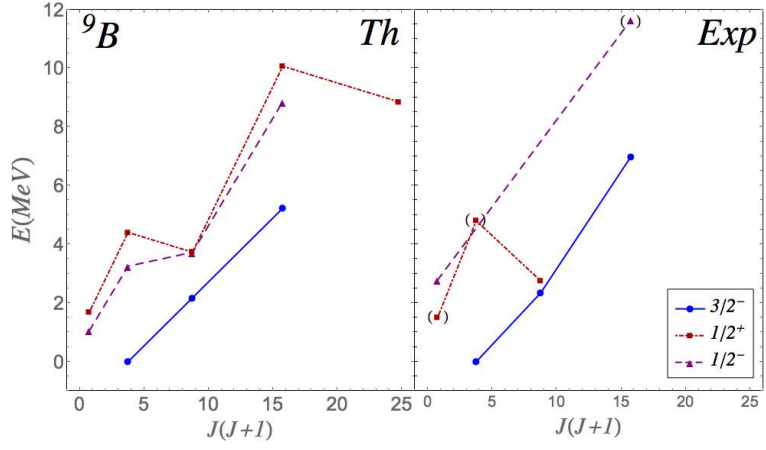} 
\caption[Rotational bands in $^9$B]
{Rotational bands in $^9$B. Reproduced from \cite{DellaRocca:2018mrt} with permission.}
\label{bandsB9}
\end{figure}

The same situation occurs for the nucleus $^9$B. In Fig.~\ref{B9} the spectrum of $^9$B is shown 
in comparison with CSM. The Coulomb displacement energies between states in $^9$Be and $^9$B are 
calculated well as shown in Table~\ref{Coulomb} \cite{DellaRocca:2018mrt}. 
Rotational bands in $^9$B are shown in Fig.~\ref{bandsB9}.

\begin{table}
\centering
\caption[Coulomb displacement energies]
{Coulomb displacement energies in MeV. 
The experimental uncertainty is estimated from 
the width of the states in $^9$Be and $^9$B.}
\label{Coulomb}
\vspace{10pt}
\begin{tabular}{ccc}
\hline
\noalign{\smallskip}
$K^P$ & CSM & Exp \\
\noalign{\smallskip}
\hline
\noalign{\smallskip}
$3/2^-$ & $1.92$ & $1.84 \pm 0.02$ \\
$1/2^-$ & $1.64$ & $1.81 \pm 0.36$ \\
$1/2^+$ & $1.46$ & $1.76 \pm 0.52$ \\
\noalign{\smallskip}
\hline
\end{tabular}
\end{table}

\subsubsection{Equilateral triangle configuration}

\begin{figure}
\centering
\includegraphics[width=4in]{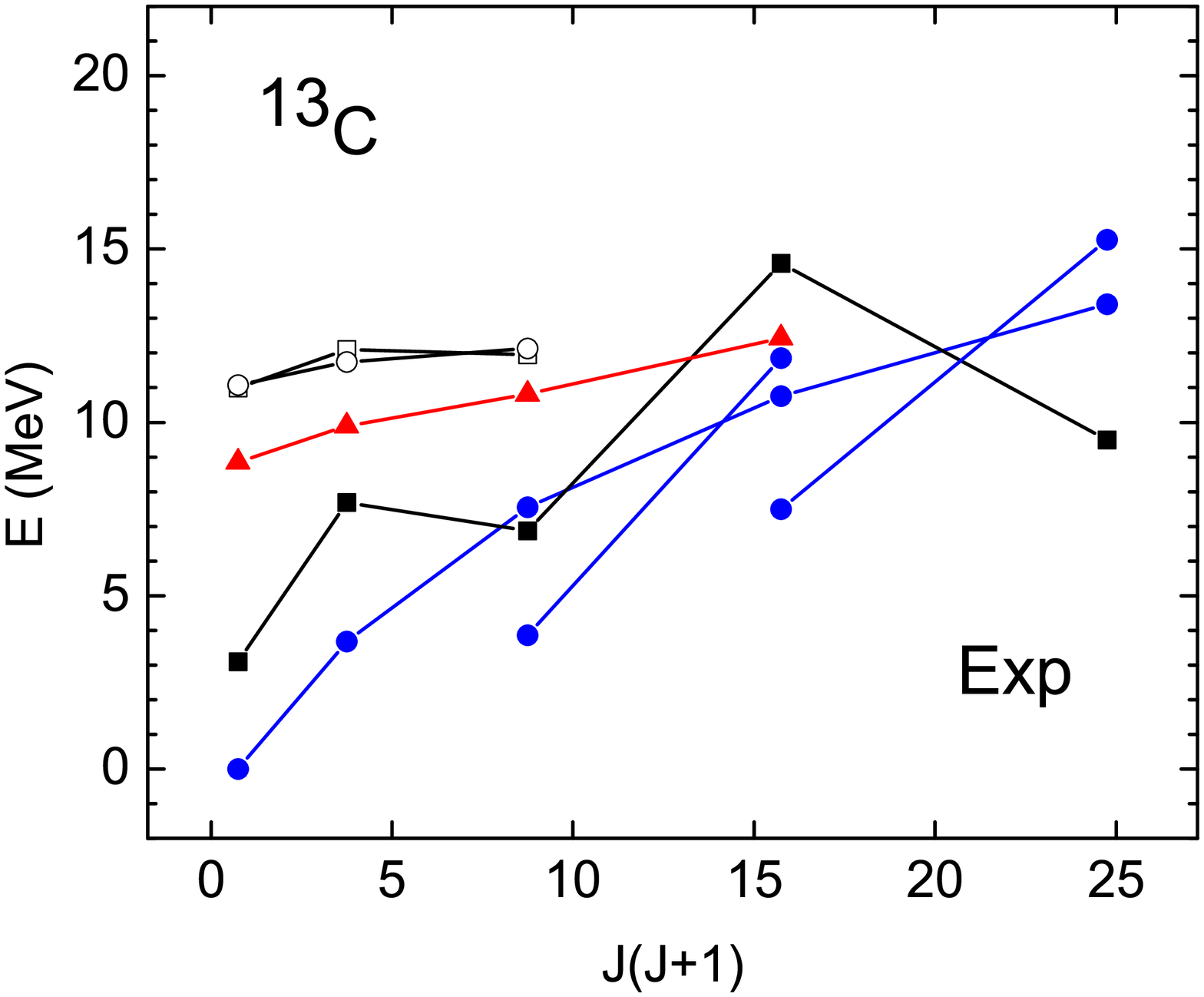} 
\includegraphics[width=4in]{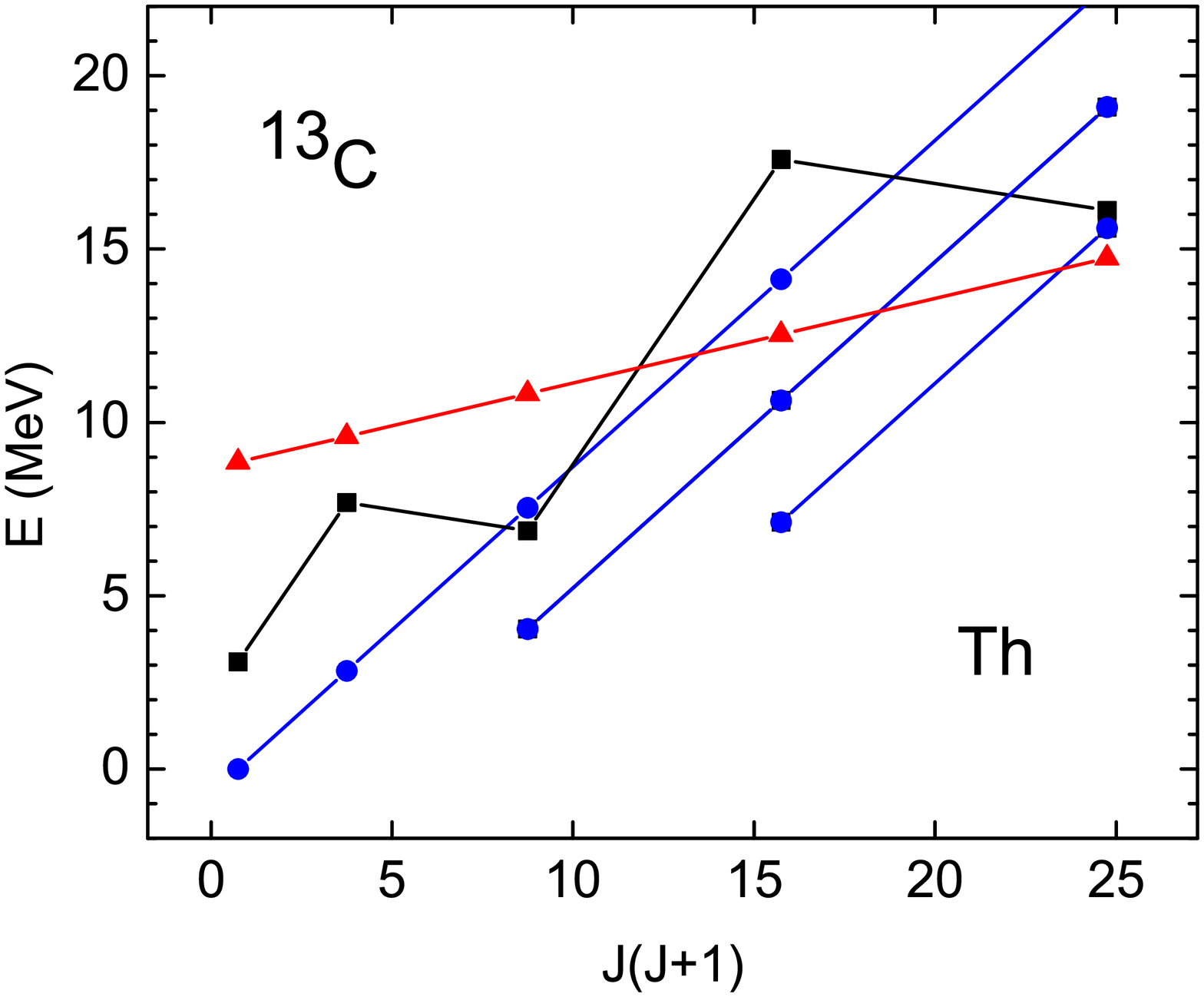} 
\caption[Rotational bands in $^{13}$C]
{Rotational bands in $^{13}$C. Energy levels are plotted as a function of $J(J+1)$. 
Rotational spectra expected on the basis of $D’_{3h}$ symmetry \cite{BI}.}
\label{bandsC13}
\end{figure}

\begin{figure}
\centering
\includegraphics[width=3.5in]{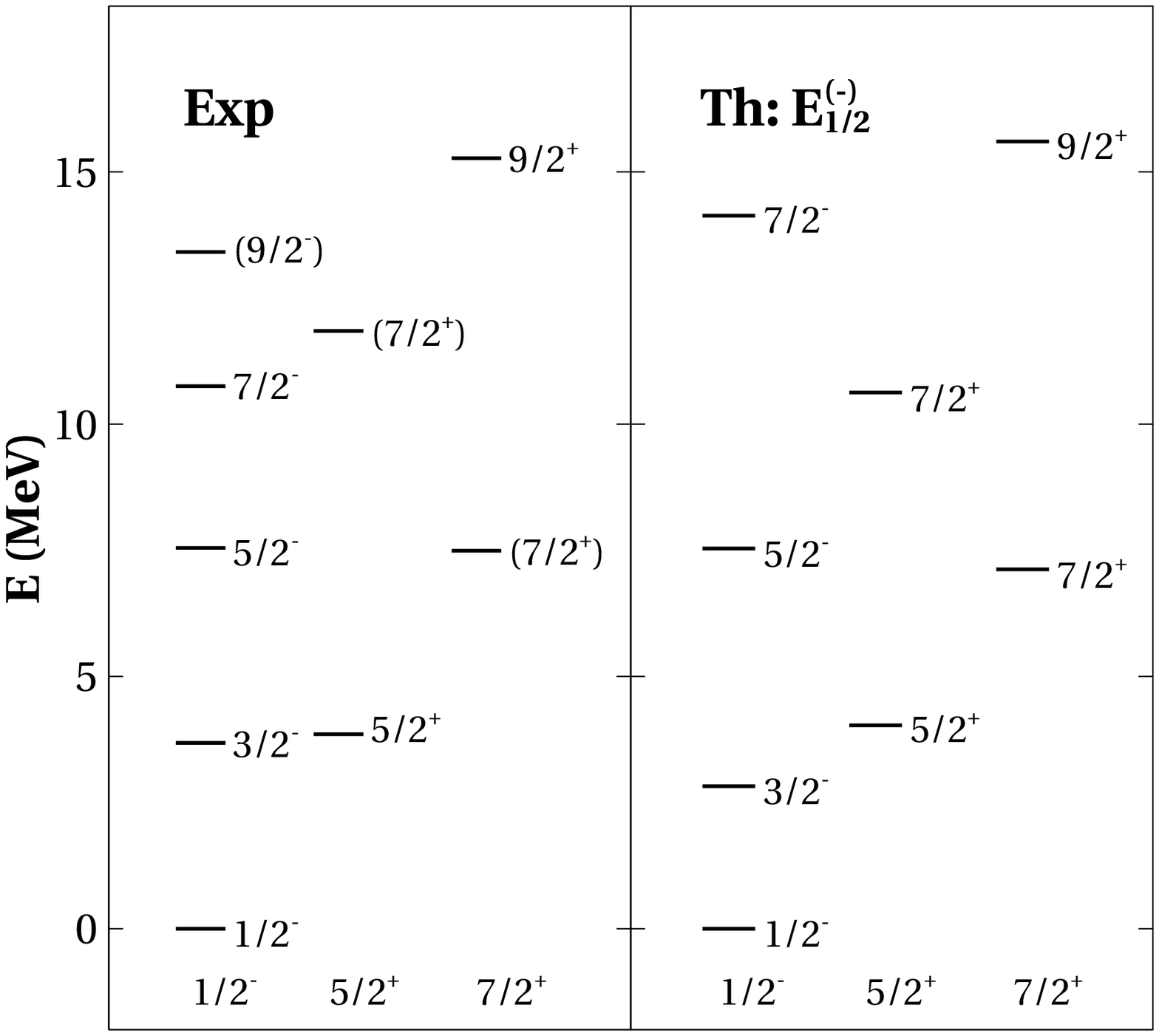} 
\includegraphics[width=3.5in]{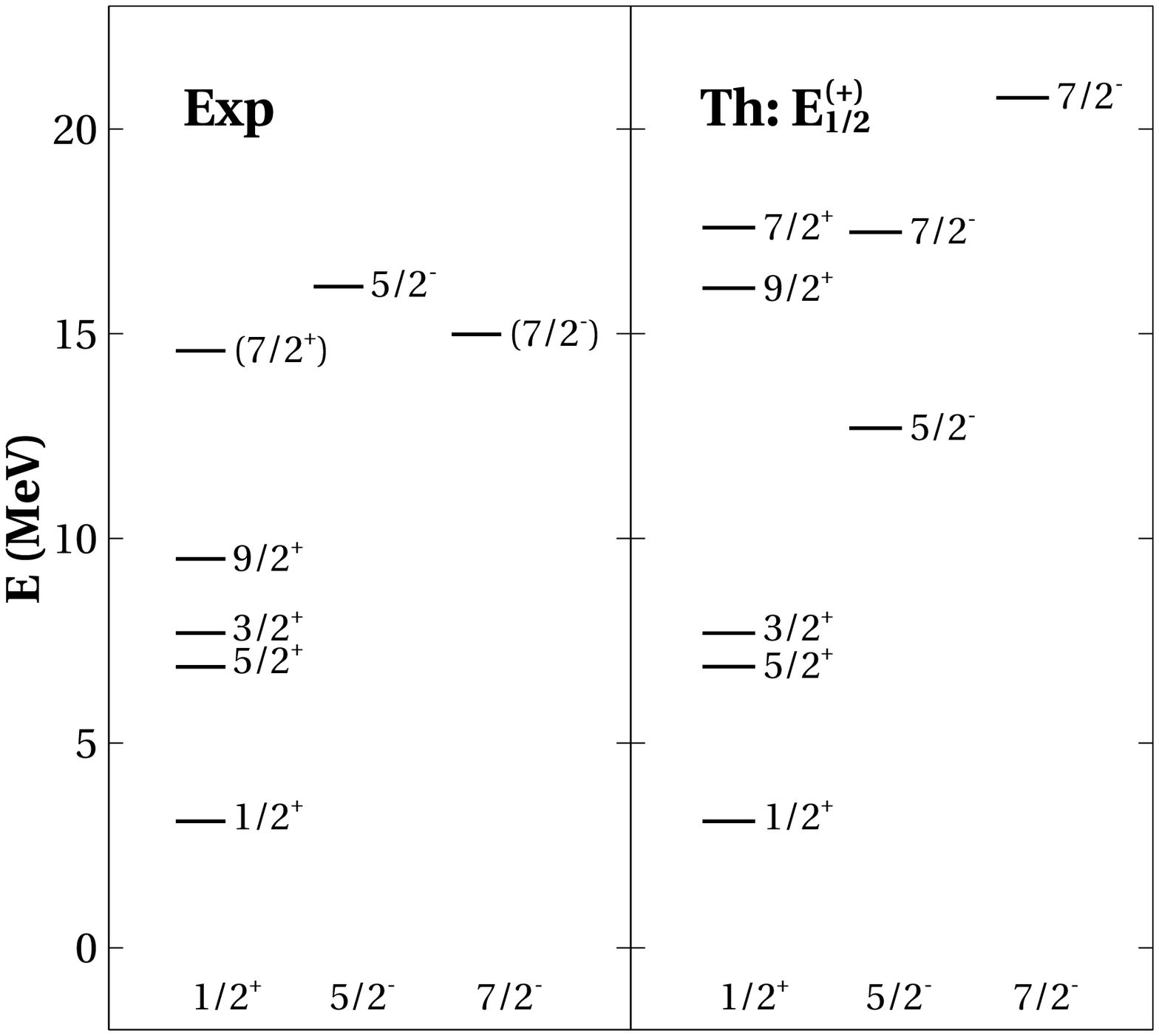} 
\caption[Energy spectrum in $^{13}$C]
{Top: Comparison between experimental and theoretical energy level diagram for the
ground-state band of $^{13}$C assigned to the representation $E_{1/2}^{(-)}$ of $D'_{3h}$. 
Bottom: Same as top but for the first excited band assigned to the representation 
$E_{1/2}^{(+)}$ of $D'_{3h}$. Taken from \cite{BI}.}
\label{C13}
\end{figure}

The rotational bands of $^{13}$C are shown in Fig.~\ref{bandsC13}, top (experiment) and bottom (CSM). 
It appears that two rotational bands with $\Omega=E_{1/2}^{(-)}$ and $E_{1/2}^{(+)}$ have been observed. 
In addition, it also appears that a vibrational band with $\Omega=E_{1/2}^{(-)}$ has been observed 
analogous to the Hoyle band in $^{12}$C. A striking result here is that the angular momentum content of 
each observed band is what expected on the basis of $D’_{3h}$ triangular symmetry, as shown in Fig.~\ref{C13}. 

\subsection{Electromagnetic transition rates}

\subsubsection{Dumbbell configuration}

Extensive calculations of electromagnetic transition rates have been done in $^9$Be and $^9$B 
\cite{DellaRocca:2018mrt}. Here we show some selected results. Electric transitions within a 
rotational band are dominated by the cluster contribution. For the ground state rotational band 
with $K^P=3/2^-$ we have
\ba
B(E2;3/2,J' \rightarrow 3/2,J) \;=\; (Ze\beta^2)^2 \frac{5}{4\pi} \left| \left< J',3/2,2,0 | J,3/2 \right> \right|^2 ~,
\ea
and quadrupole moment
\ba
Q^{(2)}(3/2,J) \;=\; \sqrt{\frac{16\pi}{5}} Ze\beta^2 \sqrt{\frac{5}{4\pi}} 
\left< J,3/2,2,0 | J,3/2 \right> \left< J,J,2,0 | J,J \right> ~.
\ea
Inserting the value of $\beta$ as determined from the moment of inertia of $^8$Be we obtain the results of
Table X. The agreement between theory and experiment is excellent and provides the strongest
argument for the cluster structure of $^9$Be seen as $^8\mbox{Be}+n$. 

Magnetic transitions within a rotational band are determined by the single-particle contribution.
For the ground state rotational band with $K^P=3/2^-$ we have
\ba
B(M1;3/2,J' \rightarrow 3/2,J) \;=\; \left| \left< J',3/2,1,0 | J,3/2 \right> \right|^2 
\left| G_{1,0}^{\rm sp}(3/2) \right|^2 ~,
\ea
and quadrupole moment
\ba
\mu^{(1)}(3/2,J) \;=\; \sqrt{\frac{4\pi}{3}}  
\left< J,3/2,1,0 | J,3/2 \right> \left< J,J,1,0 | J,J \right> G_{1,0}^{\rm sp}(3/2) ~,
\ea
where $G_{1,0}^{\rm sp}$ is given by Eq.~(\ref{Glambda}). Inserting $g_s^n=-3.8256$ gives the results of Table~\ref{emBe9}. 
The magnetic moment is well reproduced while $B(M1)$ is a factor of $\sim 2$ smaller than the experimental value. 
(The units of $B(M1)$ are those used in electron scattering \cite{PhysRevC.43.1740}.)

\begin{table}
\centering
\caption[Electromagnetic moments and transitions in $^9$Be]
{Electromagnetic moments and transitions in $^9$Be. Experimental data are taken from \cite{Tilley:2004zz} 
and theoretical CSM results from \cite{DellaRocca:2018mrt}.}
\label{emBe9}
\vspace{10pt}
\begin{tabular}{cccl}
\hline
\noalign{\smallskip}
& Exp & CSM & \\
\noalign{\smallskip}
\hline
\noalign{\smallskip}
$Q(3/2^-)$ & $5.288 \pm 0.038$ & $5.30$ & efm$^2$ \\
$B(E2;3/2^- \rightarrow 5/2^-)$ & $40.5 \pm 3.0$ & $35.9$ & e$^2$fm$^4$ \\ 
$B(E2;3/2^- \rightarrow 7/2^-)$ & $18   \pm 8  $ & $20.0$ & e$^2$fm$^4$ \\ 
\noalign{\smallskip}
\hline
\noalign{\smallskip}
$\mu(3/2^-)$ & $-1.1778 \pm 0.0009$ & $-1.13$ & $\mu_N$ \\
$B(M1;3/2^- \rightarrow 5/2^-)$ & $0.82 \pm 0.03$ & $0.35$ & e$^2$fm$^2$ \\ 
\noalign{\smallskip}
\hline 
\end{tabular}
\end{table}

\subsubsection{Equilateral triangle configuration}

Some calculations are available for electromagnetic transition rates in $^{13}$C. Table~\ref{emC13} 
shows results for electric transitions in the ground state band, representation $\Omega=E_{1/2}^{(-)}$ 
of $D'_{3h}$. One should note the large $B(E3)$ value for the transition $5/2^+ \rightarrow 1/2^-$. 

\begin{table}
\centering
\caption[$B(EL)$ values in $^{13}$C]
{$B(EL)$ values in $^{13}$C. Experimental data are taken from \cite{AJZENBERGSELOVE19911} 
and theoretical CSM results from \cite{BI}.}
\label{emC13}
\vspace{10pt}
\begin{tabular}{cccl}
\hline
\noalign{\smallskip}
& Exp & CSM & \\
\noalign{\smallskip}
\hline
\noalign{\smallskip}
$B(E2;3/2^- \rightarrow 1/2^-)$ & $6.4 \pm 1.5$ & $8.3$ & $e^{2}\mbox{fm}^{4}$ \\ 
$B(E2;5/2^- \rightarrow 1/2^-)$ & $5.6 \pm 0.4$ & $5.5$ & $e^{2}\mbox{fm}^{4}$ \\ 
$B(E3;5/2^+ \rightarrow 1/2^-)$ & $100 \pm 40$  & $42$  & $e^{2}\mbox{fm}^{6}$ \\ 
\noalign{\smallskip}
\hline
\end{tabular}
\end{table}

\subsection{Form factors in electron scattering}

\subsubsection{Dumbbell configuration}

\begin{figure}
\centering
\includegraphics[width=4.5in]{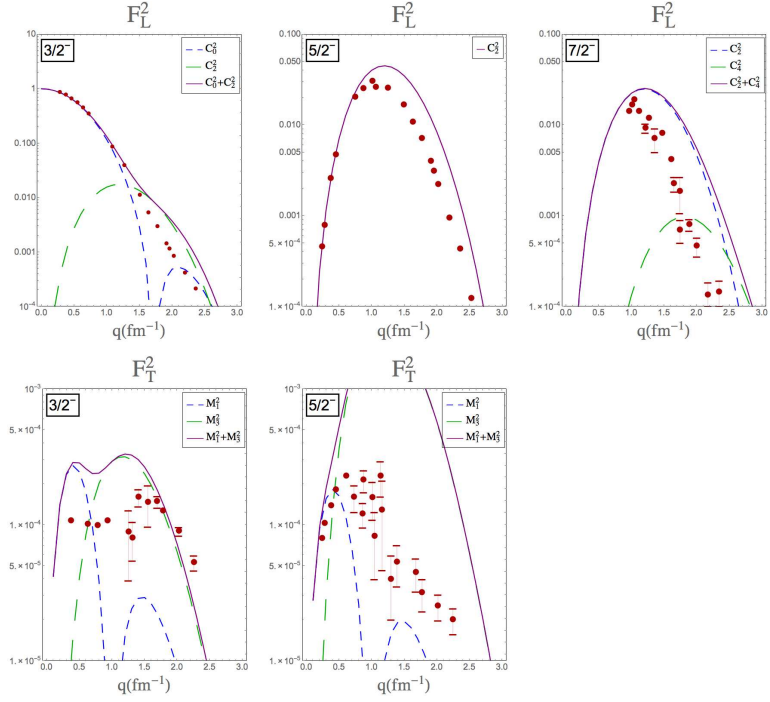} 
\caption[Form factors of $^{9}$Be]
{Comparison between calculated and experimental \cite{PhysRevC.43.1740} form factors of $^9$Be for members of the
ground-state rotational band with $K^P=3/2^-$. Top panels: Longitudinal form factors. 
Bottom panels: Transverse magnetic form factors. Experimental data from \cite{PhysRevC.43.1740}. 
Adapted from \cite{DellaRocca:2018mrt} with permission.}
\label{ffBe9}
\end{figure}

Extensive calculations of form factors have been done in $^9$Be \cite{DellaRocca:2018mrt}, including longitudinal
electric, transverse magnetic and transverse electric form factors. Here we show in Fig.~\ref{ffBe9} only
selected results. As one can see from this figure, longitudinal electric form factors dominated by
the cluster contribution are very well described by CSM, while magnetic transverse form factors
dominated by the single-particle contribution are not, especially for the ground state $J^P=3/2^-$, 
in spite of the fact that the magnetic moment is calculated very well, see Table~\ref{emBe9}. The same problem appears 
in large shell model calculations as reported in \cite{PhysRevC.43.1740}. The disagreement may be due to an inconsistency 
between experiments measuring the magnetic moments and those extracting the form factors from electron scattering. 

\subsubsection{Equilateral triangle configuration}

\begin{figure}
\centering
\includegraphics[width=4in]{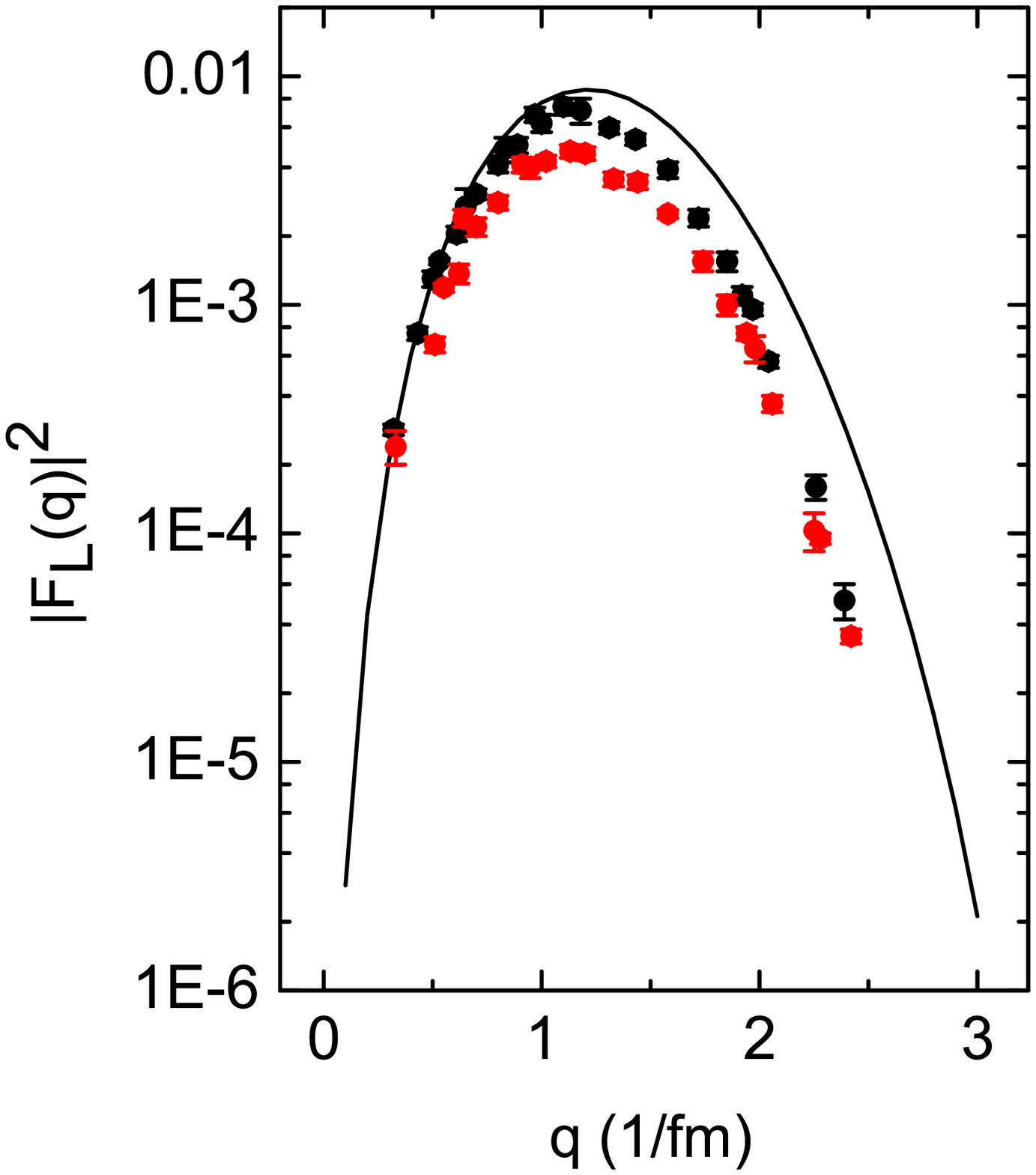} 
\caption[Form factors of $^{13}$C]
{Comparison between calculated and experimental \cite{PhysRevC.39.14} longitudinal $E2$ form factors for the 
ground-state band of $^{13}$C, $1/2^- \rightarrow 5/2^-$ (black) and $1/2^- \rightarrow 3/2^-$ (red). 
Taken from \cite{BI}.}
\label{ffC13}
\end{figure}

Only some longitudinal electric form factors have been calculated so far in $^{13}$C. Fig.~\ref{ffC13} shows
the results for $E2$ form factors for the ground state band. Particularly noteworthy is the fact that
the two form factors in Fig.~\ref{ffC13} are expected to be identical in the $D'_{3h}$ symmetry, and indeed
appear to be so.

\section{Softness and higher-order corrections}

For the two configurations discussed in the previous sections, dumbbell and equilateral triangle,
the effect of softness can be analyzed by modifying the rotational formula to
\ba
E_{\rm rot}(\Omega,K,J) &=& \varepsilon_{\Omega} + A_{\Omega} 
\left[ J(J+1) + \eta_{\Omega} J^2(J+1)^2 + b_{\Omega} K^{2} \right.
\nonumber\\
&& \hspace{2cm} \left. + a_{\Omega} \delta_{K,1/2} (-1)^{J+1/2}(J+1/2) \right] ~,
\ea
where $\varepsilon_{\Omega}$, $A_{\Omega}$, $b_{\Omega}$ and $a_{\Omega}$ have the same meaning 
as in Eq.~(\ref{erot3}), and $\eta_{\Omega}$ is a stretching parameter.
Similarly, the vibrational energy needs to be modified to
\begin{equation}
E_{\rm vib}(\Omega;v_{\Omega}) \;=\; \omega_{\Omega} v_{\Omega} + x_{\Omega} v_{\Omega}^2 ~,
\end{equation}
for a dumbbell configuration, and
\begin{equation}
E_{\rm vib}(\Omega;v_{1\Omega},v_{2\Omega},v_{3\Omega}) \;=\; \sum_{i=1}^3 \omega_{i\Omega} v_{i\Omega} 
+ \sum_{i \geq j=1}^3 x_{ij,\Omega} v_{i\Omega} v_{j\Omega} ~,
\end{equation}
for a triangular configuration. The values of $x_{ij,\Omega}$ are the anharmonicities.

\section{Summary and conclusions}

In this article, the cluster structure of light nuclei has been reviewed. In the first part, cluster
structures in $k\alpha$ nuclei with $k=2$ ($^{8}$Be), $k=3$ ($^{12}$C) and $k=4$ ($^{16}$O) have been 
analyzed in terms of the algebraic cluster model (ACM). The advantage of this model is that it produces 
explicit analytic results for energies, electromagnetic transition rates and form factors in electron
scattering. Evidence for a cluster dumbbell configuration in $^{8}$Be, an equilateral triangle
configuration in $^{12}$C and a tetrahedral configuration in $^{16}$O has been presented. This
evidence confirms early suggestions \cite{Brink,Brink:1970ufk,Eichler:1970nqw} for the occurrence of 
these configurations in $^{8}$Be,
$^{12}$C and $^{16}$O. The ACM makes use of algebraic methods adapted to the symmetry of the structure
which is $Z_2$ (dumbbell), $D_{3h}$ (equilateral triangle) and $T_d$ (tetrahedron). These symmetries are
exploited to obtain the analytic results that are used to analyze experimental data.

In the second part, cluster structures of $k\alpha+x$ nuclei are analyzed in terms of a cluster shell
model (CSM). The advantage of this model is that single particle levels in cluster potentials with
arbitrary discrete symmetry can be easily calculated. The three cases of single-particle levels in
cluster potentials with discrete symmetry $Z_2$, $D_{3h}$, $T_d$ are shown explicitly. Here again the use of
symmetry considerations plays an important role, particularly in the classification of states
through the use of the double groups $Z'_2$, $D'_{3h}$, $T'_d$. Evidence for cluster structures in the odd
nuclei $^9$Be, $^9$B ($k=2$, $x=1$) and $^{13}$C ($k=3$, $x=1$) is presented. This evidence demonstrates that
cluster structures survive the addition of one nucleon, and confirms early suggestions 
\cite{vonOertzen:1970ecu,VONOERTZEN19751,IMANISHI198729,Oertzen} that spectra of $^{8}$Be and $^{9}$B can 
be well described as $^{8}$Be plus one particle.

We emphasize that in the ACM and the CSM most results can be obtained in terms of a single parameter, 
$\beta$, which represents the distance from the center of mass of the $\alpha$ particles to the center of 
mass of the nucleus. The value of this parameter is $\sim 2$ fm for all nuclei described in this review. 
This is an astonishing result which supports the ``simplicity in complexity'' program advocated by the authors. 

In this program of investigation of cluster structures in light nuclei what remains to be done is:
(1) in even nuclei the study of $^{20}$Ne ($k=5$), $^{24}$Mg ($k=6$) and $^{28}$Si ($k=7$) suggested 
in \cite{Brink,Brink:1970ufk} to have bi-pyramidal ($k=5$), octahedral or bi-pyramidal ($k=6$) and stacked 
triangular body-centered ($k=7$) structure with symmetry $D_{3h}$, $O_h$ or $D_{2h}$, $D_{3h}$ or $D_{3v}$, 
respectively; (2) in odd nuclei, the study of $^{17}$O, $^{17}$F ($k=4$, $x=1$) and, most importantly, 
the study of $k\alpha+x$ nuclei with $x>1$, especially $^{10}$Be, $^{10}$B and $^{11}$Be suggested in 
\cite{vonOertzen:1970ecu,VONOERTZEN19751} to have a dumbbell configuration plus $x=2$ and $x=3$ particles. 
The case of $^{10}$Be and $^{11}$Be is particularly timely since many experimental studies of 
these nuclei have been done in recent times. A preliminary calculation of $^{10}$Be within the framework of the 
CSM plus residual interactions has been done, which appears to indicate that cluster structures even survive 
the addition of two nucleons.

Most importantly, the review presented here in which most results are given in explicit analytic form, 
provides benchmarks for microscopic studies of cluster structures in light nuclei.

\appendix

\section{Permutation symmetry}
\label{app}

For $k$ identical clusters, the Hamiltonian has to be invariant under their permutation.  
Therefore, the eigenstates can be classified according to the representations of the 
permutation group $S_k$. The permutation symmetry of $k$ objects is determined by the 
transposition $P(12)$ and the cyclic permutation $P(12 \cdots k)$ (see Table~\ref{sk}). 
All other permutations can be expressed in terms of these elementary ones \cite{KRAMER1966241}. 
In this appendix we review the construction of eigenfunctions of the ACM Hamiltonian with 
definite permutation symmetry for cluster composed of $k=2$, $3$ and $4$ $\alpha$-particles, 
and clarify the notation used in Tables~\ref{sk} and \ref{irreps} \cite{Bijker_2016}. 

\subsection{Dumbbell configuration}
\label{app2}

For the permutation of two objects there are two different symmetry classes  
characterized by the Young tableaux $[2]$ and $[11]$. Due to the isomorphism 
$S_2 \sim Z_2$, the three symmetry classes can also be labeled by the irreducible 
representations of the point group $Z_2$ as $[2] \sim A$ and $[11] \sim B$ (Table~\ref{irreps}). 

The permutation symmetry can be determined by considering the transposition $P(12)$ 
\ba
P(12) \left( \begin{array}{c} \psi_{A} \\ \psi_{B} \end{array} \right) 
&=& \left( \begin{array}{rr} 1 & 0 \\ 0 & -1 \end{array} \right) 
\left( \begin{array}{c} \psi_{A} \\ \psi_{B} \end{array} \right) ~.
\label{trans2}
\ea

In the ACM, the transformation properties under $S_2 \sim Z_2$ follow from those of the 
building blocks. Algebraically, the transposition can be expressed as 
\ba
P(12) \left( \begin{array}{c} s^{\dagger} \\ b^{\dagger}_{\rho} \end{array} \right) 
\;=\; U_{\rm tr} \left( \begin{array}{c} s^{\dagger} \\ b^{\dagger}_{\rho} \end{array} \right) U^{-1}_{\rm tr} 
\;=\; \left( \begin{array}{rr} 1 & 0 \\ 0 & -1 \end{array} \right) 
\left( \begin{array}{c} s^{\dagger} \\ b^{\dagger}_{\rho} \end{array} \right) ~, 
\ea
with 
\ba
U_{\rm tr} \;=\; \mbox{e}^{i \pi b^{\dagger}_{\rho} b_{\rho}} ~,  
\label{p12}
\ea 
where $b^{\dagger}_{\rho} b_{\rho}$ is a shorthand notation for $\sum_{m} b^{\dagger}_{\rho,m} b_{\rho,m}$. 
The scalar boson, $s^{\dagger}$, transforms as the symmetric representation $[2] \sim A$, whereas the 
vector Jacobi boson, $b^{\dagger}_{\rho}$, transforms as the antisymmetric representation $[11] \sim B$. 

The discrete symmetry $t$ of a given wave function can be determined by evaluating the matrix 
element 
\ba
\left< \psi_t \right| P(12) \left| \psi_t \right> \;=\; 
\left< \psi_t \right| U_{\rm tr} \left| \psi_t \right> \;=\; \pm 1 ~. 
\ea

\subsection{Equilateral triangular configuration}
\label{app3}

For the permutation of three objects there are three different symmetry classes  
characterized by the Young tableaux $[3]$, $[21]$ and $[111]$. Due to the isomorphism 
$S_3 \sim D_3$, the three symmetry classes can also be labeled by the irreducible 
representations of the dihedral group $D_3$ as $[3] \sim A_{1}$, $[21] \sim E$, 
and $[111] \sim A_2$, with dimensions 1, 2 and 1, respectively (Table~\ref{irreps}).  

In this case, the permutation symmetry can be determined by considering the transposition $P(12)$ and the 
cyclic permutation $P(123)$. The transformation properties of the three different symmetry 
classes under $P(12)$ and $P(123)$ are given by
\ba
P(12) \left( \begin{array}{c} \psi_{A_1} \\ \psi_{E_{\rho}} \\ \psi_{E_{\lambda}} \\ \psi_{A_2} \end{array} \right) 
&=& \left( \begin{array}{rrrr} 1 & 0 & 0 & 0 \\ 0 & -1 & 0 & 0 \\ 0 & 0 & 1 & 0 \\ 0 & 0 & 0 & -1 \end{array} \right) 
\left( \begin{array}{c} \psi_{A_1} \\ \psi_{E_{\rho}} \\ \psi_{E_{\lambda}} \\ \psi_{A_2} \end{array} \right) ~,
\label{trans3}
\ea
and 
\ba
P(123) \left( \begin{array}{c} \psi_{A_1} \\ \psi_{E_{\rho}} \\ \psi_{E_{\lambda}} \\ \psi_{A_2} \end{array} \right) 
&=& \left( \begin{array}{cccc} 1 & 0 & 0 & 0 \\ 0 & -\frac{1}{2} & \frac{\sqrt{3}}{2} & 0 \\ 
0 & -\frac{\sqrt{3}}{2} & -\frac{1}{2} & 0 \\ 0 & 0 & 0 & 1 \end{array} \right) 
\left( \begin{array}{c} \psi_{A_1} \\ \psi_{E_{\rho}} \\ \psi_{E_{\lambda}} \\ \psi_{A_2} \end{array} \right) ~,
\label{cyclic3}
\ea

In the ACM, the transformation properties under $S_3 \sim D_3$ follow from those of the building blocks. 
Algebraically, the transposition and cyclic permutation can be expressed in terms of 
the generators $b^{\dagger}_{i} b_{j} \equiv \sum_{m} b^{\dagger}_{i,m} b_{j,m}$ that act in 
index space ($i,j=\rho$, $\lambda$). The transposition is given by 
\ba
P(12) \left( \begin{array}{c} s^{\dagger} \\ b^{\dagger}_{\rho} \\ b^{\dagger}_{\lambda} \end{array} \right) 
\;=\; U_{\rm tr} \left( \begin{array}{c} s^{\dagger} \\ b^{\dagger}_{\rho} \\ b^{\dagger}_{\lambda} \end{array} \right) U^{-1}_{\rm tr} \;=\; \left( \begin{array}{rrr} 1 & 0 & 0 \\ 0 & -1 & 0 \\ 0 & 0 & 1 \end{array} \right) 
\left( \begin{array}{c} s^{\dagger} \\ b^{\dagger}_{\rho} \\ b^{\dagger}_{\lambda} \end{array} \right) ~, 
\ea
where $U_{\rm tr}$ is given by Eq.~(\ref{p12}), and the cyclic permutation by
\ba
P(123) \left( \begin{array}{c} s^{\dagger} \\ b^{\dagger}_{\rho} \\ b^{\dagger}_{\lambda} \end{array} \right) 
&=& U_{\rm cycl} \left( \begin{array}{c} s^{\dagger} \\ b^{\dagger}_{\rho} \\ b^{\dagger}_{\lambda} \end{array} \right) U^{-1}_{\rm cycl}
\nonumber\\ 
&=& \left( \begin{array}{ccc} 1 & 0 & 0 \\ 0 & -\frac{1}{2} & \frac{\sqrt{3}}{2} \\ 
0 & -\frac{\sqrt{3}}{2} & -\frac{1}{2} \end{array} \right)  
\left( \begin{array}{c} s^{\dagger} \\ b^{\dagger}_{\rho} \\ b^{\dagger}_{\lambda} \end{array} \right) ~, 
\ea
with
\ba
U_{\rm cycl} \;=\; \mbox{e}^{i \pi (b^{\dagger}_{\rho} b_{\rho} + b^{\dagger}_{\lambda} b_{\lambda})} \, 
\mbox{e}^{\theta (b^{\dagger}_{\rho} b_{\lambda} - b^{\dagger}_{\lambda} b_{\rho})} ~, 
\label{p123}
\ea
and $\theta=\arctan \sqrt{3}$. 
The scalar boson, $s^{\dagger}$, transforms as the symmetric representation $[3] \sim A_1$, whereas the 
two vector Jacobi bosons, $b^{\dagger}_{\rho}$ and $b^{\dagger}_{\lambda}$, 
transform as the two components of the mixed symmetry representation $[21] \sim E$. 

The discrete symmetry $t$ of a given wave function can be determined by evaluating the matrix 
elements 
\ba
\left< \psi_t \right| P(12) \left| \psi_t \right> &=&  
\left< \psi_t \right| U_{\rm tr} \left| \psi_t \right> \;=\; \pm 1 ~, 
\nonumber\\
\left< \psi_t \right| P(123) \left| \psi_t \right> &=&  
\left< \psi_t \right| U_{\rm cycl} \left| \psi_t \right> ~, 
\ea
and comparing with Eqs.~(\ref{trans3}) and (\ref{cyclic3}). 

\subsection{Tetrahedral configuration}
\label{appa4}

For the permutation of four objects there are five different symmetry classes  
characterized by the Young tableaux $[4]$, $[31]$, $[211]$, $[22]$ and $[1111]$. 
Due to the isomorphism with the tetrahedral group $S_4 \sim T_d$, the five 
symmetry classes can also be labeled by the irreducible representations of the point 
group $T_d$ as $[4] \sim A_{1}$, $[31] \sim F_2$, $[22] \sim E$, $[211] \sim F_1$ 
and $[1111] \sim A_2$, with dimensions 1, 3, 2, 3 and 1, respectively (Table~\ref{irreps}). 

The transformation properties of the five different symmetry classes under the 
transposition $P(12)$ and the cyclic permutation $P(1234)$ are given by
\ba
P(12) \left( \begin{array}{c} \psi_{A_1} \\ \psi_{E_{\rho}} \\ \psi_{E_{\lambda}} \\ \psi_{A_2} \end{array} \right) 
&=& \left( \begin{array}{rrrr} 1 & 0 & 0 & 0 \\ 0 & -1 & 0 & 0 \\ 0 & 0 & 1 & 0 \\ 0 & 0 & 0 & -1 \end{array} \right) 
\left( \begin{array}{c} \psi_{A_1} \\ \psi_{E_{\rho}} \\ \psi_{E_{\lambda}} \\ \psi_{A_2} \end{array} \right) ~,
\nonumber\\
P(12) \left( \begin{array}{c} 
\psi_{F_{2\rho}} \\ \psi_{F_{2\lambda}} \\ \psi_{F_{2\eta}} \end{array} \right) 
&=& \left( \begin{array}{rrr} -1 & 0 & 0 \\ 0 & 1 & 0 \\ 0 & 0 & 1 \end{array} \right) 
\left( \begin{array}{c} \psi_{F_{2\rho}} \\ \psi_{F_{2\lambda}} \\ \psi_{F_{2\eta}} \end{array} \right) , 
\nonumber\\
P(12) \left( \begin{array}{c} 
\psi_{F_{1\rho}} \\ \psi_{F_{1\lambda}} \\ \psi_{F_{1\eta}} \end{array} \right) 
&=& \left( \begin{array}{rrr} 1 & 0 & 0 \\ 0 & -1 & 0 \\ 0 & 0 & -1 \end{array} \right) 
\left( \begin{array}{c} \psi_{F_{1\rho}} \\ \psi_{F_{1\lambda}} \\ \psi_{F_{1\eta}} \end{array} \right) ,
\label{trans4}
\ea
and 
\ba
P(1234) \left( \begin{array}{c} \psi_{A_1} \\ \psi_{E_{\rho}} \\ \psi_{E_{\lambda}} \\ \psi_{A_2} \end{array} \right) 
&=& \left( \begin{array}{cccc} 1 & 0 & 0 & 0 \\ 0 & \frac{1}{2} & -\frac{\sqrt{3}}{2} & 0 \\ 
0 & -\frac{\sqrt{3}}{2} & -\frac{1}{2} & 0 \\ 0 & 0 & 0 & 1 \end{array} \right) 
\left( \begin{array}{c} \psi_{A_1} \\ \psi_{E_{\rho}} \\ \psi_{E_{\lambda}} \\ \psi_{A_2} \end{array} \right) ~,
\nonumber\\
P(1234) \left( \begin{array}{c} 
\psi_{F_{2\rho}} \\ \psi_{F_{2\lambda}} \\ \psi_{F_{2\eta}} \end{array} \right) 
&=& \left( \begin{array}{ccc} -\frac{1}{2} & \frac{\sqrt{3}}{2} & 0 \\ 
-\frac{1}{2\sqrt{3}} & -\frac{1}{6} & \frac{\sqrt{8}}{3} \\ 
-\frac{\sqrt{2}}{\sqrt{3}} & -\frac{\sqrt{2}}{3} & -\frac{1}{3} \end{array} \right)  
\left( \begin{array}{c} \psi_{F_{2\rho}} \\ \psi_{F_{2\lambda}} \\ \psi_{F_{2\eta}} 
\end{array} \right) ~,
\nonumber\\
P(1234) \left( \begin{array}{c} 
\psi_{F_{1\rho}} \\ \psi_{F_{1\lambda}} \\ \psi_{F_{1\eta}} \end{array} \right) 
&=& \left( \begin{array}{ccc} \frac{1}{2} & -\frac{\sqrt{3}}{2} & 0 \\ 
\frac{1}{2\sqrt{3}} & \frac{1}{6} & -\frac{\sqrt{8}}{3} \\ 
\frac{\sqrt{2}}{\sqrt{3}} & \frac{\sqrt{2}}{3} & \frac{1}{3} \end{array} \right)  
\left( \begin{array}{c} \psi_{F_{1\rho}} \\ \psi_{F_{1\lambda}} \\ \psi_{F_{1\eta}} \end{array} \right) ~.
\label{cyclic4}
\ea

In the ACM, the transformation properties under $S_4 \sim T_d$ follow from those of the building blocks. 
The transposition is given by 
\ba
P(12) \left( \begin{array}{c} s^{\dagger} \\ b^{\dagger}_{\rho} \\ b^{\dagger}_{\lambda} \\ 
b^{\dagger}_{\eta} \end{array} \right) \;=\; U_{\rm tr} 
\left( \begin{array}{c} s^{\dagger} \\ b^{\dagger}_{\rho} \\ b^{\dagger}_{\lambda} \\ 
b^{\dagger}_{\eta} \end{array} \right) U^{-1}_{\rm tr} 
\;=\; \left( \begin{array}{rrrr} 1 & 0 & 0 & 0 \\ 0 & -1 & 0 & 0 \\ 0 & 0 & 1 & 0 \\ 
0 & 0 & 0 & 1 \end{array} \right) \left( \begin{array}{c} s^{\dagger} \\ b^{\dagger}_{\rho} \\ 
b^{\dagger}_{\lambda} \\ b^{\dagger}_{\eta} \end{array} \right) ~, 
\ea
where $U_{\rm tr}$ is given by Eq.~(\ref{p12}), and the cyclic permutation by
\ba
P(1234) \left( \begin{array}{c} s^{\dagger} \\ b^{\dagger}_{\rho} \\ b^{\dagger}_{\lambda} \\ 
b^{\dagger}_{\eta} \end{array} \right) &=& U_{\rm cycl} 
\left( \begin{array}{c} s^{\dagger} \\ b^{\dagger}_{\rho} \\ b^{\dagger}_{\lambda} \\ 
b^{\dagger}_{\eta} \end{array} \right) U^{-1}_{\rm cycl}
\nonumber\\ 
&=& \left( \begin{array}{cccc} 1 & 0 & 0 & 0 \\ 0 & -\frac{1}{2} & \frac{\sqrt{3}}{2} & 0 \\ 
0 & -\frac{1}{2\sqrt{3}} & -\frac{1}{6} & \frac{\sqrt{8}}{3} \\ 
0 & -\frac{\sqrt{2}}{\sqrt{3}} & -\frac{\sqrt{2}}{3} & -\frac{1}{3} \end{array} \right)  
\left( \begin{array}{c} s^{\dagger} \\ b^{\dagger}_{\rho} \\ b^{\dagger}_{\lambda} \\ 
b^{\dagger}_{\eta} \end{array} \right) ~, 
\ea
with
\ba
U_{\rm cycl} \;=\; \mbox{e}^{i \pi (b^{\dagger}_{\rho} b_{\rho} 
+ b^{\dagger}_{\lambda} b_{\lambda}+b^{\dagger}_{\eta} b_{\eta})} \, 
\mbox{e}^{\theta_1 (b^{\dagger}_{\rho} b_{\lambda} - b^{\dagger}_{\lambda} b_{\rho})} \, 
\mbox{e}^{\theta_2 (b^{\dagger}_{\lambda} b_{\eta} - b^{\dagger}_{\eta} b_{\lambda})} ~, 
\label{p1234}
\ea
and $\theta_1=\arctan \sqrt{3}$ and $\theta_2=\arctan \sqrt{8}$. 
The scalar boson, $s^{\dagger}$, transforms as the symmetric representation $[4] \sim A_1$, whereas the 
three vector Jacobi bosons, $b^{\dagger}_{\rho}$, $b^{\dagger}_{\lambda}$ and $b^{\dagger}_{\eta}$, 
transform as the three components of the mixed symmetry representation $[31] \sim F_2$. 

The discrete symmetry $t$ of a given wave function can be determined by evaluating the matrix 
elements 
\ba
\left< \psi_t \right| P(12) \left| \psi_t \right> &=&  
\left< \psi_t \right| U_{\rm tr} \left| \psi_t \right> \;=\; \pm 1 ~, 
\nonumber\\
\left< \psi_t \right| P(1234) \left| \psi_t \right> &=&  
\left< \psi_t \right| U_{\rm cycl} \left| \psi_t \right> ~, 
\ea
and comparing with Eqs.~(\ref{trans4}) and (\ref{cyclic4}).

\section*{Acknowledgements}

This work was performed in part under U.S. Department of Energy Grant DE-FG02-91ER40608 
and in part by PAPIIT-DGAPA, UNAM Grant IN 109017. We wish to thank V. Della Rocca for 
preparing Figs.~\ref{Be8} and \ref{bands2}, and M. Gai for giving permission and providing 
files of Figs.~\ref{C12} and \ref{bands3}. 

\bibliography{RefCluster}

\begin{thebibliography}{98}
\expandafter\ifx\csname natexlab\endcsname\relax\def\natexlab#1{#1}\fi
\providecommand{\url}[1]{\texttt{#1}}
\providecommand{\href}[2]{#2}
\providecommand{\path}[1]{#1}
\providecommand{\DOIprefix}{doi:}
\providecommand{\ArXivprefix}{arXiv:}
\providecommand{\URLprefix}{URL: }
\providecommand{\Pubmedprefix}{pmid:}
\providecommand{\doi}[1]{\href{http://dx.doi.org/#1}{\path{#1}}}
\providecommand{\Pubmed}[1]{\href{pmid:#1}{\path{#1}}}
\providecommand{\bibinfo}[2]{#2}
\ifx\xfnm\relax \def\xfnm[#1]{\unskip,\space#1}\fi
%Type = Article
\bibitem[{Wheeler(1937)}]{Wheeler:1937zza}
\bibinfo{author}{J.~A. Wheeler}, \bibinfo{journal}{Phys. Rev.}
  \bibinfo{volume}{52} (\bibinfo{year}{1937}) \bibinfo{pages}{1083 -- 1106}.
  \DOIprefix\doi{10.1103/PhysRev.52.1083}.
%Type = Article
\bibitem[{Hafstad and Teller(1938)}]{PhysRev.54.681}
\bibinfo{author}{L.~R. Hafstad}, \bibinfo{author}{E.~Teller},
  \bibinfo{journal}{Phys. Rev.} \bibinfo{volume}{54} (\bibinfo{year}{1938})
  \bibinfo{pages}{681 -- 692}. \DOIprefix\doi{10.1103/PhysRev.54.681}.
%Type = Article
\bibitem[{Dennison(1954)}]{Dennison:1954zz}
\bibinfo{author}{D.~M. Dennison}, \bibinfo{journal}{Phys. Rev.}
  \bibinfo{volume}{96} (\bibinfo{year}{1954}) \bibinfo{pages}{378 -- 380}.
  \DOIprefix\doi{10.1103/PhysRev.96.378}.
%Type = Article
\bibitem[{Kameny(1956)}]{Kameny:1956zz}
\bibinfo{author}{S.~L. Kameny}, \bibinfo{journal}{Phys. Rev.}
  \bibinfo{volume}{103} (\bibinfo{year}{1956}) \bibinfo{pages}{358 -- 364}.
  \DOIprefix\doi{10.1103/PhysRev.103.358}.
%Type = Article
\bibitem[{Wildermuth and Kanellopoulos(1958)}]{WILDERMUTH1958150}
\bibinfo{author}{K.~Wildermuth}, \bibinfo{author}{T.~Kanellopoulos},
  \bibinfo{journal}{Nucl. Phys.} \bibinfo{volume}{7} (\bibinfo{year}{1958})
  \bibinfo{pages}{150 -- 162}. \DOIprefix\doi{10.1016/0029-5582(58)90245-1}.
%Type = Book
\bibitem[{Brink(1965)}]{Brink}
\bibinfo{author}{D.~M. Brink}, \bibinfo{title}{International School "Enrico
  Fermi", Course XXXVI}, \bibinfo{publisher}{SIF, Bologna},
  \bibinfo{year}{1965}.
%Type = Article
\bibitem[{Brink et~al.(1970)Brink, Friedrich, Weiguny, and
  Wong}]{Brink:1970ufk}
\bibinfo{author}{D.~M. Brink}, \bibinfo{author}{H.~Friedrich},
  \bibinfo{author}{A.~Weiguny}, \bibinfo{author}{C.~W. Wong},
  \bibinfo{journal}{Phys. Lett.} \bibinfo{volume}{33B} (\bibinfo{year}{1970})
  \bibinfo{pages}{143 -- 146}. \DOIprefix\doi{10.1016/0370-2693(70)90284-4}.
%Type = Article
\bibitem[{Horiuchi(1974)}]{10.1143/PTP.51.1266}
\bibinfo{author}{H.~Horiuchi}, \bibinfo{journal}{Prog. Theor. Phys.}
  \bibinfo{volume}{51} (\bibinfo{year}{1974}) \bibinfo{pages}{1266 -- 1268}.
  \DOIprefix\doi{10.1143/PTP.51.1266}.
%Type = Article
\bibitem[{Horiuchi(1975)}]{10.1143/PTP.53.447}
\bibinfo{author}{H.~Horiuchi}, \bibinfo{journal}{Prog. Theor. Phys.}
  \bibinfo{volume}{53} (\bibinfo{year}{1975}) \bibinfo{pages}{447 -- 460}.
  \DOIprefix\doi{10.1143/PTP.53.447}.
%Type = Article
\bibitem[{Ikeda et~al.(1980)Ikeda, Horiuchi, and Saito}]{10.1143/PTPS.68.1}
\bibinfo{author}{K.~Ikeda}, \bibinfo{author}{H.~Horiuchi},
  \bibinfo{author}{S.~Saito}, \bibinfo{journal}{Prog. Theor. Phys. Supp.}
  \bibinfo{volume}{68} (\bibinfo{year}{1980}) \bibinfo{pages}{1 -- 28}.
  \DOIprefix\doi{10.1143/PTPS.68.1}.
%Type = Article
\bibitem[{Fujiwara et~al.(1980)Fujiwara, Horiuchi, Ikeda, Kamimura, Kato,
  Suzuki, and Uegaki}]{10.1143/PTPS.68.29}
\bibinfo{author}{Y.~Fujiwara}, \bibinfo{author}{H.~Horiuchi},
  \bibinfo{author}{K.~Ikeda}, \bibinfo{author}{M.~Kamimura},
  \bibinfo{author}{K.~Kato}, \bibinfo{author}{Y.~Suzuki},
  \bibinfo{author}{E.~Uegaki}, \bibinfo{journal}{Prog. Theor. Phys. Supp.}
  \bibinfo{volume}{68} (\bibinfo{year}{1980}) \bibinfo{pages}{29 -- 192}.
  \DOIprefix\doi{10.1143/PTPS.68.29}.
%Type = Article
\bibitem[{Eichler and Faessler(1970)}]{Eichler:1970nqw}
\bibinfo{author}{J.~Eichler}, \bibinfo{author}{A.~Faessler},
  \bibinfo{journal}{Nucl. Phys. A} \bibinfo{volume}{157} (\bibinfo{year}{1970})
  \bibinfo{pages}{166 -- 176}. \DOIprefix\doi{10.1016/0375-9474(70)90106-5}.
%Type = Article
\bibitem[{Robson(1978)}]{Robson:1978vh}
\bibinfo{author}{D.~Robson}, \bibinfo{journal}{Nucl. Phys. A}
  \bibinfo{volume}{308} (\bibinfo{year}{1978}) \bibinfo{pages}{381 -- 428}.
  \DOIprefix\doi{10.1016/0375-9474(78)90558-4}.
%Type = Article
\bibitem[{Robson(1982)}]{ROBSON1982257}
\bibinfo{author}{D.~Robson}, \bibinfo{journal}{Prog. Part. Nucl. Phys.}
  \bibinfo{volume}{8} (\bibinfo{year}{1982}) \bibinfo{pages}{257 -- 299}.
  \DOIprefix\doi{10.1016/0146-6410(82)90011-4}.
%Type = Article
\bibitem[{Zhang and Rae(1993)}]{Rae_1993}
\bibinfo{author}{J.~Zhang}, \bibinfo{author}{W.~D.~M. Rae},
  \bibinfo{journal}{Nucl. Phys. A} \bibinfo{volume}{564} (\bibinfo{year}{1993})
  \bibinfo{pages}{252--270}. \DOIprefix\doi{10.1016/0375-9474(93)90520-8}.
%Type = Article
\bibitem[{Zhang et~al.(1994)Zhang, Rae, and Merchant}]{Rae_1994}
\bibinfo{author}{J.~Zhang}, \bibinfo{author}{W.~D.~M. Rae},
  \bibinfo{author}{A.~C. Merchant}, \bibinfo{journal}{Nucl. Phys. A}
  \bibinfo{volume}{575} (\bibinfo{year}{1994}) \bibinfo{pages}{61--71}.
  \DOIprefix\doi{10.1016/0375-9474(94)90137-6}.
%Type = Article
\bibitem[{von Oertzen et~al.(2006)von Oertzen, Freer, and
  Kanada-En’yo}]{VONOERTZEN200643}
\bibinfo{author}{W.~von Oertzen}, \bibinfo{author}{M.~Freer},
  \bibinfo{author}{Y.~Kanada-En’yo}, \bibinfo{journal}{Phys. Rep.}
  \bibinfo{volume}{432} (\bibinfo{year}{2006}) \bibinfo{pages}{43 -- 113}.
  \DOIprefix\doi{10.1016/j.physrep.2006.07.001}.
%Type = Article
\bibitem[{Schuck et~al.(2016)Schuck, Funaki, Horiuchi, Roepke, Tohsaki, and
  Yamada}]{Schuck_2016}
\bibinfo{author}{P.~Schuck}, \bibinfo{author}{Y.~Funaki},
  \bibinfo{author}{H.~Horiuchi}, \bibinfo{author}{G.~Roepke},
  \bibinfo{author}{A.~Tohsaki}, \bibinfo{author}{T.~Yamada},
  \bibinfo{journal}{Phys. Scr.} \bibinfo{volume}{91} (\bibinfo{year}{2016})
  \bibinfo{pages}{123001}. \DOIprefix\doi{10.1088/0031-8949/91/12/123001}.
%Type = Article
\bibitem[{Freer et~al.(2018)Freer, Horiuchi, Kanada-En'yo, Lee, and
  Mei\ss{}ner}]{RevModPhys.90.035004}
\bibinfo{author}{M.~Freer}, \bibinfo{author}{H.~Horiuchi},
  \bibinfo{author}{Y.~Kanada-En'yo}, \bibinfo{author}{D.~Lee},
  \bibinfo{author}{U.-G. Mei\ss{}ner}, \bibinfo{journal}{Rev. Mod. Phys.}
  \bibinfo{volume}{90} (\bibinfo{year}{2018}) \bibinfo{pages}{035004}.
  \DOIprefix\doi{10.1103/RevModPhys.90.035004}.
%Type = Article
\bibitem[{Freer and Fynbo(2014)}]{Freer:2014qoa}
\bibinfo{author}{M.~Freer}, \bibinfo{author}{H.~O.~U. Fynbo},
  \bibinfo{journal}{Prog. Part. Nucl. Phys.} \bibinfo{volume}{78}
  (\bibinfo{year}{2014}) \bibinfo{pages}{1 -- 23}.
  \DOIprefix\doi{10.1016/j.ppnp.2014.06.001}.
%Type = Article
\bibitem[{Freer et~al.(2007)}]{Freer:2007zz}
\bibinfo{author}{M.~Freer}, et~al., \bibinfo{journal}{Phys. Rev. C}
  \bibinfo{volume}{76} (\bibinfo{year}{2007}) \bibinfo{pages}{034320}.
  \DOIprefix\doi{10.1103/PhysRevC.76.034320}.
%Type = Article
\bibitem[{Kirsebom et~al.(2010)}]{Kirsebom:2010zz}
\bibinfo{author}{O.~S. Kirsebom}, et~al., \bibinfo{journal}{Phys. Rev. C}
  \bibinfo{volume}{81} (\bibinfo{year}{2010}) \bibinfo{pages}{064313}.
  \DOIprefix\doi{10.1103/PhysRevC.81.064313}.
%Type = Article
\bibitem[{Freer et~al.(2011)}]{Freer:2011zza}
\bibinfo{author}{M.~Freer}, et~al., \bibinfo{journal}{Phys. Rev. C}
  \bibinfo{volume}{83} (\bibinfo{year}{2011}) \bibinfo{pages}{034314}.
  \DOIprefix\doi{10.1103/PhysRevC.83.034314}.
%Type = Article
\bibitem[{Mar{\'{\i}}n-L\'ambarri et~al.(2014)Mar{\'{\i}}n-L\'ambarri, Bijker,
  Freer, Gai, Kokalova, Parker, and Wheldon}]{Marin-Lambarri:2014zxa}
\bibinfo{author}{D.~J. Mar{\'{\i}}n-L\'ambarri}, \bibinfo{author}{R.~Bijker},
  \bibinfo{author}{M.~Freer}, \bibinfo{author}{M.~Gai},
  \bibinfo{author}{T.~Kokalova}, \bibinfo{author}{D.~J. Parker},
  \bibinfo{author}{C.~Wheldon}, \bibinfo{journal}{Phys. Rev. Lett.}
  \bibinfo{volume}{113} (\bibinfo{year}{2014}) \bibinfo{pages}{012502}.
  \DOIprefix\doi{10.1103/PhysRevLett.113.012502}.
  \href{http://arxiv.org/abs/1405.7445}{\tt arXiv:1405.7445}.
%Type = Article
\bibitem[{Itoh et~al.(2011)}]{Itoh:2011zz}
\bibinfo{author}{M.~Itoh}, et~al., \bibinfo{journal}{Phys. Rev. C}
  \bibinfo{volume}{84} (\bibinfo{year}{2011}) \bibinfo{pages}{054308}.
  \DOIprefix\doi{10.1103/PhysRevC.84.054308}.
%Type = Article
\bibitem[{Freer et~al.(2012)}]{Freer:2012se}
\bibinfo{author}{M.~Freer}, et~al., \bibinfo{journal}{Phys. Rev. C}
  \bibinfo{volume}{86} (\bibinfo{year}{2012}) \bibinfo{pages}{034320}.
  \DOIprefix\doi{10.1103/PhysRevC.86.034320}.
%Type = Article
\bibitem[{Zimmerman et~al.(2013)}]{Zimmerman:2013cxa}
\bibinfo{author}{W.~R. Zimmerman}, et~al., \bibinfo{journal}{Phys. Rev. Lett.}
  \bibinfo{volume}{110} (\bibinfo{year}{2013}) \bibinfo{pages}{152502}.
  \DOIprefix\doi{10.1103/PhysRevLett.110.152502}.
  \href{http://arxiv.org/abs/1303.4326}{\tt arXiv:1303.4326}.
%Type = Article
\bibitem[{Kanada-En'yo(2007)}]{10.1143/PTP.117.655}
\bibinfo{author}{Y.~Kanada-En'yo}, \bibinfo{journal}{Prog. Theor. Phys.}
  \bibinfo{volume}{117} (\bibinfo{year}{2007}) \bibinfo{pages}{655 -- 680}.
  \DOIprefix\doi{10.1143/PTP.117.655}.
%Type = Article
\bibitem[{Chernykh et~al.(2007)Chernykh, Feldmeier, Neff, von Neumann-Cosel,
  and Richter}]{Chernykh:2007zz}
\bibinfo{author}{M.~Chernykh}, \bibinfo{author}{H.~Feldmeier},
  \bibinfo{author}{T.~Neff}, \bibinfo{author}{P.~von Neumann-Cosel},
  \bibinfo{author}{A.~Richter}, \bibinfo{journal}{Phys. Rev. Lett.}
  \bibinfo{volume}{98} (\bibinfo{year}{2007}) \bibinfo{pages}{032501}.
  \DOIprefix\doi{10.1103/PhysRevLett.98.032501}.
%Type = Article
\bibitem[{Funaki et~al.(2009)Funaki, Horiuchi, von Oertzen, Roepke, Schuck,
  Tohsaki, and Yamada}]{Funaki:2009fc}
\bibinfo{author}{Y.~Funaki}, \bibinfo{author}{H.~Horiuchi},
  \bibinfo{author}{W.~von Oertzen}, \bibinfo{author}{G.~Roepke},
  \bibinfo{author}{P.~Schuck}, \bibinfo{author}{A.~Tohsaki},
  \bibinfo{author}{T.~Yamada}, \bibinfo{journal}{Phys. Rev. C}
  \bibinfo{volume}{80} (\bibinfo{year}{2009}) \bibinfo{pages}{064326}.
  \DOIprefix\doi{10.1103/PhysRevC.80.064326}.
  \href{http://arxiv.org/abs/0912.2934}{\tt arXiv:0912.2934}.
%Type = Article
\bibitem[{Roth et~al.(2011)Roth, Langhammer, Calci, Binder, and
  Navratil}]{Roth:2011ar}
\bibinfo{author}{R.~Roth}, \bibinfo{author}{J.~Langhammer},
  \bibinfo{author}{A.~Calci}, \bibinfo{author}{S.~Binder},
  \bibinfo{author}{P.~Navratil}, \bibinfo{journal}{Phys. Rev. Lett.}
  \bibinfo{volume}{107} (\bibinfo{year}{2011}) \bibinfo{pages}{072501}.
  \DOIprefix\doi{10.1103/PhysRevLett.107.072501}.
  \href{http://arxiv.org/abs/1105.3173}{\tt arXiv:1105.3173}.
%Type = Article
\bibitem[{Navr\'atil et~al.(2000)Navr\'atil, Vary, and
  Barrett}]{PhysRevLett.84.5728}
\bibinfo{author}{P.~Navr\'atil}, \bibinfo{author}{J.~P. Vary},
  \bibinfo{author}{B.~R. Barrett}, \bibinfo{journal}{Phys. Rev. Lett.}
  \bibinfo{volume}{84} (\bibinfo{year}{2000}) \bibinfo{pages}{5728--5731}.
  \DOIprefix\doi{10.1103/PhysRevLett.84.5728}.
%Type = Article
\bibitem[{Maris(2012)}]{Maris_2012}
\bibinfo{author}{P.~Maris}, \bibinfo{journal}{J. Phys. Conf. Ser.}
  \bibinfo{volume}{402} (\bibinfo{year}{2012}) \bibinfo{pages}{012031}.
  \DOIprefix\doi{10.1088/1742-6596/402/1/012031}.
%Type = Article
\bibitem[{Epelbaum et~al.(2011)Epelbaum, Krebs, Lee, and
  Meissner}]{Epelbaum:2011md}
\bibinfo{author}{E.~Epelbaum}, \bibinfo{author}{H.~Krebs},
  \bibinfo{author}{D.~Lee}, \bibinfo{author}{U.-G. Meissner},
  \bibinfo{journal}{Phys. Rev. Lett.} \bibinfo{volume}{106}
  (\bibinfo{year}{2011}) \bibinfo{pages}{192501}.
  \DOIprefix\doi{10.1103/PhysRevLett.106.192501}.
  \href{http://arxiv.org/abs/1101.2547}{\tt arXiv:1101.2547}.
%Type = Article
\bibitem[{Epelbaum et~al.(2012)Epelbaum, Krebs, Lahde, Lee, and
  Meissner}]{Epelbaum:2012qn}
\bibinfo{author}{E.~Epelbaum}, \bibinfo{author}{H.~Krebs},
  \bibinfo{author}{T.~A. Lahde}, \bibinfo{author}{D.~Lee},
  \bibinfo{author}{U.-G. Meissner}, \bibinfo{journal}{Phys. Rev. Lett.}
  \bibinfo{volume}{109} (\bibinfo{year}{2012}) \bibinfo{pages}{252501}.
  \DOIprefix\doi{10.1103/PhysRevLett.109.252501}.
  \href{http://arxiv.org/abs/1208.1328}{\tt arXiv:1208.1328}.
%Type = Article
\bibitem[{Epelbaum et~al.(2014)Epelbaum, Krebs, L\"ahde, Lee, Mei\ss{}ner, and
  Rupak}]{PhysRevLett.112.102501}
\bibinfo{author}{E.~Epelbaum}, \bibinfo{author}{H.~Krebs},
  \bibinfo{author}{T.~A. L\"ahde}, \bibinfo{author}{D.~Lee},
  \bibinfo{author}{U.-G. Mei\ss{}ner}, \bibinfo{author}{G.~Rupak},
  \bibinfo{journal}{Phys. Rev. Lett.} \bibinfo{volume}{112}
  (\bibinfo{year}{2014}) \bibinfo{pages}{102501}.
  \DOIprefix\doi{10.1103/PhysRevLett.112.102501}.
%Type = Article
\bibitem[{Dreyfuss et~al.(2013)Dreyfuss, Launey, Dytrych, Draayer, and
  Bahri}]{Dreyfuss:2012us}
\bibinfo{author}{A.~C. Dreyfuss}, \bibinfo{author}{K.~D. Launey},
  \bibinfo{author}{T.~Dytrych}, \bibinfo{author}{J.~P. Draayer},
  \bibinfo{author}{C.~Bahri}, \bibinfo{journal}{Phys. Lett. B}
  \bibinfo{volume}{727} (\bibinfo{year}{2013}) \bibinfo{pages}{511 -- 515}.
  \DOIprefix\doi{10.1016/j.physletb.2013.10.048}.
  \href{http://arxiv.org/abs/1212.2255}{\tt arXiv:1212.2255}.
%Type = Article
\bibitem[{Bijker and Iachello(2000)}]{Bijker:2000fw}
\bibinfo{author}{R.~Bijker}, \bibinfo{author}{F.~Iachello},
  \bibinfo{journal}{Phys. Rev. C} \bibinfo{volume}{61} (\bibinfo{year}{2000})
  \bibinfo{pages}{067305}. \DOIprefix\doi{10.1103/PhysRevC.61.067305}.
  \href{http://arxiv.org/abs/nucl-th/0003007}{\tt arXiv:nucl-th/0003007}.
%Type = Article
\bibitem[{Bijker and Iachello(2002)}]{Bijker:2002ac}
\bibinfo{author}{R.~Bijker}, \bibinfo{author}{F.~Iachello},
  \bibinfo{journal}{Ann. Phys. (N.Y.)} \bibinfo{volume}{298}
  (\bibinfo{year}{2002}) \bibinfo{pages}{334 -- 360}.
  \DOIprefix\doi{10.1006/aphy.2002.6255}.
  \href{http://arxiv.org/abs/nucl-th/0203072}{\tt arXiv:nucl-th/0203072}.
%Type = Article
\bibitem[{Bijker and Iachello(2014)}]{Bijker:2014tka}
\bibinfo{author}{R.~Bijker}, \bibinfo{author}{F.~Iachello},
  \bibinfo{journal}{Phys. Rev. Lett.} \bibinfo{volume}{112}
  (\bibinfo{year}{2014}) \bibinfo{pages}{152501}.
  \DOIprefix\doi{10.1103/PhysRevLett.112.152501}.
  \href{http://arxiv.org/abs/1403.6773}{\tt arXiv:1403.6773}.
%Type = Article
\bibitem[{Bijker and Iachello(2017)}]{Bijker:2016bpb}
\bibinfo{author}{R.~Bijker}, \bibinfo{author}{F.~Iachello},
  \bibinfo{journal}{Nucl. Phys. A} \bibinfo{volume}{957} (\bibinfo{year}{2017})
  \bibinfo{pages}{154 -- 176}. \DOIprefix\doi{10.1016/j.nuclphysa.2016.08.008}.
  \href{http://arxiv.org/abs/1608.07487}{\tt arXiv:1608.07487}.
%Type = Article
\bibitem[{Kunz(1960)}]{KUNZ1960275}
\bibinfo{author}{P.~D. Kunz}, \bibinfo{journal}{Ann. Phys. (N.Y.)}
  \bibinfo{volume}{11} (\bibinfo{year}{1960}) \bibinfo{pages}{275 -- 305}.
  \DOIprefix\doi{10.1016/0003-4916(60)90111-1}.
%Type = Article
\bibitem[{Kunz(1962)}]{PhysRev.128.1343}
\bibinfo{author}{P.~D. Kunz}, \bibinfo{journal}{Phys. Rev.}
  \bibinfo{volume}{128} (\bibinfo{year}{1962}) \bibinfo{pages}{1343 -- 1351}.
  \DOIprefix\doi{10.1103/PhysRev.128.1343}.
%Type = Article
\bibitem[{Shimodaya and Hiura(1963)}]{10.1143/PTP.30.585}
\bibinfo{author}{I.~Shimodaya}, \bibinfo{author}{J.~Hiura},
  \bibinfo{journal}{Prog. Theor. Phys.} \bibinfo{volume}{30}
  (\bibinfo{year}{1963}) \bibinfo{pages}{585 -- 600}.
  \DOIprefix\doi{10.1143/PTP.30.585}.
%Type = Article
\bibitem[{Neudatchin and Smirnov(1969)}]{Neudatchin1969}
\bibinfo{author}{V.~C. Neudatchin}, \bibinfo{author}{Y.~F. Smirnov},
  \bibinfo{journal}{Prog. Nucl. Phys.} \bibinfo{volume}{10}
  (\bibinfo{year}{1969}) \bibinfo{pages}{275}.
%Type = Article
\bibitem[{Golovanova and Neudatchin(1971)}]{Golovanova1971}
\bibinfo{author}{N.~F. Golovanova}, \bibinfo{author}{V.~C. Neudatchin},
  \bibinfo{journal}{Sov. J. Nucl. Phys.} \bibinfo{volume}{13}
  (\bibinfo{year}{1971}) \bibinfo{pages}{718}.
%Type = Article
\bibitem[{Abe et~al.(1973)Abe, Hiura, and Tanaka}]{10.1143/PTP.49.800}
\bibinfo{author}{Y.~Abe}, \bibinfo{author}{J.~Hiura},
  \bibinfo{author}{H.~Tanaka}, \bibinfo{journal}{Prog. Theor. Phys.}
  \bibinfo{volume}{49} (\bibinfo{year}{1973}) \bibinfo{pages}{800 -- 824}.
  \DOIprefix\doi{10.1143/PTP.49.800}.
%Type = Article
\bibitem[{Okabe et~al.(1977)Okabe, Abe, and Tanaka}]{10.1143/PTP.57.866}
\bibinfo{author}{S.~Okabe}, \bibinfo{author}{Y.~Abe},
  \bibinfo{author}{H.~Tanaka}, \bibinfo{journal}{Prog. Theor. Phys.}
  \bibinfo{volume}{57} (\bibinfo{year}{1977}) \bibinfo{pages}{866 -- 881}.
  \DOIprefix\doi{10.1143/PTP.57.866}.
%Type = Article
\bibitem[{Okabe and Abe(1979)}]{10.1143/PTP.61.1049}
\bibinfo{author}{S.~Okabe}, \bibinfo{author}{Y.~Abe}, \bibinfo{journal}{Prog.
  Theor. Phys.} \bibinfo{volume}{61} (\bibinfo{year}{1979})
  \bibinfo{pages}{1049 -- 1064}. \DOIprefix\doi{10.1143/PTP.61.1049}.
%Type = Article
\bibitem[{Feldmeier and Schnack(2000)}]{Feldmeier:2000cn}
\bibinfo{author}{H.~Feldmeier}, \bibinfo{author}{J.~Schnack},
  \bibinfo{journal}{Rev. Mod. Phys.} \bibinfo{volume}{72}
  (\bibinfo{year}{2000}) \bibinfo{pages}{655 -- 688}.
  \DOIprefix\doi{10.1103/RevModPhys.72.655}.
  \href{http://arxiv.org/abs/cond-mat/0001207}{\tt arXiv:cond-mat/0001207}.
%Type = Article
\bibitem[{Roth et~al.(2004)Roth, Neff, Hergert, and Feldmeier}]{Roth:2004ua}
\bibinfo{author}{R.~Roth}, \bibinfo{author}{T.~Neff},
  \bibinfo{author}{H.~Hergert}, \bibinfo{author}{H.~Feldmeier},
  \bibinfo{journal}{Nucl. Phys. A} \bibinfo{volume}{745} (\bibinfo{year}{2004})
  \bibinfo{pages}{3 -- 33}. \DOIprefix\doi{10.1016/j.nuclphysa.2004.08.024}.
  \href{http://arxiv.org/abs/nucl-th/0406021}{\tt arXiv:nucl-th/0406021}.
%Type = Article
\bibitem[{Neff and Feldmeier(2004)}]{Neff:2003ib}
\bibinfo{author}{T.~Neff}, \bibinfo{author}{H.~Feldmeier},
  \bibinfo{journal}{Nucl. Phys. A} \bibinfo{volume}{738} (\bibinfo{year}{2004})
  \bibinfo{pages}{357 -- 361}. \DOIprefix\doi{10.1016/j.nuclphysa.2004.04.061}.
  \href{http://arxiv.org/abs/nucl-th/0312130}{\tt arXiv:nucl-th/0312130}.
%Type = Article
\bibitem[{Neff et~al.(2005)Neff, Feldmeier, and Roth}]{Neff:2005pvm}
\bibinfo{author}{T.~Neff}, \bibinfo{author}{H.~Feldmeier},
  \bibinfo{author}{R.~Roth}, \bibinfo{journal}{Nucl. Phys. A}
  \bibinfo{volume}{752} (\bibinfo{year}{2005}) \bibinfo{pages}{321 -- 324}.
  \DOIprefix\doi{10.1016/j.nuclphysa.2005.02.092}.
%Type = Article
\bibitem[{Kanada-En'yo and Horiuchi(2001)}]{10.1143/PTPS.142.205}
\bibinfo{author}{Y.~Kanada-En'yo}, \bibinfo{author}{H.~Horiuchi},
  \bibinfo{journal}{Prog. Theor. Phys. Supp.} \bibinfo{volume}{142}
  (\bibinfo{year}{2001}) \bibinfo{pages}{205 -- 263}.
  \DOIprefix\doi{10.1143/PTPS.142.205}.
%Type = Article
\bibitem[{Kanada-En'yo et~al.(2003)Kanada-En'yo, Kimura, and
  Horiuchi}]{KANADAENYO2003497}
\bibinfo{author}{Y.~Kanada-En'yo}, \bibinfo{author}{M.~Kimura},
  \bibinfo{author}{H.~Horiuchi}, \bibinfo{journal}{Comptes Rendus Physique}
  \bibinfo{volume}{4} (\bibinfo{year}{2003}) \bibinfo{pages}{497 -- 520}.
  \DOIprefix\doi{10.1016/S1631-0705(03)00062-8}.
%Type = Article
\bibitem[{Kanada-En'yo and Horiuchi(2003)}]{Kanada-Enyo:2003fhn}
\bibinfo{author}{Y.~Kanada-En'yo}, \bibinfo{author}{H.~Horiuchi},
  \bibinfo{journal}{Phys. Rev. C} \bibinfo{volume}{68} (\bibinfo{year}{2003})
  \bibinfo{pages}{014319}. \DOIprefix\doi{10.1103/PhysRevC.68.014319}.
  \href{http://arxiv.org/abs/nucl-th/0301059}{\tt arXiv:nucl-th/0301059}.
%Type = Article
\bibitem[{von Oertzen(1970)}]{vonOertzen:1970ecu}
\bibinfo{author}{W.~von Oertzen}, \bibinfo{journal}{Nucl. Phys. A}
  \bibinfo{volume}{148} (\bibinfo{year}{1970}) \bibinfo{pages}{529 -- 547}.
  \DOIprefix\doi{10.1016/0375-9474(70)90646-9}.
%Type = Article
\bibitem[{von Oertzen and Bohlen(1975)}]{VONOERTZEN19751}
\bibinfo{author}{W.~von Oertzen}, \bibinfo{author}{H.~Bohlen},
  \bibinfo{journal}{Phys. Rep.} \bibinfo{volume}{19} (\bibinfo{year}{1975})
  \bibinfo{pages}{1 -- 61}. \DOIprefix\doi{10.1016/0370-1573(75)90054-X}.
%Type = Article
\bibitem[{Imanishi and von Oertzen(1987)}]{IMANISHI198729}
\bibinfo{author}{B.~Imanishi}, \bibinfo{author}{W.~von Oertzen},
  \bibinfo{journal}{Phys. Rep.} \bibinfo{volume}{155} (\bibinfo{year}{1987})
  \bibinfo{pages}{29 -- 136}. \DOIprefix\doi{10.1016/0370-1573(87)90101-3}.
%Type = Article
\bibitem[{von Oertzen(1996)}]{Oertzen}
\bibinfo{author}{W.~von Oertzen}, \bibinfo{journal}{Z. Phys. A}
  \bibinfo{volume}{354} (\bibinfo{year}{1996}) \bibinfo{pages}{37}.
%Type = Article
\bibitem[{Della~Rocca et~al.(2017)Della~Rocca, Bijker, and
  Iachello}]{DellaRocca:2017qkx}
\bibinfo{author}{V.~Della~Rocca}, \bibinfo{author}{R.~Bijker},
  \bibinfo{author}{F.~Iachello}, \bibinfo{journal}{Nucl. Phys. A}
  \bibinfo{volume}{966} (\bibinfo{year}{2017}) \bibinfo{pages}{158 -- 184}.
  \DOIprefix\doi{10.1016/j.nuclphysa.2017.06.032}.
%Type = Article
\bibitem[{Della~Rocca and Iachello(2018)}]{DellaRocca:2018mrt}
\bibinfo{author}{V.~Della~Rocca}, \bibinfo{author}{F.~Iachello},
  \bibinfo{journal}{Nucl. Phys. A} \bibinfo{volume}{973} (\bibinfo{year}{2018})
  \bibinfo{pages}{1 -- 32}. \DOIprefix\doi{10.1016/j.nuclphysa.2018.02.003}.
%Type = Article
\bibitem[{Bijker(2016)}]{Bijker_2016}
\bibinfo{author}{R.~Bijker}, \bibinfo{journal}{Phys. Scr.} \bibinfo{volume}{91}
  (\bibinfo{year}{2016}) \bibinfo{pages}{073005}.
  \DOIprefix\doi{10.1088/0031-8949/91/7/073005}.
%Type = Article
\bibitem[{Nilsson(1955)}]{Nilsson:1955fn}
\bibinfo{author}{S.~G. Nilsson}, \bibinfo{journal}{Kong. Dan. Vid. Sel. Mat.
  Fys. Med.} \bibinfo{volume}{29} (\bibinfo{year}{1955}) \bibinfo{pages}{1 --
  69}.
%Type = Article
\bibitem[{Morse(1929)}]{Morse_1929}
\bibinfo{author}{P.~M. Morse}, \bibinfo{journal}{Phys. Rev.}
  \bibinfo{volume}{34} (\bibinfo{year}{1929}) \bibinfo{pages}{57--64}.
  \DOIprefix\doi{10.1103/PhysRev.34.57}.
%Type = Book
\bibitem[{Iachello and Levine(1995)}]{IachelloLevine}
\bibinfo{author}{F.~Iachello}, \bibinfo{author}{R.~D. Levine},
  \bibinfo{title}{Algebraic Theory of Molecules}, \bibinfo{publisher}{Oxford
  University Press}, \bibinfo{year}{1995}.
%Type = Article
\bibitem[{Iachello(1981)}]{IACHELLO1981581}
\bibinfo{author}{F.~Iachello}, \bibinfo{journal}{Chem. Phys. Lett.}
  \bibinfo{volume}{78} (\bibinfo{year}{1981}) \bibinfo{pages}{581 -- 585}.
  \DOIprefix\doi{10.1016/0009-2614(81)85262-1}.
%Type = Book
\bibitem[{Iachello(1994)}]{kamran1994lie}
\bibinfo{author}{F.~Iachello}, \bibinfo{title}{in Lie Algebras, Cohomology, and
  New Applications to Quantum Mechanics}, Contemporary mathematics 160, Eds. N.
  Kamran and P. Olver, \bibinfo{publisher}{American Mathematical Society},
  \bibinfo{year}{1994}.
%Type = Article
\bibitem[{Kramer and Moshinsky(1966)}]{KRAMER1966241}
\bibinfo{author}{P.~Kramer}, \bibinfo{author}{M.~Moshinsky},
  \bibinfo{journal}{Nucl. Phys.} \bibinfo{volume}{82} (\bibinfo{year}{1966})
  \bibinfo{pages}{241 -- 274}. \DOIprefix\doi{10.1016/0029-5582(66)90001-0}.
%Type = Book
\bibitem[{Herzberg(1991)}]{herzberg1989molecular}
\bibinfo{author}{G.~Herzberg}, \bibinfo{title}{Molecular Spectra and Molecular
  Structure Volume III - Electronic Structure of Polyatomic Molecules},
  \bibinfo{publisher}{Krieger Publishing Company}, \bibinfo{year}{1991}.
%Type = Book
\bibitem[{Wilson et~al.(1955)Wilson, Cross, and Decius}]{wilson1955molecular}
\bibinfo{author}{E.~B. Wilson}, \bibinfo{author}{P.~C. Cross},
  \bibinfo{author}{J.~C. Decius}, \bibinfo{title}{Molecular Vibrations},
  \bibinfo{publisher}{McGraw-Hill}, \bibinfo{year}{1955}.
%Type = Article
\bibitem[{Bijker(shed)}]{RB}
\bibinfo{author}{R.~Bijker}, \bibinfo{journal}{{Computer programs ACM and
  TDMOL}}  (\bibinfo{year}{unpublished}).
%Type = Article
\bibitem[{Sick and McCarthy(1970)}]{Sick:1970ma}
\bibinfo{author}{I.~Sick}, \bibinfo{author}{J.~S. McCarthy},
  \bibinfo{journal}{Nucl. Phys. A} \bibinfo{volume}{150} (\bibinfo{year}{1970})
  \bibinfo{pages}{631 -- 654}. \DOIprefix\doi{10.1016/0375-9474(70)90423-9}.
%Type = Article
\bibitem[{Tilley et~al.(2004)Tilley, Kelley, Godwin, Millener, Purcell, Sheu,
  and Weller}]{Tilley:2004zz}
\bibinfo{author}{D.~R. Tilley}, \bibinfo{author}{J.~H. Kelley},
  \bibinfo{author}{J.~L. Godwin}, \bibinfo{author}{D.~J. Millener},
  \bibinfo{author}{J.~E. Purcell}, \bibinfo{author}{C.~G. Sheu},
  \bibinfo{author}{H.~R. Weller}, \bibinfo{journal}{Nucl. Phys. A}
  \bibinfo{volume}{745} (\bibinfo{year}{2004}) \bibinfo{pages}{155 -- 362}.
  \DOIprefix\doi{10.1016/j.nuclphysa.2004.09.059}.
%Type = Article
\bibitem[{Barker et~al.(1968)Barker, Hay, and Treacy}]{Barker_1968}
\bibinfo{author}{F.~C. Barker}, \bibinfo{author}{H.~J. Hay},
  \bibinfo{author}{P.~B. Treacy}, \bibinfo{journal}{Aust. J. Phys.}
  \bibinfo{volume}{21} (\bibinfo{year}{1968}) \bibinfo{pages}{239--258}.
  \DOIprefix\doi{10.1071/PH680239}.
%Type = Article
\bibitem[{Barker(1969)}]{Barker_1969}
\bibinfo{author}{F.~C. Barker}, \bibinfo{journal}{Aust. J. Phys.}
  \bibinfo{volume}{22} (\bibinfo{year}{1969}) \bibinfo{pages}{293--316}.
  \DOIprefix\doi{10.1071/PH690293}.
%Type = Article
\bibitem[{Bijker and D{\'{\i}}az-Caballero(2017)}]{Bijker_2017}
\bibinfo{author}{R.~Bijker}, \bibinfo{author}{O.~A. D{\'{\i}}az-Caballero},
  \bibinfo{journal}{Phys. Scr.} \bibinfo{volume}{92} (\bibinfo{year}{2017})
  \bibinfo{pages}{124001}. \DOIprefix\doi{10.1088/1402-4896/aa9242}.
%Type = Article
\bibitem[{Datar et~al.(2013)}]{Datar:2013pbd}
\bibinfo{author}{V.~M. Datar}, et~al., \bibinfo{journal}{Phys. Rev. Lett.}
  \bibinfo{volume}{111} (\bibinfo{year}{2013}) \bibinfo{pages}{062502}.
  \DOIprefix\doi{10.1103/PhysRevLett.111.062502}.
  \href{http://arxiv.org/abs/1305.1094}{\tt arXiv:1305.1094}.
%Type = Article
\bibitem[{Kelley et~al.(2017)Kelley, Purcell, and Sheu}]{KELLEY201771}
\bibinfo{author}{J.~Kelley}, \bibinfo{author}{J.~Purcell},
  \bibinfo{author}{C.~Sheu}, \bibinfo{journal}{Nucl. Phys. A}
  \bibinfo{volume}{968} (\bibinfo{year}{2017}) \bibinfo{pages}{71 -- 253}.
  \DOIprefix\doi{10.1016/j.nuclphysa.2017.07.015}.
%Type = Article
\bibitem[{Tilley et~al.(1993)Tilley, Weller, and Cheves}]{TILLEY19931}
\bibinfo{author}{D.~Tilley}, \bibinfo{author}{H.~Weller},
  \bibinfo{author}{C.~Cheves}, \bibinfo{journal}{Nucl. Phys. A}
  \bibinfo{volume}{564} (\bibinfo{year}{1993}) \bibinfo{pages}{1 -- 183}.
  \DOIprefix\doi{10.1016/0375-9474(93)90073-7}.
%Type = Article
\bibitem[{Iachello(2011)}]{FI}
\bibinfo{author}{F.~Iachello}, \bibinfo{journal}{Riv. Nuovo Cimento}
  \bibinfo{volume}{34} (\bibinfo{year}{2011}) \bibinfo{pages}{617 -- 642}.
%Type = Article
\bibitem[{Morinaga(1956)}]{PhysRev.101.254}
\bibinfo{author}{H.~Morinaga}, \bibinfo{journal}{Phys. Rev.}
  \bibinfo{volume}{101} (\bibinfo{year}{1956}) \bibinfo{pages}{254 -- 258}.
  \DOIprefix\doi{10.1103/PhysRev.101.254}.
%Type = Article
\bibitem[{Caprio et~al.(2008)Caprio, Cejnar, and Iachello}]{CAPRIO20081106}
\bibinfo{author}{M.~Caprio}, \bibinfo{author}{P.~Cejnar},
  \bibinfo{author}{F.~Iachello}, \bibinfo{journal}{Ann. Phys. (N.Y.)}
  \bibinfo{volume}{323} (\bibinfo{year}{2008}) \bibinfo{pages}{1106 -- 1135}.
  \DOIprefix\doi{10.1016/j.aop.2007.06.011}.
%Type = Article
\bibitem[{Larese and Iachello(2011)}]{Larese_2011}
\bibinfo{author}{D.~Larese}, \bibinfo{author}{F.~Iachello},
  \bibinfo{journal}{J. Mol. Struct.} \bibinfo{volume}{1006}
  (\bibinfo{year}{2011}) \bibinfo{pages}{611--628}.
  \DOIprefix\doi{10.1016/j.molstruc.2011.10.016.}
%Type = Article
\bibitem[{Larese et~al.(2013)Larese, P\'erez-Bernal, and
  Iachello}]{Larese_2013}
\bibinfo{author}{D.~Larese}, \bibinfo{author}{F.~P\'erez-Bernal},
  \bibinfo{author}{F.~Iachello}, \bibinfo{journal}{J. Mol. Struct.}
  \bibinfo{volume}{1051} (\bibinfo{year}{2013}) \bibinfo{pages}{310--327}.
  \DOIprefix\doi{10.1016/j.molstruc.2013.08.020}.
%Type = Article
\bibitem[{Volkov(1965)}]{Volkov:1965zz}
\bibinfo{author}{A.~B. Volkov}, \bibinfo{journal}{Nucl. Phys.}
  \bibinfo{volume}{74} (\bibinfo{year}{1965}) \bibinfo{pages}{33 -- 58}.
  \DOIprefix\doi{10.1016/0029-5582(65)90244-0}.
%Type = Article
\bibitem[{Andersen et~al.(1970)Andersen, Dickmann, and
  Dietrich}]{Andersen:1970tov}
\bibinfo{author}{B.~L. Andersen}, \bibinfo{author}{F.~Dickmann},
  \bibinfo{author}{K.~Dietrich}, \bibinfo{journal}{Nucl. Phys. A}
  \bibinfo{volume}{159} (\bibinfo{year}{1970}) \bibinfo{pages}{337 -- 366}.
  \DOIprefix\doi{10.1016/0375-9474(70)90712-8}.
%Type = Article
\bibitem[{Scharnweber et~al.(1971)Scharnweber, Greiner, and
  Mosel}]{Scharnweber:1971qpn}
\bibinfo{author}{D.~Scharnweber}, \bibinfo{author}{W.~Greiner},
  \bibinfo{author}{U.~Mosel}, \bibinfo{journal}{Nucl. Phys. A}
  \bibinfo{volume}{164} (\bibinfo{year}{1971}) \bibinfo{pages}{257 -- 278}.
  \DOIprefix\doi{10.1016/0375-9474(71)90212-0}.
%Type = Book
\bibitem[{Koster(1963)}]{koster1963properties}
\bibinfo{author}{G.~F. Koster}, \bibinfo{title}{Properties of the thirty-two
  point groups}, Massachusetts Institute of Technology Press Research
  Monograph, \bibinfo{publisher}{M.I.T. Press}, \bibinfo{year}{1963}.
%Type = Article
\bibitem[{Bijker and Iachello(2019)}]{BI}
\bibinfo{author}{R.~Bijker}, \bibinfo{author}{F.~Iachello},
  \bibinfo{journal}{arXiv:1902.00451}  (\bibinfo{year}{2019}).
%Type = Book
\bibitem[{Hamermesh(1964)}]{hamermesh1964group}
\bibinfo{author}{M.~Hamermesh}, \bibinfo{title}{Group Theory and Its
  Application to Physical Problems}, \bibinfo{publisher}{Addison-Wesley},
  \bibinfo{year}{1964}.
%Type = Book
\bibitem[{Preston and Bhaduri(1975)}]{nla.cat-vn298793}
\bibinfo{author}{M.~A. Preston}, \bibinfo{author}{R.~K. Bhaduri},
  \bibinfo{title}{Structure of the nucleus}, \bibinfo{publisher}{Addison-Wesley
  Pub. Co., Advanced Book Program, Reading, Mass.}, \bibinfo{year}{1975}.
%Type = Book
\bibitem[{Shalit and Talmi(1963)}]{shalit1963nuclear}
\bibinfo{author}{A.~Shalit}, \bibinfo{author}{I.~Talmi},
  \bibinfo{title}{Nuclear Shell Theory}, Pure and Applied Physics,
  \bibinfo{publisher}{Academic Press}, \bibinfo{year}{1963}.
%Type = Book
\bibitem[{Brussaard and Glaudemans(1977)}]{brussaard1977shell}
\bibinfo{author}{P.~J. Brussaard}, \bibinfo{author}{P.~W.~M. Glaudemans},
  \bibinfo{title}{Shell Model Applications in Nuclear Spectroscopy},
  \bibinfo{publisher}{Elsevier}, \bibinfo{year}{1977}.
%Type = Article
\bibitem[{De~Forest and Walecka(1966)}]{DeForest:1966ycn}
\bibinfo{author}{T.~De~Forest, Jr.}, \bibinfo{author}{J.~D. Walecka},
  \bibinfo{journal}{Adv. Phys.} \bibinfo{volume}{15} (\bibinfo{year}{1966})
  \bibinfo{pages}{1 -- 109}. \DOIprefix\doi{10.1080/00018736600101254}.
%Type = Article
\bibitem[{Glickman et~al.(1991)Glickman, Bertozzi, Buti, Dixit, Hersman,
  Hyde-Wright, Hynes, Lourie, Norum, Kelly, Berman, and
  Millener}]{PhysRevC.43.1740}
\bibinfo{author}{J.~P. Glickman}, \bibinfo{author}{W.~Bertozzi},
  \bibinfo{author}{T.~N. Buti}, \bibinfo{author}{S.~Dixit},
  \bibinfo{author}{F.~W. Hersman}, \bibinfo{author}{C.~E. Hyde-Wright},
  \bibinfo{author}{M.~V. Hynes}, \bibinfo{author}{R.~W. Lourie},
  \bibinfo{author}{B.~E. Norum}, \bibinfo{author}{J.~J. Kelly},
  \bibinfo{author}{B.~L. Berman}, \bibinfo{author}{D.~J. Millener},
  \bibinfo{journal}{Phys. Rev. C} \bibinfo{volume}{43} (\bibinfo{year}{1991})
  \bibinfo{pages}{1740 -- 1757}. \DOIprefix\doi{10.1103/PhysRevC.43.1740}.
%Type = Article
\bibitem[{Ajzenberg-Selove(1991)}]{AJZENBERGSELOVE19911}
\bibinfo{author}{F.~Ajzenberg-Selove}, \bibinfo{journal}{Nucl. Phys. A}
  \bibinfo{volume}{523} (\bibinfo{year}{1991}) \bibinfo{pages}{1 -- 196}.
  \DOIprefix\doi{10.1016/0375-9474(91)90446-D}.
%Type = Article
\bibitem[{Millener et~al.(1989)Millener, Sober, Crannell, O'Brien, Fagg,
  Kowalski, Williamson, and Lapik\'as}]{PhysRevC.39.14}
\bibinfo{author}{D.~J. Millener}, \bibinfo{author}{D.~I. Sober},
  \bibinfo{author}{H.~Crannell}, \bibinfo{author}{J.~T. O'Brien},
  \bibinfo{author}{L.~W. Fagg}, \bibinfo{author}{S.~Kowalski},
  \bibinfo{author}{C.~F. Williamson}, \bibinfo{author}{L.~Lapik\'as},
  \bibinfo{journal}{Phys. Rev. C} \bibinfo{volume}{39} (\bibinfo{year}{1989})
  \bibinfo{pages}{14 -- 46}. \DOIprefix\doi{10.1103/PhysRevC.39.14}.

\end{thebibliography}

%\clearpage

%\tableofcontents
%\listoffigures
%\listoftables

\end{document}